\documentclass[letterpaper,12pt]{article}
\usepackage{amssymb, amsmath}
\usepackage{amsthm}
\usepackage{algorithm}
\usepackage{algorithmic}
\usepackage{color}
\usepackage{pstricks-add}
\usepackage{pstricks}
\usepackage{url}
\usepackage{subfigure}
\usepackage{a4}
\newtheorem{theorem}{Theorem}
\newtheorem{corollary}[theorem]{Corollary}
\newtheorem{fact}[theorem]{Fact}
\newtheorem{lemma}[theorem]{Lemma}

\newcommand{\comment}[1]{}

\begin{document}
\title{On irreversible dynamic monopolies in general graphs
%Bounding the number of tolerable faults in majority-based systems
%On
%the minimum size of
%irreversible
%dynamos
%dynamic monopolies
%in general graphs
\footnote{The authors are supported in part by the National
Science Council of Taiwan under grant
97-2221-E-002-096-MY3 and Excellent
Research Projects of National Taiwan University under grant 98R0062-05.}
}

\author{
Ching-Lueh Chang
\footnote{Department of Computer Science and Information Engineering, National Taiwan University, Taipei, Taiwan. Email: d95007@csie.ntu.edu.tw}
\and Yuh-Dauh Lyuu
\footnote{Department of Computer Science and Information Engineering, National Taiwan University, Taipei, Taiwan. Email: lyuu@csie.ntu.edu.tw}
\footnote{Department of Finance, National Taiwan University, Taipei, Taiwan.}
}

\maketitle

\begin{abstract}
%Let $G(V,E)$ be a simple directed graph,
%$\phi^\text{strict}(v)=\lceil\rceil$
Consider the
following
coloring process
in
%played on
a
%network ${\cal N}(G,\phi)$
%composed
%consisting
%of a
simple directed graph $G(V,E)$ with positive indegrees.
%and a function $\phi:V\to \mathbb{N}$ with
%$1\le\phi(v)\le \text{deg}^\text{in}(v),$ $v\in V$.
Initially, a set $S$ of vertices are white, whereas all the others are black.
Thereafter,
%Subsequently,
a black vertex is colored white whenever
%the
%majority
more than half
of its
in-neighbors are white.
%A black vertex without in-neighbors
%can never be colored white.
%remains black forever.
The coloring process ends when no additional vertices can be colored white.
If all vertices
end up
%turn out
%are
white,
%at the end,
we
call
%say that
$S$
%is
an irreversible
%dynamo
dynamic monopoly (or dynamo for short)
%of
%$G$
under the strict-majority scenario.
%${\cal N}(G,\phi)$.
%Similarly, 
An irreversible dynamo under the simple-majority scenario is defined
similarly except that a black vertex is colored white when at least half
of its in-neighbors are white.
%Following Chang and Lyuu~\cite{CL08},
We derive upper bounds of $(2/3)\,|\,V\,|$ and $|\,V\,|/2$
on the minimum sizes of irreversible dynamos
under the
strict and
the
simple-majority scenarios,
%of ${\cal N}(G,\phi^\text{strict})$ and ${\cal N}(G,\phi^\text{simple})$
respectively.
%The bounds hold whenever $G$ has a positive minimum indegree.
For the special case when $G$ is
an undirected connected
graph,
% without isolated vertices,
we prove the existence of an irreversible dynamo with size at most
$\lceil |\,V\,|/2 \rceil$
under the strict-majority scenario.
%upper bounds of $\lceil |\,V\,|/2 \rceil$ and $\lfloor |\,V\,|/2 \rfloor$ are given on the minimum sizes of irreversible dynamos under the strict and the simple-majority scenarios, respectively.
%The $(2/3)\,|\,V\,|$ upper bound is further refined
%to $(d+2)\,|\,V\,|/(2d+2)$
%for $(2d)$-regular graphs.
%Under either the strict or simple majority scenario,
Let $\epsilon>0$ be any constant.
We also show that,
unless $\text{NP}\subseteq \text{TIME}(n^{O(\ln \ln n)}),$
%Finally,
%we show that
no
polynomial-time,
%$o(\ln |\,V\,|)$-approximation
$((1/2-\epsilon)\ln |\,V\,|)$-approximation
algorithms exist for
finding the minimum irreversible dynamo under either the strict or
the simple-majority scenario.
%, unless $\text{NP}\subseteq \text{TIME}(n^{O(\ln \ln n)})$.
%\comment{
The inapproximability results hold even for
bipartite
graphs with diameter
at most $8$.
%}
%$O(1)$.
\end{abstract}

\section{Introduction}
%Let $G(V,E)$ be a simple directed graph (or digraph for short)~\cite{Wes01}.
Let $G(V,E)$ be a simple directed graph (or digraph for
short) with positive indegrees.
%and $S\subseteq V$.
A simple undirected graph is interpreted as a directed one
where each edge is accompanied by the edge in the opposite direction.
In this paper, all
%directed and undirected
graphs are simple
and have positive indegrees.
%Suppose that all vertices in $S$ are initially white, whereas
%all the others are black.
%Consider
%This paper considers
%In this paper, we consider
%the following coloring process.
The following coloring process extends that of Flocchini et al.~\cite{FGS01}
by taking digraphs into consideration.
Initially, all vertices in a set $S\subseteq V$ are white, whereas all the others are black.
Thereafter, a black vertex is colored white when more than half of its
in-neighbors are white.
%A black vertex without in-neighbors remains black forever.
The coloring
process
proceeds asynchronously
until no additional vertices can be colored white.
If all vertices end up white, then $S$ is called an irreversible
dynamo under the strict-majority
%scenario~\cite{Pel02}.
scenario.
An irreversible dynamo under the simple-majority scenario is defined
similarly except that a black vertex is colored white
when at least half of its in-neighbors are white.
Tight or nearly tight bounds on
the minimum size of irreversible dynamos
are known
when $G$ is
a
toroidal mesh~\cite{FLLPS04, PZ05}, torus cordalis, torus
serpentinus~\cite{FLLPS04},
butterfly, wrapped butterfly, cube-connected cycle,
hypercube, DeBruijn, shuffle-exchange,
complete tree, ring~\cite{FKRRS03, LPS99}
and
chordal ring~\cite{FGS01}.

% the paragraph is now scattered

% kind of hard to describe

Chang and Lyuu~\cite{CL09}
%are the first to
%derive
show that
%every digraph
%simple directed graph
$G(V,E)$
%with a positive minimum indegree
has an irreversible dynamo of size at most $(23/27)\,|\,V\,|$
under the strict-majority scenario.
%which is the first bound that holds for all simple undirected
%graphs without isolated vertices.
%Sensibly,
%Their $(23/27)\,|\,V\,|$ bound also applies for every undirected graph
%$G$ without isolated vertices.
%including all those in the previous paragraph.
%In fact,
%all graphs in the previous paragraph are simple undirected
%ones without isolated vertices.
%When every connected component of $G=(V,E)$ is an undirected triangle,
%it is easy to see that the minimum size of irreversible dynamos of $G$ has
%size $(2/3)\,|\,V\,|$ under the strict-majority scenario.
%Hence the $23/27$ constant in Chang and Lyuu's bound cannot be improved to
%below $2/3$.
%This paper improves
%In this paper, we improve
%Chang and Lyuu's
This paper
%We
improves their
$(23/27)\,|\,V\,|$ bound
%by showing that every
%digraph
%with a positive minimum in-degree has an irreversible dynamo of size
%at most $0.7732\,|\,V\,|$ under the strict-majority scenario.
to
%only
$(2/3)\,|\,V\,|$.
Moreover,
if $G$ is undirected and connected, our
$(2/3)\,|\,V\,|$ upper
bound can be further lowered
to
%only
%$(2/3)\,|\,V\,|$.
$\lceil|\,V\,|/2\rceil$.
%When every connected component of $G(V,E)$ is an undirected triangle,
%it is easy to see that the minimum size of irreversible dynamos of $G$
%is
%precisely
%has size
%exactly
%$(2/3)\,|\,V\,|$ under the strict-majority scenario.
%Hence the $2/3$ constant in our $(2/3)\,|\,V\,|$ bound can no more be improved.
%We also refine the $(2/3)\,|\,V\,|$ bound to
%$(d+2)\,|\,V\,|/(2d+2)$ for $(2d)$-regular graphs.
Under the simple-majority scenario,
%our $0.7732\,|\,V\,|$ and $(2/3)\,|\,V\,|$ bounds are lowered to
%$0.727\,|\,V\,|$ and $|\,V\,|/2,$ respectively.
%we derive
%upper bounds of $0.727\,|\,V\,|$ and $|\,V\,|/2$ for the minimum size of
%irreversible dynamos, assuming that $G$ is a digraph with a positive minimum
%in-degree a
we show that every digraph
%with a positive minimum indegree
has an
irreversible dynamo of size at most $|\,V\,|/2$.
%Moreover, every undirected graph $G(V,E)$ has an irreversible dynamo of size at most $\lfloor |\,V\,|/2\rfloor$ under the simple-majority scenario.
%Again, the $1/2$ constant cannot be lowered because the minimum size of
%irreversible dynamos is exactly $|\,V\,|/2$ for
%complete undirected graphs $G(V,E)$ with an even number of vertices.
%In fault-tolerant computing,
In the literature on fault-tolerant computing,
%a processor is assumed to become faulty
%when the strict or the simple majority
%of its neighboring processors are faulty.
%So
an irreversible dynamo
is interpreted as
%models
a set of
processors
%entities
whose
faulty behavior
%failure
leads all
%entities
processors
to
erroneous results~\cite{FGS01, FKRRS03, FLLPS04, LPS99, Pel02}.
%to behave incorrectly
%in majority-based
%systems~\cite{FGS01, FKRRS03, FLLPS04, LPS99, Pel02}.
Under this interpretation,
%Therefore,
%So
our upper bounds
%put limits on
limit
the number of
%arbitrarily
adversarially
%(and perhaps maliciously)
placed faulty
processors
that any system can guarantee to
tolerate without inducing erroneous results on all processors.
%For example, consider
%our result that every undirected graph $G(V,E)$ without
%isolated vertices has an irreversible dynamo of size $|\,V\,|/2$.
%It can be interpreted as saying that no systems represented by $G$
%can guarantee to tolerate an arbitrary set of $|\,V\,|/2$
%faulty processors without inducing erroneous results on all processors,
%assuming the simple-majority.
%For example, the $\lceil |\,V\,|/2\rceil$ bound implies that it is
%impossible to tolerate faulty behavior from any set of more than half of the
%processors.
\comment{ % omit justifying why we don't consider isolated vertices
The assumption for an undirected $G$ to have no isolated vertices
is justifiable because isolated vertices can be considered separately.
For the directed case,
the assumption that $G$ has a positive minimum indegree
is also reasonable because
it is obeyed by many graphs.
} % omit justifying why we don't consider isolated vertices

Under several randomized mechanisms for coloring the vertices,
%Using the theory of submodularity,
Kempe, Kleinberg and Tardos~\cite{KKT03, KKT05} and Mossel and Roch~\cite{MR07}
show $(1-(1/e)-\epsilon)$-approximation algorithms
for allocating a given number of seeds to color the most vertices white,
where $\epsilon>0$ is an arbitrary constant.
%and inapproximability results
%for
%several
%a large class of
%related problems.
Kempe, Kleinberg and Tardos~\cite{KKT03}
%They
also show inapproximability results for allocating seeds
%to color the most vertices white
in digraphs
to color the most vertices white.
%where each vertex has an arbitrary threshold
%But our results are independent of theirs.
%But their approach does not apply to our problem.
This paper considers
%We also consider
the
related
computational
problem of finding
a minimum irreversible dynamo given an undirected graph,
%Our problem
which
arises naturally because an extensive literature has been
investigating
the minimum size of irreversible dynamos~\cite{FGS01, FKRRS03,
FLLPS04, LPS99, Pel02}.
%an irreversible dynamo of the minimum size.
%Given a graph $G(V,E),$
We show that,
unless $\text{NP}\subseteq \text{TIME}(n^{O(\ln \ln n)}),$
%we show that
no
polynomial-time,
%$o(\ln |\,V\,|)$-approximation
$((1/2-\epsilon)\ln |\,V\,|)$-approximation
algorithms exist for
the minimum irreversible dynamo,
%assuming
%under
either
under
the strict
%and
or
the
simple-majority
scenario.
The inapproximability results hold even for
bipartite
graphs
with diameter
at most $8$.
%$O(1)$.
%i.e., bipartite
%graphs where every two vertices
%are connected by a path of length $O(1)$.
%, unless $\text{NP}\subseteq \text{TIME}(n^{O(\ln \ln n)})$.
%We do so by
%reducing the problem of finding the minimum dominating set
%to that of finding the minimum irreversible dynamo.
%Then
%showing that any approximation algorithm for the minimum irreversible dynamo
%can be used to give an approximation algorithm for the minimum dominating
%set~\cite{Pap94}.
%To do so,
%\comment{ % no need?
In proving our inapproximability results,
we make use of Feige's~\cite{Fei98} famous result on the inapproximability
%of finding the minimum dominating set of an undirected graph.
of finding
a minimum dominating set
%a dominating set of the minimum size
in an undirected graph.
%} % no need?

Variants on the coloring process appear in the literature.
%statistical physics.
%The coloring process is similar to Watts'~\cite{Wat02} model on
%global cascades, which is suitable for describing the decision-making
%processes in many social and economical systems.
%Watts' model is more general in that
%vertices can adopt different thresholds
%each vertex
%The differences are that
%Watts~\cite{Wat02} argues that individuals in social and economical systems
%typically take their actions based on the fraction of their neighboring
%individuals choosing that action.
%\comment{ % quite detailed description of Watts' work
Given two alternative actions,
Watts~\cite{Wat02} argues that an individual in a social or economical
system typically
chooses an
%decides whether to adopt a particular
alternative
based on the fraction of
%its
the
neighboring individuals
%choosing
adopting
%that
%alternative.
it.
%Based on this, Watts proposes a model of global cascades.
%, which differs from
%our coloring process in two aspects.
%First, the vertices in Watts' model do not always use 
%Watts models the individuals and the links between them by
%a sparse, undirected and random graph.
%In Watts' model,
%there is a sparse, undirected and random graph.
Watts' model assumes a sparse, undirected and random graph.
%and
%There is also
%a random set of vertices are initially white in a sparse, undirected, random
%graph.
%All the other vertices are black in the beginning.
%In the model,
%Unlike in this paper,
There is
also
a random variable distributed in $[\,0,1\,],$
from which every vertex
independently
draws a
%threshold for itself.
ratio.
Initially, a
uniformly
random set of vertices are
%colored
white,
leaving
%whereas
all the
others
%are
black.
%each
%vertex holds
%a
%random
%threshold.
%that is frozen after being drawn from a probability distribution.
%drawn
%randomly
%for each vertex.
%from a certain probability distribution.
%Initially, a random set of vertices are colored white, whereas all the others are
%black.
%Then
%At any time,
Thereafter,
a black vertex
becomes white
%if
when
the fraction of its white
neighbors exceeds
%its threshold.
the above ratio.
Finally,
the coloring process ends when no additional vertices can be colored white.
%So Watts' model is more general than ours in that different vertices may
%have different thresholds.
%In Watts' model, the vertices do not necessarily use a majority-
%However, we consider arbitrary digraphs,
%whereas Watts considers undirected, sparse, random graphs.
%The underlying graph in Watts' model are undirected, sparse and random.
%Watts focuses on
%how
%the characteristics of the underlying graph
Watts gives theoretical and numerical results on
the fraction of white
vertices at the end.
%to be susceptible to large cascades, i.e., 
Gleeson and Cahalane~\cite{GC07} extend Watts' work by
%giving
deriving
an analytical
solution for the fraction of white vertices at the end
%, assuming that
%a uniformly random set of vertices are initially white
%the graph exhibits a tree-like structure.
%in a tree-like graph.
in tree-like graphs.
%} % quite detailed description of Watts' work
Samuelsson and Socolar~\cite{SS06}
study
%the effects of
%a
%uniformly
%random set of initially white vertices under
a more general
process
%setting
called
the unordered binary avalanche,
which allows
%a wider range of
coloring mechanisms
beyond
%besides
the threshold-driven ones.
%rules for coloring.
%However,
%we do not assume the initially white vertices to distribute
%randomly among the graph.
Unlike the works mentioned
%in this paragraph,
above,
we do not assume that the initially white vertices are uniformly and
randomly distributed.

This paper is organized as follows.
Section~\ref{definitionssection} gives the definitions.
Sections~\ref{directedsection}--\ref{undirectedsection}
%Sections~\ref{strictmajoritysection}--\ref{simplemajoritysection}
present upper bounds on the minimum
size of irreversible dynamos
%under the strict
%and the simple-majority scenarios, respectively.
for directed and undirected graphs, respectively.
%Section~\ref{simplemajoritysection}
Section~\ref{inapproximabilitysection} presents inapproximability results on
finding minimum irreversible dynamos.

\section{Definitions}\label{definitionssection}
%Let $G(V,E)$ be a simple digraph.
Let $G(V,E)$ be a simple directed graph (or digraph for short)~\cite{Wes01}
with positive indegrees.
For $v\in V,$
we denote by $N^\text{in}(v)\subseteq V\setminus \{v\}$
%and
%(resp.,
%$N^\text{out}(v)\subseteq V\setminus \{v\}$
%)
the set of vertices
incident on an edge coming into
%(resp., going from)
$v$.
Similarly, $N^\text{out}(v)\subseteq V\setminus \{v\}$ is
the set of vertices incident on an edge going from $v$.
%$N^\text{out}(v)\subseteq V\setminus \{v\}$.
%Then we
Define $\text{deg}^\text{in}(v)=|\,N^\text{in}(v)\,|$ and
$\text{deg}^\text{out}(v)=|\,N^\text{out}(v)\,|$ as the indegree and outdegree
of $v,$ respectively.
For $X,Y\subseteq V,$ we write $e(X,Y)=|\,(X\times Y)\cap E\,|,$ i.e., the
number of edges going from a vertex in $X$ to one in $Y$.
%A vertex $v$ is called a source if $\text{deg}^\text{in}(v)=0$.
An undirected graph is a directed one
%where
with
%where
%each pair of edges in opposite
%directions simultaneously exist
%each
every
edge
%is bidirectional.
%is
accompanied by
%the
an
edge in the opposite direction.
For a vertex $v$ of an undirected graph,
%both $\text{deg}^\text{in}(v)$ and
%$\text{deg}^\text{out}(v)$ are denoted $\text{deg}(v)$.
%Furthermore,
we
%let
define
%$\text{deg}(v)=\text{deg}^\text{in}(v)=\text{deg}^\text{out}(v)$
$\text{deg}(v)=\text{deg}^\text{in}(v)$
and
%$N(v)=N^\text{in}(v)=N^\text{out}(v)$.
$N(v)=N^\text{in}(v)$ without loss of generality.
Furthermore,
%we
%write
define
%For convenience, denote
$N^*(v)=N(v)\cup \{v\}$.
For
%an undirected connected graph $G(V,E)$ and
%any two vertices $x,y\in V,$ let $d(x,y)$ be their
any two vertices $x$ and $y$ of an undirected connected graph,
let $d(x,y)$ be their
%their
distance,
i.e., the number of edges on a shortest path
%from $x$ to $y$.
between $x$ and $y$.
For any $v\in V$ and nonempty $U\subseteq V,$
denote $d(v,U)=\min_{u\in U}\, d(v,u)$ for convenience.
%for the set containing $v$ and $v$'s neighbors.
%Clearly,
%an equivalent definition says
%$\text{deg}(v)=\text{deg}^\text{out}(v)$ and $N(v)=N^\text{out}(v)$.
%are also true.
%$\text{deg}(v)$ denotes both $\text{deg}^\text{in}(v)$ and
%$\text{deg}^\text{out}(v)$.
%Furthermore, $N(v)$ is written for both $N^\text{in}(v)$ and
%$N^\text{out}(v)$.
For any $V^\prime\subseteq V,$ the subgraph
%$G^\prime$
of $G$ induced by $V^\prime$
is
%given
denoted
by
$G[\,V^\prime\,]=(V^\prime, E\cap (V^\prime\times V^\prime))$.
That is, $G[\,V^\prime\,]$
has
%the set
%the vertices in
%$V^\prime$
%of vertices
%and
all the edges in $E$
with both endpoints in $V^\prime$.
For
%clarity,
emphasis,
we may sometimes write
%$N^\text{in}(v)$ and
%$N^\text{out}(v)$ as $N_G^\text{in}(v)$ and $N_G^\text{out}(v),$
$N_G^\text{in}(v),$
%and
$N_G^\text{out}(v)$
and $N_G^*(v)$
for $N^\text{in}(v),$
%and
$N^\text{out}(v)$
and $N^*(v),$
respectively.
Similarly, we may write $\text{deg}_G^\text{in}(v),$
%and
$\text{deg}_G^\text{out}(v)$
and $d_G(x,y)$
for $\text{deg}^\text{in}(v),$
%and
$\text{deg}^\text{out}(v)$
and $d(x,y),$
respectively.
%to explicitly name the graph referred to.
All
%By default,
%directed and undirected
graphs
in this paper
are
%all
%assumed to be
simple
and have positive indegrees.
%Hereafter, graphs are all simple.
%Hereafter, digraphs refer to simple directed graphs.
%adjacent
%to $v$ and $\text{deg}(v)\equiv |N(v)|$.
%A function $\phi:V\to \mathbb{N}$ with $1\le \phi(v)\le
%\text{deg}^\text{in}(v)$ controls the extent to which

A network
${\cal N}(G,\phi)$
consists of
%is modeled by
a digraph $G(V,E)$ with positive indegrees and
a function $\phi:V\to \mathbb{N}$.
% with $1\le \phi(v)\le
%\text{deg}^\text{in}(v)$ if $\text{deg}^\text{in}(v)>0$ and $\phi(v)=\infty$
%otherwise.
%Hereafter, digraphs refer to simple directed graphs.
%We are interested in the following coloring process in ${\cal N}(G,\phi)$.
The coloring process in ${\cal N}(G,\phi)$ proceeds asynchronously.
Initially, a set $S\subseteq V$ of
vertices, called the seeds,
%seeds
are white whereas all the others
%vertices
%in $V\setminus S$
are black.
Thereafter,
a vertex $v$ becomes white when at least $\phi(v)$ of the vertices in
$N^\text{in}(v)$ are white.
The coloring process ends when no additional vertices can be colored white.
%We are interested in the minimum number of vertices, denoted
%$\text{min-seed}(G,\phi)$
%When many vertices can be colored white at the same time,
%it does not matter which is colored first for all the vertices in $c()$
%Sensibly, a non-seed vertex $v$ with $\text{deg}^\text{in}(v)=0$ remains
%black throughout the process.
Let $c(S,G,\phi)\subseteq V$ be the set of vertices that are white at the end
given that $S$ is the
%initial
set of seeds.
Define $\text{min-seed}(G,\phi)=\min_{U\subseteq V,
c(U,G,\phi)=V}\, |\,U\,|,$ namely, the minimum number of seeds needed to color
all vertices white at the end.
Clearly,
%when many vertices can be colored white at the same time,
it does not matter
%which is colored first
%for the vertices in $c(S,G,\phi)$ to become white eventually.
in what sequences the vertices are colored white
as they will end up with the same $c(S,G,\phi)$.
\comment{ % maybe write it as a Fact
Clearly, if
%Note that for
$S\subseteq T$ and $U\subseteq c(S,G,\phi),$
then
$c(S,G,\phi)\subseteq c(T,G,\phi)$ and $c(S,G,\phi)=c(S\cup U,G,\phi)$.
} % maybe write it as a Fact

We are
%particularly
interested
in
%when
$\phi$
being
%is
one of the following functions:
%the strict-majority and
\begin{itemize}
%\addtolength{\itemsep}{-0.2\baselineskip}
%\addtolength{\itemsep}{-0.5\baselineskip}
\item Strict majority: $\phi^\text{strict}(v)=\lceil(\text{deg}^\text{in}(v)+1)/2\rceil$;
%if $\text{deg}^\text{in}(v)>0$ and $\phi^\text{strict}(v)=\infty$ otherwise;
%$\phi_G^\text{strict}(v)=\lceil(\text{deg}^\text{in}(v)+1)/2\rceil,$
so a vertex $v$
%with $\text{deg}^\text{in}(v)>0$
is colored white when more than half of the vertices in
$N^\text{in}(v)$ are white.
\item Simple majority: $\phi^\text{simple}(v)=\lceil\text{deg}^\text{in}(v)/2\rceil$;
%if $\text{deg}^\text{in}(v)>0$ and $\phi^\text{simple}(v)=\infty$ otherwise;
%$\phi_G^\text{non-strict}(v)=\lceil\text{deg}^\text{in}(v)/2\rceil,$
so a vertex $v$
%with $\text{deg}^\text{in}(v)>0$
is colored white when at least half of the vertices in
$N^\text{in}(v)$ are white.
\end{itemize}
%For convenience, we say that
%the coloring process is under the strict-majority (resp., simple-majority)
%scenario
%if $\phi_G^\text{strict}$ (resp., $\phi_G^\text{simple}).
A set $S\subseteq V$
%with $c(S,G,\phi^\text{strict})=V$
is
called
%said to be
an
irreversible dynamic monopoly (or irreversible dynamo for short)
of
%$G$
${\cal N}(G,\phi^\text{strict})$
%under the strict-majority scenario
if $c(S,G,\phi^\text{strict})=V$~\cite{Pel02}.
Similarly,
it
%$S\subseteq V$
is an irreversible dynamo of
%$G$
${\cal N}(G,\phi^\text{simple})$
%under the simple-majority scenario
if $c(S,G,\phi^\text{simple})=V$.
We may sometimes
write $\phi_G^\text{strict}$ and $\phi_G^\text{simple}$
instead of $\phi^\text{strict}$ and $\phi^\text{simple}$
%when it is needed for clarity.
to emphasize the role of $G$.
\comment{ % maybe no need to mention his terms
In Peleg's~\cite{Pel02} terms,
$\phi^\text{strict}$
%$\phi_G^\text{strict}$
is
called
the $(\text{PW}, \text{SN})$ model
and
$\phi^\text{simple}$
%$\phi_G^\text{non-strict}$
the $(\text{PB}, \text{SN})$ model.
%The coloring process with $\phi^\text{strict}$
%$\phi^\text{non-strict}$
%is called (PS,SN) and (PW,SI) models for irreversible dynamos
} % maybe no need to mention his terms

%We are interested in four special cases on $\phi$ that are studied
%extensively in the literature~\ref{FLLS98, LPS99, FGS01, Pel02}.
%In the literature,
%the case where $\phi(v)$ is near $\text{deg}^{in}(v)/2$
%has received special attention.
%the coloring process is studied extensively in undirected graphs,
%assuming
%Much of the literature assumes
\comment{ % Found that PB,SN is the same as PW,SI for irreversible dynamos!
We are interested when $\phi$ is one of the following four functions.
\begin{enumerate}
%\addtolength{\itemsep}{-0.2\baselineskip}
\item $\phi^\text{PW,SN}(v)=\lceil(\text{deg}^\text{in}(v))/2\rceil$.
\item $\phi^\text{PB,SN}(v)=\lceil(\text{deg}^\text{in}(v)+1)/2\rceil$.
\item $\phi^\text{PW,SI}(v)=\lceil(\text{deg}^\text{in}(v)+1)/2\rceil$.
%\item $\phi^\text{PB,SI}(v)=\lceil(\text{deg}^\text{in}(v)+2)/2\rceil$.
\end{enumerate}
} % Found that PB,SN is the same as PW,SI for irreversible dynamos!
\comment{ % omit the story because it seems to confuse the readers
In much of the literature,
%that
a vertex of a simple undirected graph
is colored white when the majority of its neighbors are white.
But the exact definitions differ in
%whether majority means the strict one
%and
whether the neighbors of a vertex $v$ includes $v$ itself
and whether majority is the strict one.
Peleg~\cite{Pel02} describes these choices below.
\begin{enumerate}
%\addtolength{\itemsep}{-0.2\baselineskip}
\item In the Self-Not-included (SN) model, the neighbors of a vertex $v$
excludes $v$ itself.
\item In the Self-Included (SI) model, the neighbors of a vertex $v$
includes $v$ itself.
\item In the Prefer-White (PW) model,
a vertex becomes white when at least half of its neighbors are white.
\item In the Prefer-Black model (PW), a vertex becomes white when more than half
of its neighbors are white.
\end{enumerate}
Together, four models are described
\begin{enumerate}
\item $\phi^\text{PW,SN}(v)=$
\end{enumerate}
%Four special cases on $\phi$ are extensively studied in the literature.
%For convenience, we define
%$V_G^{[2]}\equiv \{v\in V\mid \text{deg}_G(v)=2\}$
%and write $G^{[2]}\equiv (V_G^{[2]}, E\cap (V_G^{[2]}\times V_G^{[2]}))$ for the
%subgraph of $G$ induced by $V_G^{[2]}$.
%by $\text{deg}_G(v)$
%the degree of $v$ in $G$ by $\text{deg}_G(v)$.
%The degree of a vertex $v\in V$ is denoted $\text{deg}(v)$.
} % omit the story because it seems to confuse the readers

Given an undirected graph $G(V,E),$
the problem
{\sc irreversible dynamo (strict majority)} asks for
%the minimum
%an
%irreversible dynamo
%of the minimum size
a minimum irreversible dynamo
under the strict-majority scenario.
Similarly, {\sc irreversible dynamo (simple majority)} asks for
%the
one
under the simple-majority scenario.
An $\ell$-approximation algorithm for each of the above problems
outputs an irreversible dynamo with size at most $\ell$ times the minimum.
%The {\sc dominating set} problem asks for a minimum set of vertices
A dominating set of an undirected graph $G(V,E)$ is a set of vertices
sharing at least one vertex with
%with nonempty intersection with
$N_G^*(v)$ for each $v\in V$~\cite{Wes01}.
Given an undirected graph $G(V,E),$
an $\ell$-approximation algorithm for the {\sc dominating set} problem
outputs a dominating set of $G$ with size at most $\ell$ times the minimum.
Recall that
%As usual in complexity theory,
an algorithm is said to run in polynomial
time
if its running time is
%at most
polynomial in the length of its
input~\cite{Pap94}.

\comment{ % maybe just say this in plain English
The following result is due to Chang and Lyuu~\cite{CL09}.
\begin{fact}(\cite{CL09})\label{statedinCL08upper}
%(\cite{CL08})
%Let $G(V,E)$ be connected and $|V|\ge 2$.
%There exists a polynomial-time algorithm that computes a $(23/27)|V|$-size
%self-ignoring monopoly with strict majority.
%Then a self-ignoring monopoly of size
%at most $(23/27)\, |V|$ can be computed in time polynomial in $|V|$.
%For every strongly connected digraph $G(V,E)$ with $|\,V\,|\ge 2,$
%$$\text{min-seed}(G,\phi^\text{strict})\le (23/27)\, |\,V\,|.$$
Let $G(V,E)$ be a strongly connected digraph with $|\,V\,|\ge 2$.
Then
%an irreversible dynamo
$$\text{min-seed}\left(G,\phi^\text{strict}\right)\le \frac{23\cdot |\,V\,|}{27}.$$
%Every strongly connected digraph $G(V,E)$ with $|\,V\,|\ge 2$
%has an irreversible dynamo of size at most $(23/27)\, |\,V\,|$
%under the
%strict-majority scenario.
\end{fact}
} % maybe just say this in plain English

% no need to derive weaker results

%In fact, the $(23/27)\,|\,V\,|$ upper bound can be restored by taking
%$\phi=\phi^\text{strict}$ and $p=2/3$ in Fact~\ref{CL08detailed}~\cite{CL08}.
%We will sometimes use the following straightforward fact.
The following fact is straightforward.

\begin{fact}\label{monotone}
For any network ${\cal N}(G(V,E),\phi)$ and
any $S,T\subseteq V,$
%$S,T\subseteq V,$
$$c\left(S,G,\phi\right)\subseteq c\left(S\cup T,G,\phi\right).$$
\end{fact}

%An immediate corollary follows from Fact~\ref{CL08detailed}.
%\begin{corollary}(\cite{CL08})\label{CL08SIMON}
%Given a graph $G(V,E)$ without isolated vertices,
%a self-ignoring monopoly of size at most
%$$\frac{2\cdot |V|}{3} + \frac{5\cdot \left|\{v\in V\mid
%\text{deg}_G(v)=2\}\right|}{27}
%+ \frac{11\cdot\left|\{v\in V\mid \text{deg}_G(v)\neq 2\}\right|}{81}$$
%can be computed
%by an algorithm ALG
%in time polynomial in $|V|$.
%\end{corollary}
%The algorithm in Corollary~\ref{CL08SIMON} will be denoted ALG.

\section{Irreversible dynamos of directed graphs}
\label{directedsection}

%upper bounds on the sizes of
%Let $k$ be a positive integer, $G$ be a simple digraph with positive
%indegrees and $w:E\to{\mathbb{R}}^+$.
%For
%a digraph $G$ with positive indegrees,
%$w:E\to{\mathbb{R}}^+,$
%a positive integer $k,$
%a partition of $V$ into $k+1$ sets
%$V_1,\ldots,V_{k+1},$
%$V=\bigcup_{i=1}^{k+1} V_i$.
%$V=\dotcup_{i=1}^{k+1} V_i$.
%$V=\bigdotcup_{i=1}^{k+1} V_i$.
%$V={\mathaccent{\cdot}{\bigcup}}$.
%$V={\mathaccent\cdot\bigcup}_{i=1}^{k+1} V_i$.
%$V={\dot\bigcup}_{i=1}^{k+1} V_i$.
%$V=\sqcup_{i=1}^{k+1} V_i$.
%$V=\bigsqcup_{i=1}^{k+1} V_i$.

Let $G(V,E)$ be a
%simple
digraph with positive indegrees,
%$w:E\to{\mathbb{R}}^+,$
%and
$k$ be a positive integer and $\phi_{k/(k+1)}(v)\equiv
\text{deg}^\text{in}(v)\cdot k/(k+1)$.
%We are interested in
This section derives upper bounds on
$\text{min-seed}(G,\phi_{k/(k+1)})$.
% where $\phi_{k/(k+1)}(v)$.
As corollaries, we obtain upper bounds on the minimum sizes of irreversible
dynamos under the strict and the simple-majority scenarios.
%For simplicity, we write $k/(k+1)$ instead of $\phi_{k/(k+1)}$.
%For convenience,
%As $\phi_{k/(k+1)}\equiv k/(k+1),$
%we will simply write $k/(k+1)$ for $\phi_{k/(k+1)}$ throughout this
%section.
%Let
For a partition
% and
%$(V_1,\ldots,V_{k+1})$ be a partition of $V$.
$V=\bigcup_{i=1}^{k+1}\, V_i$
% be a partition.
of $V,$
%into $k+1$ sets.
% and $i\in [\,k+1\,]$.
define
\begin{eqnarray*}
%\mu\left(G,w,V_1,\ldots,V_{k+1},i\right)
%&\equiv&
%\left|\,\, c\left(V\setminus V_i, G, w, \phi_{k/(k+1)}
%\frac{k}{k+1}\right)
%\,\,\right|,\\
\eta\left(G,V_1,\ldots,V_{k+1}\right)
&\equiv&
\sum_{i=1}^{k+1}
\,
%\mu\left(G,w,V_1,\ldots,V_{k+1},i\right).
\left|\, c\left(V\setminus V_i, G,
%\frac{k}{k+1}
\phi_{k/(k+1)}
\right)\,\right|.
%\left|\, c\left(\cup_{j\in [\,k+1\,] \setminus \{i\}} V_j, G, w,
%\frac{k}{k+1}
%\phi_{k/(k+1)}\right) \cap V_i \,\right|.
\end{eqnarray*}

%Clearly, $\mu(G,w,V_1,\ldots,V_{k+1},i)=|\,V\,|$
%if and only if $V_i\subseteq c(\cup_{j\in [\,k+1\,]\setminus \{i\}} V_j, G,
%w, k/(k+1))$.
%Furthermore, $\psi(G,w,V_1,\ldots,V_{k+1})=(k+1)\,|\,V\,|$ if and only if
%$\mu(G,w,V_1,\ldots,V_{k+1},i)=|\,V\,|$ for all $i\in [\,k+1\,]$.

An easy lemma follows.

\begin{lemma}\label{fullpotentialcondition}
Let $G$ be a
%simple
digraph with positive indegrees,
%$w:E\to{\mathbb{R}}^+,$
$k$ be a positive integer and
%$(V_1,\ldots,V_{k+1})$ be a partition of $V$.
$V=\bigcup_{i=1}^{k+1}\, V_i$ be a partition.
% into $k+1$ sets.
Then the
following conditions are equivalent:
\begin{enumerate}
\item~\label{equivalent1} $\eta(G,V_1,\ldots,V_{k+1})=(k+1)\,|\,V\,|$.
\item~\label{equivalent2} $c(V\setminus V_i,G,\phi_{k/(k+1)})=V$ for all
$i\in [\,k+1\,]$.
\item~\label{equivalent3} $V_i\subseteq c(V\setminus V_i, G,
\phi_{k/(k+1)})$ for all $i\in [\,k+1\,]$.
\end{enumerate}
%$\eta(G,w,V_1,\ldots,V_{k+1})=(k+1)\,|\,V\,|$ if and only if
%$V_i\subseteq c(\cup_{j\in [\,k+1\,]\setminus \{i\}} V_j, G,
%w, k/(k+1))$ for all $i\in [\,k+1\,]$.
\end{lemma}
\begin{proof}
%Clearly, $\eta(G,w,V_1,\ldots,V_{k+1})=(k+1)\,|\,V\,|$ if and only if
%$\mu(G,w,V_1,\ldots,V_{k+1},i)=|\,V\,|$ for all $i\in [\,k+1\,]$.
Items~\ref{equivalent1}--\ref{equivalent2} are
%clearly
equivalent
by noting that $|\,c(V\setminus V_i,G,\phi_{k/(k+1)})\,|\leq |\,V\,|$ for
each
$i\in [\,k+1\,]$.
As $V\setminus V_i
\subseteq
c(V\setminus V_i, G, \phi_{k/(k+1)}),$
$c(V\setminus V_i,G, \phi_{k/(k+1)})=V$
if and only if
$V_i\subseteq c(V\setminus V_i, G,
\phi_{k/(k+1)})$.
% for $i\in [\,k+1\,]$.
\end{proof}

%Clearly, $\eta(G,w,V_1,\ldots,V_{k+1})=|\,V\,|$
%if and only if

The next lemma allows us to iteratively modify
a partition of $V$ until one with
$\eta(G,V_1,\ldots,V_{k+1})=(k+1)\,|\,V\,|$
is obtained.

\begin{lemma}\label{mainlemma}
Let $k$ be a positive integer.
Given a
%simple
digraph $G$ with positive indegrees
% $w:E\to{\mathbb{R}}^+$
and
a partition $V=\bigcup_{i=1}^{k+1}\, V_i$
%of $V$ into
%$k+1$
%sets $V_1,\ldots,V_{k+1}$
with $\eta(G,V_1,\ldots,V_{k+1})<(k+1)\,|\,V\,|,$
%such that $V_i\not\subseteq
%c(\cup_{j\in \{1,\ldots,k+1\}\setminus \{i\}}, G, w, \phi_{k/(k+1)}),$
a
%new
partition
%of $V$ into
%$k+1$
%sets $V^\prime_1,\ldots,V^\prime_{k+1}$
$V=\bigcup_{i=1}^{k+1}\, V^\prime_i$
%can be found in polynomial time such that
satisfying
\begin{eqnarray*}
\eta\left(G,V^\prime_1,\ldots,V^\prime_{k+1}\right)
>\eta\left(G,V_1,\ldots,V_{k+1}\right)
%\label{mainlemmainequality}
\end{eqnarray*}
can be found in polynomial time.
\end{lemma}
\begin{proof}
By
the equivalence of
%items~\ref{equivalent1} and \ref{equivalent3} of
Lemma~\ref{fullpotentialcondition}(\ref{equivalent1})~and~(\ref{equivalent3}),
%As $\eta(G,w,V_1,\ldots,V_{k+1})<(k+1)\, |\,V\,|,$
there exists
%Find in polynomial time
an $i^*\in [\,k+1\,]$ with
%$c(\cup_{j\in [\,k+1\,] \setminus \left\{i^*\right\}}, G, w,
% \phi_{k/(k+1)})\neq V$.
%Equivalently, as $\cup_{j\in [\,k+1\,] \setminus \{i^*\}} V_j
%\subseteq
%c(\cup_{j\in [\,k+1\,] \setminus \{i^*\}}, G, w, \phi_{k/(k+1)})\neq V,$
\begin{eqnarray*}
V_{i^*}\not\subseteq c\left(V\setminus V_{i^*}, G,
%\frac{k}{k+1}
\phi_{k/(k+1)}
\right).
%\label{selectionofparttodepriveavertex}
\end{eqnarray*}
%whose existence is guaranteed by
%the equivalence of
%items~\ref{equivalent1}
%and \ref{equivalent3} of
%Lemma~\ref{fullpotentialcondition}(\ref{equivalent1})
%and~\ref{equivalent3}).
Take any
%$v\in V_{i^*}\setminus c(V\setminus V_{i^*}, G, w, \phi_{k/(k+1)})$.
$$v\in V_{i^*}\setminus c\left(V\setminus V_{i^*}, G,
%\frac{k}{k+1}
\phi_{k/(k+1)}
\right).$$
%Clearly, (one of the choices of) $i^*$ and $v$ can be found in polynomial
% time by computing
%$c(V\setminus V_i, G, w, \phi_{k/(k+1)})$ for all $i\in [\,k+1\,]$.
% so
%Then
%because every vertex in $V\setminus V_{i^*}$ must be active
%with $V\setminus V_{i^*}$ as the set of seeds,
%by
%in polynomial time.
Clearly,
%Now
%By
%the activation process and
%the picking of $u,$
%and the activation process,
%As $u\notin c(V\setminus V_{i^*}, G, w, \phi_{k/(k+1)}),$
\begin{eqnarray*}
\left|\,N^\text{in}(v)\cap\left(V\setminus
V_{i^*}\right)\,\right|
< \frac{k}{k+1}\cdot |\,V\,|.
%\sum_{(u,v)\in E, u\in V\setminus V_{i^*}}\, w\left((u,v)\right)
%< \frac{k}{k+1}\cdot \sum_{(u,v)\in E}\, w\left((u,v)\right)
%\label{selectionofdeprivedvertex}
\end{eqnarray*}
%by the activation process.
%which
%Then
%Now
%find
%in polynomial time
This and the fact that
%$(V_1,\ldots,V_{i^*-1},V_{i^*+1},\ldots,V_{k+1})$
%is a partition of $V\setminus V_{i^*}$
%into $k$ sets
$V\setminus V_{i^*}=\bigcup_{i\in [\,k+1\,]\setminus \{i^*\}}\, V_i$
is a partition of $V\setminus V_{i^*}$ into $k$ sets
%imply
show
%implies
the existence of
a $j^*\in [\,k+1\,]\setminus \{i^*\}$ with
$|\,N^\text{in}(v)\cap V_{j^*}\,| < (1/(k+1))\,|\,V\,|$.
%\begin{eqnarray*}
%\sum_{(u,v)\in E, u\in V_{j^*}}\, w\left((u,v)\right)
%< \frac{1}{k+1}\cdot \sum_{(u,v)\in
%E}\, w\left((u,v)\right).
%\end{eqnarray*}
%whose existence follows from inequality~(\ref{selectionofdeprivedvertex})
%and the fact that
%$V\setminus V_{i^*}=\bigcup_{i\in [\,k+1\,]\setminus \{i^*\}}\, V_i$
%is a partition of $V\setminus V_{i^*}$ into $k$ sets.
%Clearly, $j^*$ can be found in polynomial time by evaluating
%$\sum_{(u,v)\in E}\, w((u,v))$ and $\sum_{(u,v)\in E, u\in V_j}\, w((u,v))$
%for all $j\in [\,k+1\,],$ followed by checking.
Equivalently,
%to inequality~(\ref{thedestinationpart}),
\begin{eqnarray}
%\left|\,N^\text{in}(u)\cap V_{j^*}\,\right| < \frac{|\,V\,|}{k+1}.
\left|\,N^\text{in}(v)\setminus V_{j^*}\,\right| > \frac{k}{k+1}\cdot
% |\,V\,|.
%\sum_{(u,v)\in E, u\in V\setminus V_{j^*}}\, w\left((u,v)\right)
%>
%\frac{k}{k+1}\cdot \sum_{(u,v)\in E}\, w\left((u,v)\right).
\label{fewneighborspart}
\end{eqnarray}
%for some $j^*\in [\,k+1\,]$.
%Clearly, a triple $(i^*, u, j^*)\in [\,k+1\,]\times V\times[\,k+1\,]$
% satisfying
%inequalities~(\ref{selectionofparttodepriveavertex})--(\ref{fewneighborspart})
%can be found in polynomial time
%by trying out all the possible $(k+1)\cdot|\,V\,|\cdot(k+1)$ combinations.
%Clearly, $i^*, v$ and $j^*$ can be found in polynomial time by trying out
%all their possible values in $[\,k+1\,],$ $V$ and $[\,k+1\,]$
%and verifying
%inequalities~(\ref{selectionofparttodepriveavertex})--(\ref{fewneighborspart}).
%\comment{ % incorporate above the fact that their choice are easy
Clearly,
%(one of the choices of)
$i^*$ and $v$ can be found in polynomial time
by calculating
$c(V\setminus V_i, G, \phi_{k/(k+1)})$ for all $i\in [\,k+1\,]$.
Then $j^*$ can be found
in polynomial time
by
%after
evaluating
$|\,N^\text{in}(v)\cap V_j\,|$
%$\sum_{(u,v)\in E, u\in V_j}\, w((u,v))$ and
%$\sum_{(u,v)\in E}\, w((u,v))$ and
%$\sum_{(u,v)\in E, u\in V_j}\, w((u,v))$
for all $j\in [\,k+1\,]$.
%as well as $\sum_{(u,v)\in E}\, w((u,v))$.
%in polynomial time.
%} % incorporate above the fact that their choice are easy

Now let $V^\prime_{i^*}\equiv
V_{i^*}\setminus \{v\},$ $V^\prime_{j^*}\equiv V_{j^*}\cup \{v\}$
and $V^\prime_h\equiv V_h$ for $h\in [\,k+1\,]\setminus \{i^*,j^*\}$.
Clearly,
%$(V^\prime_1,\ldots,V^\prime_{k+1})$
$V=\bigcup_{i=1}^{k+1} V^\prime_i$
is a partition
of $V$.
%As $i^*, u$ and $j^*$ in the previous paragraph
%can be found
Trivially,
for $h\in [\,k+1\,]\setminus \{i^*,j^*\},$
%for $i\in [\,k+1\,]\setminus \{i^*,j^*\},$
%$$V\setminus V^\prime_i
%=V\setminus V_i,$$
%for $j\in [\,k+1\,]\setminus \{i^*,j^*\}$.
%which yields
$$c\left(V\setminus V^\prime_h,G,
%\frac{k}{k+1}
\phi_{k/(k+1)}
\right)
=c\left(V\setminus V_h,G,
%\frac{k}{k+1}
\phi_{k/(k+1)}
\right).$$
%Hence
Therefore,
\begin{eqnarray}
&& \eta\left(G,V^\prime_1,\ldots,V^\prime_{k+1}\right)
-\eta\left(G,V_1,\ldots,V_{k+1}\right)\nonumber\\
&=&
\left|\,c\left(V\setminus V^\prime_{i^*},G,
%\frac{k}{k+1}
\phi_{k/(k+1)}
\right)\,\right|
-\left|\,c\left(V\setminus V_{i^*},G,
%\frac{k}{k+1}
\phi_{k/(k+1)}
\right)\,\right|\nonumber\\
&+&\left|\,c\left(V\setminus V^\prime_{j^*},
G,
%\frac{k}{k+1}
\phi_{k/(k+1)}
\right)\,\right|
%\nonumber\\
%&-&
%\left|\,c\left(V\setminus V_{i^*},G,w,
%\frac{k}{k+1}
%\phi_{k/(k+1)}\right)\,\right|
-\left|\,c\left(V\setminus V_{j^*},G,
%\frac{k}{k+1}
\phi_{k/(k+1)}
\right)\,\right|.
%\nonumber\\
%&=&
%\left|\, c\left(\cup_{j\in [\,k+1\,]
%\setminus \left\{i^*\right\}} V^\prime_j, G, w,
%\frac{k}{k+1}
%\phi_{k/(k+1)}\right)\cap V^\prime_{i^*}\,\right|
%+\left|\, c\left(\cup_{j\in [\,k+1\,]
%\setminus \left\{j^*\right\}} V^\prime_j, G, w,
%\phi_{k/(k+1)}
%\frac{k}{k+1}
%\right)\cap V^\prime_{j^*}\,\right|\nonumber\\
%-
%\left|\, c\left(\cup_{j\in [\,k+1\,]
%\setminus \left\{i^*\right\}} V_j, G, w, \phi_{k/(k+1)}
%\frac{k}{k+1}
%\right)\cap V_{i^*}\,\right|
%-\left|\, c\left(\cup_{j\in [\,k+1\,]
%\setminus \left\{j^*\right\}} V_j, G, w, \phi_{k/(k+1)}
%\frac{k}{k+1}
%\right)\cap V_{j^*}\,\right|.
\label{changeinpotential}
\end{eqnarray}
By the choice of $v,$
\begin{eqnarray}
v\notin c\left(
V\setminus V_{i^*}, G,
\phi_{k/(k+1)}
%\frac{k}{k+1}
\right).
\label{notinthepart}
\end{eqnarray}
As $v\notin V^\prime_{i^*},$
%However,
\begin{eqnarray}
v\in c\left(
V\setminus V^\prime_{i^*},
G,
%\frac{k}{k+1}
\phi_{k/(k+1)}
\right).
\label{movedsointhepart}
\end{eqnarray}
%because $u\notin V^\prime_{i^*}$.
Relations~(\ref{notinthepart})--(\ref{movedsointhepart})
and the easily verifiable
fact
%$V^\prime_{i^*}\subseteq V_{i^*}$
$V\setminus V_{i^*}\subseteq V\setminus V^\prime_{i^*}$
imply
%Therefore,
%the inclusion in
%relation~(\ref{growlargerside}) is proper, i.e.,
\begin{eqnarray}
c\left(V\setminus V_{i^*},
G,
%\frac{k}{k+1}
\phi_{k/(k+1)}
\right)
\subsetneq
c\left(V\setminus V^\prime_{i^*},
G,
%\frac{k}{k+1}
\phi_{k/(k+1)}
\right).
%c\left(V\setminus V^\prime_{i^*},G,w,\frac{k}{k+1}\right)
%> \mu\left(G,w,V_1,\ldots,V_{k+1},i^*\right).
\label{largersidegrows}
\end{eqnarray}

As $V^\prime_{j^*}=V_{j^*}\cup\{v\}$ and $v\notin N^\text{in}(v),$
%By
inequality~(\ref{fewneighborspart})
gives
%$$\left|\,N^\text{in}(u)\setminus V_{j^*} \,\right| > \frac{k}{k+1}\cdot
% |\,V\,|, $$
%Therefore,
%$$\left|\,N^\text{in}(u)\cap \left(\cup_{j\in [\,k+1\,] \setminus
%\left\{j^*\right\}} V_j\right)
%\,\right|
%>
%\frac{k}{k+1}\cdot |\,V\,|,
%$$
%implying
\begin{eqnarray*}
\left|\,N^\text{in}(v)\setminus V^\prime_{j^*} \,\right|
=\left|\,N^\text{in}(v)\setminus V_{j^*} \,\right|
> \frac{k}{k+1}\cdot |\,V\,|.
%\sum_{(u,v)\in E, u\in V\setminus V^\prime_{j^*}}\, w\left((u,v)\right)
%=\sum_{(u,v)\in E, u\in V\setminus V_{j^*}}\, w\left((u,v)\right)
%> \frac{k}{k+1}\cdot \sum_{(u,v)\in E}\, w\left((u,v)\right),
%\label{middlestepdontknowhowtonameit}
\end{eqnarray*}
%where the equality holds because $V^\prime_{j^*}=V_{j^*}\cup\{v\}$ and
% $(v,v)\notin E$.
Consequently,
$v\in c(V\setminus
V^\prime_{j^*},
G, \phi_{k/(k+1)})$
and, therefore,
\begin{eqnarray}
c\left(
V\setminus V^\prime_{j^*},
G,
%\frac{k}{k+1}
\phi_{k/(k+1)}
\right)=
%u\in
c\left(
\{v\}\cup\left(
V\setminus V^\prime_{j^*}\right),
G,
%\frac{k}{k+1}
\phi_{k/(k+1)}
\right).
%=c\left(\bigcup_{j\in [\,k+1\,] \setminus \left\{j^*\right\}} V^\prime_j,
%G,w, \frac{k}{k+1}\right).
\label{doesnotmatterbyonevertex}
\end{eqnarray}
%As $V^\prime_{j^*}=V_{j^*}\cup\{v\},$
Clearly,
%and $V^\prime_h=V_h$ for $h\in [\,k+1\,]\setminus\{i^*,j^*\},$
$$\{v\}\cup\left(
V\setminus V^\prime_{j^*}\right)
=
V\setminus V_{j^*}.$$
%Hence
This and Eq.~(\ref{doesnotmatterbyonevertex}) give
\begin{eqnarray}
c\left(V\setminus V^\prime_{j^*},
G,
%\frac{k}{k+1}
\phi_{k/(k+1)}
\right)
=c\left(V\setminus V_{j^*},
G,
%\frac{k}{k+1}
\phi_{k/(k+1)}
\right).
\label{smallersidenotsmaller}
\end{eqnarray}
%The proof is complete by
Inequalities~(\ref{changeinpotential}),
%--(\ref{smallersidenotsmaller})
%and
(\ref{largersidegrows})
%--
and
(\ref{smallersidenotsmaller})
complete the proof.
%establish inequality~(\ref{mainlemmainequality}).
%The facts that
%$V^\prime_i\neq V_i$ for only two indices $i$
%To find a partition $(V^\prime_1,\ldots,V^\prime_{k+1})$
%satisfying inequality~(\ref{mainlemmainequality})
%in polynomial time,
%simply compute
%$$\eta(G,w,V^\prime_1,\ldots,V^\prime_{k+1})$$
%with $V^\prime_i=V_i\setminus \{u\},$
%$V^\prime_j=V_j\cup \{u\}$ and $V^\prime_h=V_j$ for all
%$h\in[\,k+1\,]\setminus \{i,j\}$
%for
%all
%$i,j\in[\,k+1\,]$ and $v\in V$
\comment{ % not needed 20091217
As $V^\prime_{i^*}=V_{i^*}\setminus \{u\},$
%By construction,
%\begin{eqnarray*}
%V\setminus V_{i^*}
%\subseteq
%V\setminus V^\prime_{i^*},
%\label{growlargerside}
%\end{eqnarray*}
%implying
\begin{eqnarray}
c\left(V\setminus V_{i^*},
G,w,
%\frac{k}{k+1}
\phi_{k/(k+1)}
\right)

\subseteq
c\left(V\setminus V^\prime_{i^*},
G,w,
%\frac{k}{k+1}
\phi_{k/(k+1)}
\right)
\label{growlargerside}
\end{eqnarray}
%Furthermore,
} % not needed 20091217
\end{proof}

%Below is
%We now come to
The main result of this section follows.

\begin{theorem}\label{maintheoremdirected}
%Let $k$ be a positive integer, $G$ be a simple digraph with positive
%indegrees and $w:E\to{\mathbb{R}}^+$.
Given a
%simple
digraph $G(V,E)$ with positive
indegrees
and a positive integer $k,$
%Then
a set $S\subseteq V$ with $c(S,G,\phi_{k/(k+1)})=V$ and $|\,S\,|\le
(k/(k+1))\, |\,V\,|$ can be found in polynomial time.
\end{theorem}
\begin{proof}
By repeated applications of Lemma~\ref{mainlemma},
a partition
%of $V$ into
%$k+1$
%sets $V_1,\ldots,V_{k+1}$
$V=\bigcup_{i=1}^{k+1}\, V_i$
with $$\eta\left(G,V_1,\ldots,V_{k+1}\right)=(k+1)\,|\,V\,|$$
can be found in polynomial time.
By
the equivalence of
%items~\ref{equivalent1}--\ref{equivalent2} of
Lemma~\ref{fullpotentialcondition}(\ref{equivalent1})~and~(\ref{equivalent2}),
$c(V\setminus V_i,G,\phi_{k/(k+1)})=V$ for all $i\in [\,k+1\,]$.
%Equivalently,
%$c(
%V\setminus V_i,G,w,\frac{k}{k+1})
%=V$ for all $i\in [\,k+1\,]$.
%Finally,
Now take $S$ to be
%one of the
a
smallest set
%of the form $V\setminus V_i$.
among $V\setminus V_1,\ldots,V\setminus V_{k+1}$.
Clearly, $|\,S\,|\le (k/(k+1))\,|\,V\,|$.
\end{proof}

%Several corollaries are immediate.
Several
%special cases
theorems
are immediate.

%\comment{ % we don't consider the fraction of activated vertices here
\begin{theorem}
For
%Given
any digraph $G(V,E)$ with positive
indegrees,
% and $\delta\in[\,0,1\,],$
$$\text{\rm min-seed}\left(G,\phi^\text{\rm simple}\right)
\le
\left\lfloor\frac{|\,V\,|}{2}\right\rfloor.$$
\end{theorem}
\begin{proof}
Take $k=1$ in Theorem~\ref{maintheoremdirected}.
\end{proof}
%} % we don't consider the fraction of activated vertices here

\begin{theorem}\label{twothirdsfordigraphs}
For any
digraph $G(V,E)$ with positive
indegrees,
$$\text{\rm min-seed}\left(G,\phi^\text{\rm strict}\right)
\le
\left\lfloor\frac{2\cdot |\,V\,|}{3}\right\rfloor.$$
\end{theorem}
\begin{proof}
%Clearly, $\text{min-seed}(G,\phi^\text{strict}_\text{maj})\le \lceil
%\delta\,|\,V\,|\rceil$.
Take $k=2$ in Theorem~\ref{maintheoremdirected} and note that
$\phi^\text{strict}(v)\le (2/3)\,\text{deg}^\text{in}(v)$ for
all $v\in V$.
\end{proof}

% it's been improved anyway!

\section{Irreversible dynamos of undirected graphs}
%\section{The undirected case}
\label{undirectedsection}
We now
%Next we
turn to irreversible dynamos of undirected connected graphs.
%Fix throughout the rest of this section
%an undirected connected graph $G(V,E)$ and $v^*\in V$.
%Let $d:V\times V\to\mathbb{N}$ map any two vertices $x,y\in V$ to
Let $G(V,E)$ be an undirected connected graph.
A cut is
%a
an unordered
pair $(S,V\setminus S)$
%where
with
$S\subseteq V$.
%which is
We call a cut $(S,V\setminus S)$
%It is
%said to be
%Call
%We say that
%a set $S\subseteq V$
proper if
\begin{eqnarray}
\left|\,N(v)\cap S\,\right|&\le& \left|\,N(v)\setminus S\,\right|,
\,\,\,\,\forall v\in S,\label{properone}\\
\left|\,N(v)\setminus S\,\right|&\le& \left|\,N(v)\cap
S\,\right|,\,\,\,\,\forall v\in V\setminus S,\label{propertwo}
\end{eqnarray}
and improper otherwise.
So a proper cut is such that
no vertex has more neighbors in the side ($S$ or
$V\setminus S$)
it belongs to
than in the
%cut
side
it does not.
%every vertex on one side (i.e., $S$ or
%$V\setminus S$) has more or equally many neighbors on the other side.
%than on the same side.
The following fact is implicit in~\cite[pp.\ 303--304]{Pap94}.

\begin{fact}(\cite[pp.\ 303--304]{Pap94})\label{localadjustment}
Given an undirected graph $G(V,E)$
%$S\subseteq V$
%is not
and an improper cut
$(S,V\setminus S),$
%proper.
a proper cut
%$T\subseteq V$
$(T,V\setminus T)$
with $e(T,V\setminus
T)>e(S,V\setminus S)$ can be found in polynomial time.
% polynomial in $|\,V\,|$.
\end{fact}

%For any cut $(S,V\setminus S),$
A vertex $v\in V$ is said to be bad with respect to (abbreviated w.r.t.)
%$S\subseteq V$
a cut $(S,V\setminus S)$
if $|\,N(v)\cap S\,|=|\,N(v)\setminus S\,|$; it is good w.r.t.\
$(S,V\setminus S)$ otherwise.
%Clearly, a vertex is bad w.r.t.\
%$S$
%$(S,V\setminus S)$
%if and only if it is bad w.r.t.\
%$V\setminus S$.
%$(V\setminus S,S)$.
A connected component
%$G^\prime(V^\prime,E^\prime)$
of $G[\,S\,]$ or $G[\,V\setminus S\,]$ is
%said to be
bad
w.r.t.\
%$S$
$(S,V\setminus S)$
if
{\em all} its vertices are bad w.r.t.\
%$V\setminus S$;
$(S,V\setminus S)$;
%there exists a $v\in V^\prime$ with $|\,N(v)\cap S\,|=|\,N(v)\setminus S\,|$
it is good
w.r.t.\
$(S,V\setminus S)$
otherwise.
%Again, a connected component of $G[\,S\,]$ or $G[\,V\setminus S\,]$ is bad
%w.r.t.\
%$S$
%$(S,V\setminus S)$
%if and only if it is bad
%w.r.t.\
%$V\setminus S$.
%$(V\setminus S,S)$.
The set of connected components of $G[\,S\,]$
%and
%(resp., $G[\,V\setminus S\,]$)
that are bad
w.r.t. $(S,V\setminus S)$
%are
is
denoted ${\cal B}(S)$.
%and
%(resp., ${\cal B}(V\setminus S)$).
Similarly, ${\cal B}(V\setminus S)$ is the set of
bad (w.r.t. $(S,V\setminus S)$)
connected components
of $G[\,V\setminus S\,]$.
%respectively.
For $v^*\in V$ and
%a proper set
$S\subseteq V,$
define
\begin{eqnarray}
&&
\psi\left(S,v^*\right)\nonumber\\
&\equiv&
e\left(S,V\setminus S\right)\cdot |\,V\,|^2
%\nonumber\\
%&&- \left(\left|\,{\cal B}(S)\,\right| + \left|\,{\cal B}(V\setminus S)\,\right|\right)\cdot
%|\,V\,|^4\nonumber\\
%&&
- \left[\sum_{\hat{G}(\hat{V},\hat{E})\in {\cal B}(S)}\,
%\min_{v\in\hat{V}}\, d\left(v,v^*\right)
d\left(v^*,\hat{V}\right)
+\sum_{\hat{G}(\hat{V},\hat{E})\in {\cal B}(V\setminus S)}\,
%\min_{v\in\hat{V}}\, d\left(v,v^*\right)
d\left(v^*,\hat{V}\right)
\right].
\,\,\,\,\,\,\,\,\,\,
\label{potential}
\end{eqnarray}
%An immediate property of $\psi(\cdot,\cdot)$ is stated below.
%Below we show that for any $v^*\in V,$
%$\psi(S,v^*)$ increases as $e(S,V\setminus S)$ increases.

For a fixed $v^*\in V,$ we will keep refining cuts by increasing their
$\psi(\cdot,v^*)$-values until
%we obtain one
a cut
suitable for creating an
irreversible dynamo results.
One way to increase the $\psi(\cdot,v^*)$-values is to find larger cuts, as
shown below.

\begin{lemma}\label{cutsizedominate}
Let $G(V,E)$ be an undirected connected graph and $v^*\in V$.
%For any
If two
cuts $(A,V\setminus A)$ and $(B,V\setminus B)$
%with
satisfy
$e(A,V\setminus A)>e(B,V\setminus B),$
then
$$\psi\left(A,v^*\right)>\psi\left(B,v^*\right).$$
\end{lemma}
\begin{proof}
In Eq.~(\ref{potential}),
the
$e(S,V\setminus S)$ term is multiplied by $|\,V\,|^2>|\,V\,|\,(|\,V\,|-1)$.
%in
%whereas
But the summations within the brackets of Eq.~(\ref{potential}) evaluate to be
at most $|\,V\,|\,(|\,V\,|-1)$
because
$|\,{\cal B}(S)\,|+|\,{\cal B}(V\setminus S)\,|\le |\,V\,|$ and
$d(v^*,U)\le |\,V\,|-1$ for any $\emptyset\subsetneq U\subseteq V$.
\end{proof}

The following lemma is straightforward.

\begin{lemma}\label{takeawayavertexleft}
For an undirected connected graph $G(V,E)$ and $v\in V,$
every connected component of $G[\,V\setminus \{v\}\,]$
%has
shares
a vertex
%in
with
$N_G(v)$.
%adjacent to $v$ in $G$.
\end{lemma}
\begin{proof}
%If a connected component of $G[\,V\setminus \{v\}\,]$ has no vertices in
%$N_G(v),$
%then no vertices of the connected component can reach $v$ by a path in $G,$
%a contradiction as $G$ is connected.
%Trivial.
As $G$ is connected,
any $u\in V\setminus \{v\}$ can reach $v$ by a path $P$ in $G$.
Starting from $u$ and going along with $P,$
a vertex in $N_G(v)$ must be reached before arriving at $v$.
%Hence $u$ is in the same component of $G[\,V\setminus \{v\}\,]$
%as some vertex in $N_G(v)$.
Hence the connected component of $G[\,V\setminus \{v\}\,]$
containing $u$ must have a vertex in $N_G(v)$.
%Before reaching $v,$ however, a vertex in $N_G(v)$ must be reached
\end{proof}

The next lemma shows that
moving a bad vertex $v$ across a cut
does not change
the cut size.
%of
%the
%a
%cut.
%remains when a bad vertex is moved acr.

\begin{lemma}\label{cutsizeremainsundermovingbadverticeslemma}
Let $G(V,E)$ be an undirected connected graph, $(S,V\setminus S)$ be a cut
and $v\in S$ be bad w.r.t.\ $(S,V\setminus S)$.
Then
\begin{eqnarray}
e\left(S,V\setminus S\right)=e\left(S\setminus \{v\},\left(V\setminus
S\right)\cup \{v\}\right).\label{cutsizeremainsundermovingbadvertices}
\end{eqnarray}
\end{lemma}
\begin{proof}
As $v$ is bad w.r.t.\ $(S,V\setminus S),$
$|\,N(v)\cap S\,|=|\,N(v)\setminus S\,|$.
Now Eq.~(\ref{cutsizeremainsundermovingbadvertices}) holds
because (1) the edges incident on $v$ contribute $|\,N(v)\setminus S\,|$
to the lefthand side and $|\,N(v)\cap S\,|$ to the righthand side and (2)
all other edges contribute the same amount to either side.
\end{proof}

The next lemma shows that moving a vertex $v$ across a cut does not
change whether a connected component without vertices in
%$N_G(v)\cup \{v\}$
$N_G^*(v)$
is bad.

%\comment{ % let me incorporate this into the lemma below
\begin{lemma}\label{goodbadremainsifnotadjacenttov}
Let $G(V,E)$ be an undirected connected graph,
$(S,V\setminus S)$ be a proper cut,
$H(V_H,E_H)$ be a connected component
of $G[\,S\,]$
or $G[\,V\setminus S\,],$
$v\in S$
and
%$(N_G(v)\cup \{v\})\cap V_H=\emptyset$.
$N_G^*(v)\cap V_H=\emptyset$.
Then
%$H$ is a bad connected component
%w.r.t.\ $(S\setminus \{v\},V\setminus (S\setminus \{v\}))$
%if and only if it is
%a
%bad
%connected component
%w.r.t.\ $(S,V\setminus S)$.
\begin{enumerate}
\item\label{badonSsideremains} $H\in {\cal B}(S\setminus \{v\})$ if and only if $H\in {\cal B}(S)$.
\item\label{badonVSsideremains} $H\in {\cal B}((V\setminus S)\cup \{v\})$ if and only if $H\in {\cal
B}(V\setminus S)$.
\end{enumerate}
\end{lemma}
\begin{proof}
%Suppose that $H$ is a connected component of $G[\,S\,]$.
%Then
As
%$(N_G(v)\cup \{v\})\cap V_H=\emptyset$
$N_G^*(v)\cap V_H=\emptyset$
and
$H$ is a connected component of $G[\,S\,]$ or
$G[\,V\setminus S\,],$
%and $(N_G(v)\cup \{v\})\cap V_H=\emptyset,$
$H$
must remain
%is also
a connected component of $G[\,S\setminus \{v\}\,]$
%because $(N_G(v)\cup \{v\})\cap V_H=\emptyset$.
or $G[\,(V\setminus S)\cup \{v\}\,],$ respectively.
%If $G^\prime$ is a connected component of $G[\,S\,]$ (resp., $G[\,V\setminus
%S\,]$), then it is also one of $G[\,S\setminus \{v\}\,]$ (resp.,
%$G[\,V\setminus (S\setminus \{v\})\,]$) because
%$(N_G(v)\cup \{v\})\cap V^\prime=\emptyset$.
As
%$(N_G(v)\cup \{v\})\cap V_H=\emptyset,$
$N_G^*(v)\cap V_H=\emptyset,$
every $u\in V_H$ satisfies
$|\,N_G(u)\cap S\,|=|\,N_G(u)\cap (S\setminus \{v\})\,|$
and $|\,N_G(u)\setminus S\,|=|\,N_G(u)\setminus (S\setminus \{v\})\,|$;
so $u$ is bad w.r.t.\ $(S,V\setminus S)$ if and only if it is bad w.r.t.\
%$(S\setminus \{v\},V\setminus (S\setminus \{v\}))$.
$(S\setminus \{v\},(V\setminus S)\cup \{v\})$.
Therefore,
$H$ is bad w.r.t.\ $(S,V\setminus S)$ if and only if it is bad
w.r.t.\ $(S\setminus \{v\},(V\setminus S)\cup \{v\})$.
\end{proof}
%} % let me incorporate this into the lemma below

\begin{figure}
\centering
\subfigure[Below the
%dotted
%horizontal
gray
line are the connected components of
$G{[\,S\,]},$
namely $G^\prime,G_2,\ldots,G_k$.
%shown as dashed rectangles.
One of the components, $G^\prime,$ contains a vertex $v$ whose neighbors are
shown as double circles.
As in the proof of Lemma~\ref{changesofbadcomponentsSside}, the connected components
obtained by
removing $v$ from $G^\prime$ are
%denoted
$G_{11},\ldots,G_{1h}$.
Above the
%dotted
gray
%horizontal
line
are the connected components of $G{[\,V\setminus S\,]},$
i.e., $\hat{G}_1,\ldots,\hat{G}_\ell$.
%shown as double
%rectangles.
]
%[Subscription a]
{ % begin subfigure
\label{fig:1a}
\begin{pspicture}(-2,-1.7)(9.1,5) % the height is actually 6.1, I set it to 5 for otherwise the picture will lie too low & confuse with the page number
%(-2,-1.7)(9.1,6.1)
%\pspolygon[linestyle=dashed](-0.7,-1.2)(3.1,-1.2)(3.1,2)(-0.7,2)
\psframe[framearc=0.3](-0.7,-1.7)(3.1,2)
\rput(2.7,1.6){\rnode{G1}{$G^\prime$}}

\cnodeput[](1.2,0.8){v}{$v$}
\cnodeput[doubleline=true](0,0){v1}{}
\psdots(1,-0.5)(1.2,-0.5)(1.4,-0.5)
\cnodeput[doubleline=true](2.4,0){vh}{}

%\pspolygon[linestyle=dotted](-0.5,-1)(0.5,-1)(0.5,0.5)(-0.5,0.5)
\psframe[framearc=0.3](-0.5,-1.5)(0.5,0.5)
\rput(0, -0.5){\rnode{G11}{$G_{11}$}}
%\pspolygon[linestyle=dotted](1.9,-1)(2.9,-1)(2.9,0.5)(1.9,0.5)
\psframe[framearc=0.3](1.9,-1.5)(2.9,0.5)
\rput(2.4, -0.5){\rnode{G1h}{$G_{1h}$}}
%\pspolygon[linestyle=dashed](3.4,-1)(4.4,-1)(4.4,0.5)(3.4,0.5)
\psframe[framearc=0.3](3.4,-1.5)(4.4,0.5)
\rput(3.9, -0.5){\rnode{G2}{$G_2$}}
\psdots(4.9,-0.5)(5.1,-0.5)(5.3,-0.5)
%\pspolygon[linestyle=dashed](5.8,-1)(6.8,-1)(6.8,0.5)(5.8,0.5)
\psframe[framearc=0.3](5.8,-1.5)(6.8,0.5)
\rput(6.3, -0.5){\rnode{Gk}{$G_k$}}

\ncline[]{v}{v1}
\ncline[]{v}{vh}

\psline[linecolor=gray,linewidth=0.1]{-}(-2,2.3)(9.1,2.3) % the horizontal line
\rput(9.1,1.3){\rnode{lowerhalfisS}{$S$}}
\rput(9.1,3.3){\rnode{upperhalfisVS}{$V\setminus S$}}
%\ncline[]{<->}{lowerhalfisS}{upperhalfisVS}

%\pspolygon[doubleline=true](-0.7,2.6)(3.1,2.6)(3.1,5.3)(-0.7,5.3)
%\rput(2.7,4.9){\rnode{Gbar}{$\bar{G}$}}

%\cnodeput[](1.2,3.2){vclone}{$v$}

%\pspolygon[doubleline=true](-0.5,3.3)(0.5,3.3)(0.5,4.8)(-0.5,4.8)
%\pspolygon[doubleline=true](-0.5,3.3)(0.5,3.3)(0.5,4.8)(-0.5,4.8)
\psframe[framearc=0.3](-0.5,4.1)(0.5,6.1)
%\cnodeput[doubleline=true](0,3.8){v1anotherside}{$w$}
\cnodeput[doubleline=true](0,4.6){v1anotherside}{}
\rput(0, 5.1){\rnode{G1anotherside}{$\hat{G_{1}}$}}
\psdots(1,5.1)(1.2,5.1)(1.4,5.1)
%\pspolygon[doubleline=true](1.9,3.3)(2.9,3.3)(2.9,4.8)(1.9,4.8)
\psframe[framearc=0.3](1.9,4.1)(2.9,6.1)
\cnodeput[doubleline=true](2.4,4.6){vhanotherside}{}
\rput(2.4, 5.1){\rnode{Gtanotherside}{$\hat{G_{t}}$}}

%\pspolygon[doubleline=true](3.4,3.3)(4.4,3.3)(4.4,4.8)(3.4,4.8)
\psframe[framearc=0.3](3.4,4.1)(4.4,6.1)
\rput(3.9, 5.1){\rnode{G1anotherside}{$\hat{G_{t+1}}$}}
\psdots(4.9,5.1)(5.1,5.1)(5.3,5.1)
%\pspolygon[doubleline=true](5.8,3.3)(6.8,3.3)(6.8,4.8)(5.8,4.8)
\psframe[framearc=0.3](5.8,4.1)(6.8,6.1)
\rput(6.3, 5.1){\rnode{G1anotherside}{$\hat{G_\ell}$}}

%\ncline[]{vclone}{v1anotherside}
%\ncline[]{vclone}{vhanotherside}
\ncline[]{v}{v1anotherside}
\ncline[]{v}{vhanotherside}
%\ncline[]{vclone}{v1}
%\ncline[]{vclone}{vh}

%\ncline{->}{v}{vclone}
\end{pspicture}
} % end subfigure

% do something dirty: let's have two graphs!

\subfigure
[Continuing from Fig.~\ref{fig:1a}, the connected components of
$G{[\,S\setminus \{v\}\,]}$
and
%$G{[\,V\setminus (S\setminus \{v\})\,]}$
$G{[\,(V\setminus S)\cup \{v\}\,]}$
are shown
% as dashed and dotted rectangles,
below and above the
%dotted
gray
%horizontal
line, respectively.
In the proof of Lemma~\ref{changesofbadcomponentsSside},
$G_{11},\ldots,G_{1h}$ are shown to be good w.r.t.\
$(S\setminus \{v\},(V\setminus S)\cup \{v\})$
by showing that $v$'s neighbors in $S$ are good w.r.t.\
$(S\setminus \{v\},(V\setminus S)\cup \{v\})$.
%Below the dotted horizontal line are the connected components of $G[\,S\,]$
%shown in dashed rectangles.
%One of the components, $G^\prime,$ contains a vertex $v$ whose neighbors are
%shown in double circles.
%As in the proof of Lemma~\ref{changesofbadcomponentsSside}, the connected components
%obtained by
%removing $v$ from $G^\prime$ are denoted $G_{11},\ldots,G_{1h}$.
%Above the dotted horizontal line
%are the connected components of $G[\,V\setminus S\,]$ shown in double
%rectangles.
]
%[Subcaption b.]
{
% begin subfigure 2
\label{fig:1b}
\begin{pspicture}(-2,-1.5)(9.1,6.3)
%\pspolygon[linestyle=dashed](-0.7,-1.2)(3.1,-1.2)(3.1,2)(-0.7,2)
%\rput(2.7,1.6){\rnode{G1}{$G^\prime$}}

%\cnodeput[linestyle=dotted](1.2,0.8){v}{$v$}
\cnodeput[doubleline=true](0,0){v1}{}
\psdots(1,-0.5)(1.2,-0.5)(1.4,-0.5)
\cnodeput[doubleline=true](2.4,0){vh}{}

%\pspolygon[linestyle=dashed](-0.5,-1)(0.5,-1)(0.5,0.5)(-0.5,0.5)
\psframe[framearc=0.3](-0.5,-1.5)(0.5,0.5)
\rput(0, -0.5){\rnode{G11}{$G_{11}$}}
%\pspolygon[linestyle=dashed](1.9,-1)(2.9,-1)(2.9,0.5)(1.9,0.5)
\psframe[framearc=0.3](1.9,-1.5)(2.9,0.5)
\rput(2.4, -0.5){\rnode{G1h}{$G_{1h}$}}
%\pspolygon[linestyle=dashed](3.4,-1)(4.4,-1)(4.4,0.5)(3.4,0.5)
\psframe[framearc=0.3](3.4,-1.5)(4.4,0.5)
\rput(3.9, -0.5){\rnode{G2}{$G_2$}}
\psdots(4.9,-0.5)(5.1,-0.5)(5.3,-0.5)
%\pspolygon[linestyle=dashed](5.8,-1)(6.8,-1)(6.8,0.5)(5.8,0.5)
\psframe[framearc=0.3](5.8,-1.5)(6.8,0.5)
\rput(6.3, -0.5){\rnode{Gk}{$G_k$}}

%\ncline[linestyle=dotted]{v}{v1}
%\ncline[linestyle=dotted]{v}{vh}

\psline[linecolor=gray,linewidth=0.1]{-}(-2,2.3)(9.1,2.3) % the horizontal line
\rput(9.1,1.3){\rnode{lowerhalfisS}{$S\setminus \{v\}$}}
\rput(9.1,3.3){\rnode{upperhalfisVS}{$(V\setminus S)\cup \{v\}$}}
%\ncline[]{<->}{lowerhalfisS}{upperhalfisVS}

%\pspolygon[doubleline=true](-0.7,2.6)(3.1,2.6)(3.1,5.8)(-0.7,5.8)
\psframe[framearc=0.3](-0.7,2.6)(3.1,6.3)
\rput(2.7,3){\rnode{Gbar}{$\bar{G}$}}

\cnodeput[](1.2,3.8){vclone}{$v$}

%\pspolygon[](-0.5,3.3)(0.5,3.3)(0.5,4.8)(-0.5,4.8)
%\pspolygon[linestyle=dotted](-0.5,3.3)(0.5,3.3)(0.5,4.8)(-0.5,4.8)
\psframe[framearc=0.3](-0.5,4.1)(0.5,6.1)
%\cnodeput[doubleline=true](0,3.8){v1anotherside}{$w$}
\cnodeput[doubleline=true](0,4.6){v1anotherside}{}
\rput(0, 5.1){\rnode{G1anotherside}{$\hat{G_{1}}$}}
\psdots(1,5.1)(1.2,5.1)(1.4,5.1)
%\pspolygon[linestyle=dotted](1.9,3.3)(2.9,3.3)(2.9,4.8)(1.9,4.8)
\psframe[framearc=0.3](1.9,4.1)(2.9,6.1)
\cnodeput[doubleline=true](2.4,4.6){vhanotherside}{}
\rput(2.4, 5.1){\rnode{Gtanotherside}{$\hat{G_{t}}$}}

%\pspolygon[doubleline=true](3.4,3.3)(4.4,3.3)(4.4,4.8)(3.4,4.8)
\psframe[framearc=0.3](3.4,4.1)(4.4,6.1)
\rput(3.9, 5.1){\rnode{G1anotherside}{$\hat{G_{t+1}}$}}
\psdots(4.9,5.1)(5.1,5.1)(5.3,5.1)
%\pspolygon[doubleline=true](5.8,3.3)(6.8,3.3)(6.8,4.8)(5.8,4.8)
\psframe[framearc=0.3](5.8,4.1)(6.8,6.1)
\rput(6.3, 5.1){\rnode{G1anotherside}{$\hat{G_\ell}$}}

\ncline[]{vclone}{v1anotherside}
\ncline[]{vclone}{vhanotherside}
%\ncline[linestyle=dotted]{v}{v1anotherside}
%\ncline[linestyle=dotted]{v}{vhanotherside}
\ncline[]{vclone}{v1}
\ncline[]{vclone}{vh}

%\ncline{->}{v}{vclone}
\end{pspicture}
} % end subfigure 2

\caption{Figs.~\ref{fig:1a}--\ref{fig:1b} show how the connected components
of $G[\,S\,]$ and $G[\,V\setminus S\,]$ change as a vertex $v$ is moved from
$S$ to $V\setminus S$.
The notations $G^\prime,$ $\bar{G},$ $G_{11},\ldots,G_{1h}$
and $\hat{G}_1,\ldots,\hat{G}_\ell$
are
%consistent with
from
the proofs of
Lemmas~\ref{changesofbadcomponentsSside} and \ref{componentsonVSside}.
% abandon old again
% old caption
}
\label{movingvertexfigure}
\end{figure}
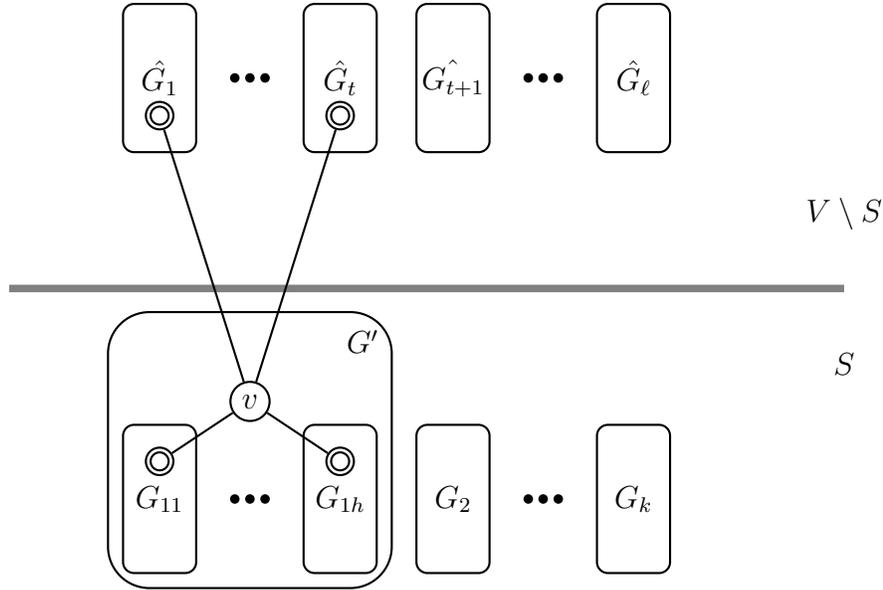
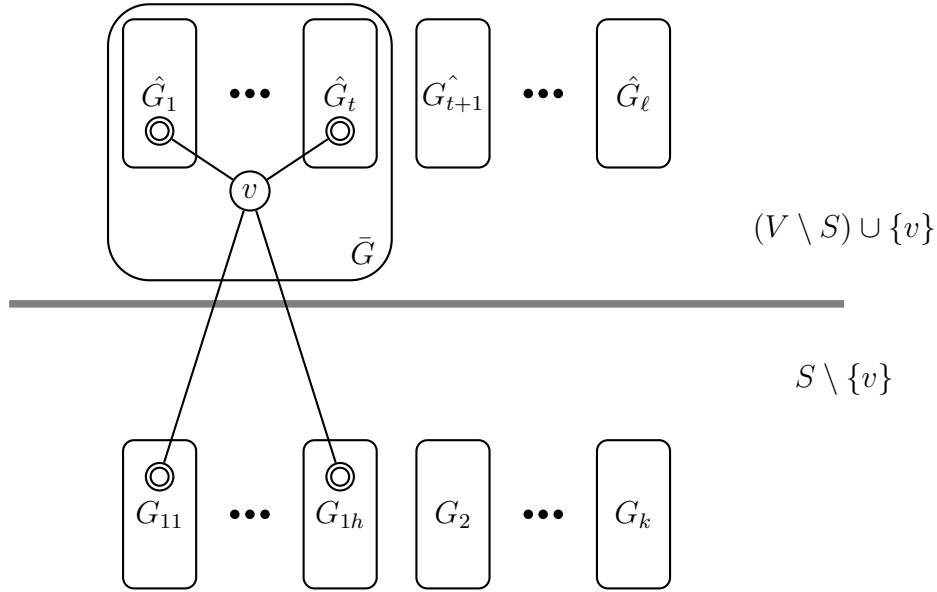

%For a proper cut $(S,V\setminus S)$ and $v^*,$
%We will keep increasing the $\psi(\cdot,v^*)$-values of cuts by
%moving one vertex at a time.
%As $\psi(\cdot,\cdot)$ is affected by the bad components w.r.t.\ the
%respective cuts,
Given a proper cut $(S,V\setminus S),$
%Next,
%we
%need to
the next two lemmas
analyze
how bad components
%w.r.t.\ $(S,V\setminus S)$
evolve when a vertex is moved away from $S$.
%as done below.
%To visualize the proofs of the lemmas,
%it is helpful to refer to
%For visualizations please refer to
See
Fig.~\ref{movingvertexfigure} for illustration.
%visualize the proofs of the lemmas.

%a $v\in S$ to $V\setminus S$.
%away from a connected component
%$\bar{G}(\bar{V},\bar{E})$
%of $G[\,S\,]$ breaks the component
%Such a movement
%may break the connected component of $G[\,S\,]$ containing $v$ into
%several new components.
%But the new components will all be good w.r.t.\ the new cut $(S\setminus \{v\},
%V\setminus (S\setminus \{v\}))$.
%Furthermore, other connected components
%But only bad (w.r.t.\ $(S,V\setminus S)$)
%connected components of $G[\,S\,]$ can remain bad (w.r.t.\ $(S\setminus
%\{v\}, V\setminus (S\setminus \{v\}))$),
% after
%$v$ is moved,
%as shown below.
%The main intuition behind the following lemma is that the new
%connected components will be good w.r.t.\ the new cut (with $v$ moved to
%$V\setminus S$).
%Thus the summations in
%Eq.~(\ref{potential}) will not increase
%Then
%By having as many good components as possible,
%we will be able to bound
%the summations in Eq.~(\ref{potential}) because.

\begin{lemma}\label{changesofbadcomponentsSside}
Let $G(V,E)$ be an undirected connected graph,
%$S\subseteq V$ be proper,
$(S,V\setminus S)$ be a proper cut,
%$G_1(V_1,E_1), \ldots, G_k(V_k,E_k)$ be the
$G^\prime(V^\prime,E^\prime)$ be a
%$G^\prime(V^\prime, E^\prime)$ be a
%bad
connected component of $G[\,S\,]$
%with $v\in V^\prime$.
and
$v\in V^\prime$.
%$v\in V_1$.
Then
%the
%connected components of $G[\,S\setminus \{v\}\,]$
%are $G_2, \ldots, G_k$ plus zero or more
%connected components that are good
%w.r.t.\ $(S\setminus \{v\},V\setminus (S\setminus \{v\}))$.
${\cal B}(S\setminus \{v\})
\subseteq
{\cal B}(S)\setminus \{G^\prime\}$.
\end{lemma}
\begin{proof}
%Denote by
Let
$G_1(V_1,E_1)=G^\prime(V^\prime,E^\prime), \ldots, G_k(V_k,E_k)$
be
%be the
the
%$G^\prime(V^\prime, E^\prime)$ be a
%bad
connected components of $G[\,S\,]$.
For
%each
$u\in N_G(v)\cap S,$
$$\left|\,N_G(u)\cap S\,\right|\le \left|\,N_G(u)\setminus S\,\right|$$
because
%$S$
$(S,V\setminus S)$
is proper.
Consequently, for every $u\in N_G(v)\cap S,$
\begin{eqnarray}
&&\left|\,N_G(u)\cap \left(S\setminus \{v\}\right)\,\right|\nonumber\\
&=& \left|\,N_G(u)\cap S\,\right|-1\nonumber\\
&\le& \left|\,N_G(u)\setminus S\,\right|-1\nonumber\\
&<& \left|\,N_G(u)\setminus S\,\right|+1\nonumber\\
&=& \left|\,N_G(u)\setminus \left(S\setminus \{v\}\right)\,\right|,
%\,\,\,\,\forall u\in N_G(v)\cap S.
\label{becomesgoodafterremoval}
\end{eqnarray}
%i.e., $u$ is good w.r.t.\ $S\setminus \{v\}$.
where both equalities follow from $v\in S$ and $u\in N_G(v)\cap S$.

Let $G_{11},\ldots,G_{1h}$ be the connected components of
$G[\,V_1\setminus\{v\}\,],$ where $h\ge 0$ ($h=0$ if and only if
$V_1=\{v\}$).
Clearly, the connected components of $G[\,S\setminus \{v\}\,]$
are $G_{11},\ldots,G_{1h},$
%and
$G_2,\ldots,G_k$.
%To complete the proof, we need only
%We
%proceed to
%show that all of $G_{11},\ldots,G_{1h}$
%are good w.r.t.\ $(S\setminus \{v\},V\setminus(S\setminus \{v\}))$.
With $G_1=G^\prime$ playing the role of $G$ in Lemma~\ref{takeawayavertexleft},
each of $G_{11},\ldots,G_{1h}$ has a vertex in
$N_{G^\prime}(v)=N_G(v)\cap S$.
%the set of $v$'s neighbors in $G^\prime,$ i.e.,
%$N_G(v)\cap S$.
%$N_G(v)\cap V_1=N_G(v)\cap S$.
%the set of $v$'s neighbors in $G_1$.
Hence
%by inequality~(\ref{becomesgoodafterremoval}),
%all vertices in $N_G(v)\cap S$ are good w.r.t.\ $S\setminus \{v\}$.
each of
%Hence
$G_{11},\ldots,G_{1h}$
%are all good w.r.t.\ $S\setminus \{v\}$.
has a vertex satisfying inequality~(\ref{becomesgoodafterremoval}),
implying that
%forcing
$G_{11},\ldots,G_{1h}$
are
all
%to be
good w.r.t.\
%$(S\setminus \{v\},V\setminus(S\setminus \{v\}))$.
$(S\setminus \{v\},(V\setminus S)\cup \{v\})$.
%implying that all of $G_{11},\ldots,G_{1h}$ are good w.r.t.\ $S\setminus
%\{v\}$ because
%As we have seen that each vertex in $N_G(v)\cap S$ is good w.r.t.\
%$S\setminus \{v\},$
%all of $G_{11},\ldots,G_{1h}$ are good w.r.t.\
%$S\setminus \{v\}$.
Therefore,
${\cal B}(S\setminus \{v\})\subseteq \{G_2,\ldots,G_k\}$.

As ${\cal B}(S\setminus \{v\})\subseteq \{G_2,\ldots,G_k\},$
%Fix $2\le i\le k$ arbitrarily.
%To complete the proof, we need only
it remains to
show that
%for $2\le i\le k,$
%$G_i$ is bad w.r.t.\ $(S\setminus \{v\},V\setminus (S\setminus \{v\}))$
%if and only if it is bad w.r.t.\ $(S,V\setminus S)$.
$G_i\in {\cal B}(S\setminus \{v\})$
%if and
only if $G_i\in {\cal B}(S),$
for all $2\le i\le k$.
By
%item~\ref{badonSsideremains} of
Lemma~\ref{goodbadremainsifnotadjacenttov}(\ref{badonSsideremains}) with $G_i$ playing the role of
$H,$
we need only check that
%$(N_G(v)\cup \{v\})\cap V_i=\emptyset$
$N_G^*(v)\cap V_i=\emptyset$
for $2\le i\le k,$
which is true because $v\in V_1$ and
$G_1,\ldots,G_k$ are disjoint connected components
of $G[\,S\,]$.
\comment{ % invoke previous lemma for this
As
$v\in V_1,$
the fact that
$G_1$ and $G_i$ are distinct connected components of $G[\,S\,]$ implies
$v\notin N_G(u)$ for each $u\in V_i$.
%$V_i$ is a different connected component of $G[\,S\,],$
%For each $2\le i\le k,$
%$(N_G(v)\cup\{v\})\cap V_i=\emptyset$.
Hence every $u\in V_i$ satisfies
$|\,N_G(u)\cap S\,|=|\,N_G(u)\cap (S\setminus \{v\})\,|$
and $|\,N_G(u)\setminus S\,|=|\,N_G(u)\setminus (S\setminus \{v\})\,|$;
%\begin{eqnarray}
%|\,N(u)\cap V_i\,|&=&|\,N(u)\cap (V_i\setminus \{v\})\,|,\\
%|\,N(u)\setminus V_i\,|&=&|\,N(u)\setminus (V_i\setminus \{v\})\,|.
%\end{eqnarray}
so $u$ is bad w.r.t.\
%$(S,V\setminus S)$
$(S\setminus \{v\},V\setminus (S\setminus \{v\}))$
if and only if it is bad w.r.t.\
%$(S\setminus \{v\},V\setminus (S\setminus \{v\}))$.
$(S,V\setminus S)$.
Therefore,
%for, otherwise,
%$V_1$ and $V_i$ are in the same connected component of $G[\,S\,]$.
%Hence by Lemma~\ref{goodbadremainsifnotadjacenttov},
%for $2\le i\le k,$
%$G_i\in {\cal B}(S\setminus \{v\})$ implies $G_i\in {\cal B}(S)$.
%$G_i$ is bad w.r.t.\ $(S,V\setminus S)$
%if and only if it is bad w.r.t.\ $(S\setminus \{v\},V\setminus (S\setminus \{v\}))$.
$G_i\in {\cal B}(S\setminus \{v\})$ if and only if $G_i\in {\cal B}(S)$.
} % invoke previous lemma for this
\end{proof}

\comment{ % no more need this simple corollary
\begin{corollary}\label{remainsgoodorbadspecificforS}
Let $G(V,E)$ be an undirected connected graph,
$(S,V\setminus S)$ be a proper cut,
$G_1(V_1,E_1), \ldots, G_k(V_k,E_k)$ be the
connected components of $G[\,S\,]$
%$G^\prime(V^\prime,E)$
and
$v\in V_1$.
Then for $2\le i\le k,$
%${\cal B}(S\setminus \{v\})\subsetneq {\cal B}(S)$.
$G_i$ is bad w.r.t.\ $(S,V\setminus S)$
if and only if it is bad w.r.t.\ $(S\setminus \{v\},V\setminus (S\setminus \{v\}))$.
\end{corollary}
\begin{proof}
For each $2\le i\le k,$
$(N_G(v)\cup\{v\})\cap V_i=\emptyset$ for, otherwise,
$V_1$ and $V_i$ are in the same connected component of $G[\,S\,]$.
Hence by Lemma~\ref{goodbadremainsifnotadjacenttov},
for $2\le i\le k,$
%$G_i\in {\cal B}(S\setminus \{v\})$ implies $G_i\in {\cal B}(S)$.
$G_i$ is bad w.r.t.\ $(S,V\setminus S)$
if and only if it is bad w.r.t.\ $(S\setminus \{v\},V\setminus (S\setminus \{v\}))$.
\end{proof}
} % no more need this simple corollary

\comment{ % can be deleted now since the above lemma is lengthened
\begin{corollary}\label{listthebadofS}
Let $G(V,E)$ be an undirected connected graph,
$(S,V\setminus S)$ be a proper cut,
$G_1(V_1,E_1), \ldots, G_k(V_k,E_k)$ be the
connected components of $G[\,S\,],$
%and
$v\in V_1$ and $1\le t\le k$.
If ${\cal B}(S)=\{G_1,\ldots,G_t\},$
then
${\cal B}(S\setminus \{v\})=\{G_2,\ldots,G_t\}$.
\end{corollary}
\begin{proof}
Immediate from Lemma~\ref{changesofbadcomponentsSside}
and Corollary~\ref{remainsgoodorbadspecificforS}.
\end{proof}
} % can be deleted now since the above lemma is lengthened

% # of bad components is no longer part of the potential

\begin{corollary}
Let $G(V,E)$ be an undirected connected graph,
%$S\subseteq V$ be proper,
$(S,V\setminus S)$ be a proper cut,
%$G^\prime(V^\prime, E^\prime)$ be a bad (w.r.t.\ $(S,V\setminus S)$)
%connected component of $G[\,S\,]$
$G^\prime(V^\prime, E^\prime)\in {\cal B}(S),$
$v\in V^\prime$ and $v^*\in V$.
Then
\begin{eqnarray}
\sum_{\hat{G}(\hat{V},\hat{E})\in {\cal B}(S\setminus \{v\})}\,
%\min_{u\in \hat{V}}\, d\left(u,v^*\right)
d\left(v^*,\hat{V}\right)
%=
\le
\left(\sum_{\hat{G}(\hat{V},\hat{E})\in {\cal B}(S)}\,
%\min_{u\in \hat{V}}\, d\left(u,v^*\right)
d\left(v^*,\hat{V}\right)
\right)
-
%\min_{u\in V^\prime}\, d\left(u,v^*\right).
d\left(v^*,V^\prime\right).
\label{sumofdistancesSside}
\end{eqnarray}
\end{corollary}
\begin{proof}
Immediate from
%Corollary~\ref{listthebadofS}.
Lemma~\ref{changesofbadcomponentsSside}.
\comment{ % can be thought of as said before
%Immediate from Lemma~\ref{changesofbadcomponentsSside}.
Let the bad (w.r.t.\ $(S,V\setminus S)$)
connected components of $G[\,S\,]$ be $G_1=G^\prime,\ldots,G_k$.
By Lemma~\ref{changesofbadcomponentsSside},
the bad (w.r.t.\ $(S\setminus \{v\},V\setminus (S\setminus \{v\}))$)
connected components of $G[\,S\setminus \{v\}\,]$ are
$G_2,\ldots,G_k$.
As a result,
${\cal B}(S)=\{G_1,\ldots,G_k\}$ whereas ${\cal B}(S\setminus
\{v\})=\{G_2,\ldots,G_k\}$.
} % can be thought of as said before
\end{proof}

\comment{ % omit the symmetric for now
The following lemma is symmetric to Lemma~\ref{changesofbadcomponentsSside}.

\begin{lemma}\label{changesofbadcomponentsVSside}
Let $G(V,E)$ be an undirected connected graph,
%$S\subseteq V$ be proper,
$(S,V\setminus S)$ be a proper cut,
$G_1(V_1,E_1), \ldots, G_k(V_k,E_k)$ be the
%$G^\prime(V^\prime, E^\prime)$ be a
%bad
connected components of $G[\,V\setminus S\,]$
and
%$v\in V^\prime$.
$v\in V_1$.
Then
the
%every bad
connected components of $G[\,(V\setminus S)\setminus \{v\}\,]$
%is also a bad connected component of $G[\,S\,]$.
are $G_2, \ldots, G_k$ plus zero or more
connected components that are good w.r.t.\ $(V\setminus S)\setminus \{v\}$.
\end{lemma}
} % omit the symmetric for now

For a proper cut $(S,V\setminus S)$ and $v\in S,$
%$v\in V\setminus (S\setminus \{v\})$
%$G[\,V\setminus (S\setminus \{v\})\,]$
$G[\,(V\setminus S)\cup \{v\}\,]$
has a unique connected component
$\bar{G}(\bar{V},\bar{E})$
that contains $v$.
%with $v\in \bar{V}$
%because $v\in V\setminus
%(S\setminus \{v\})$.
%All other connected components of $G[\,V\setminus (S\setminus \{v\})\,]$
%are those of $G[V\setminus S],$
%as shown below.
%Except for $\bar{G},$
%Below we show that $\bar{G}$ is the only
Every
other
connected component of
%$G[\,V\setminus (S\setminus \{v\})\,]$
$G[\,(V\setminus S)\cup \{v\}\,]$
that
%may be
is
bad w.r.t.\
%$(S\setminus \{v\}, V\setminus (S\setminus \{v\}))$
$(S\setminus \{v\}, (V\setminus S)\cup \{v\})$
%but not
must also be bad
w.r.t.\
$(S,V\setminus S),$
as shown below.

\begin{lemma}\label{componentsonVSside}
Let $G(V,E)$ be an undirected connected graph,
%$S\subseteq V$ be proper
$(S,V\setminus S)$ be a proper cut,
%and $v\in S$.
%Then each connected component of $G[\,V\setminus (S\setminus \{v\})\,],$
%except for the one containing $v,$ is a connected component of
%$G[\,V\setminus S\,]$.
%$G_1(V_1,E_1),\ldots,G_k(V_k,E_k)$ be
%Let
%the connected components of $G[\,V\setminus S\,],$
%and
$v\in S$
and
%and $t\ge 0,$
%where
%.
%$G_1(V_1,E_1),\ldots,G_t(V_t,E_t)$ be the connected components of
%$G[\,V\setminus S\,]$
%Suppose that
$\bar{G}(\bar{V},\bar{E})$ be the connected component of
%$G[V\setminus (S\setminus \{v\})]$
$G[\,(V\setminus S)\cup \{v\}\,]$
%with $v\in \bar{V}$.
that contains $v$.
%If
%each of $G_1,\ldots,G_t$ has a vertex
%in $N_G(v),$ whereas none of $G_{t+1},\ldots,G_k$ does.
%$V_i\cap N_G(v)\neq\emptyset$ for $1\le i\le t$
%but
%not
%$V_i\cap N_G(v)=\emptyset$
%for $t+1\le i\le k$.
Then
%the connected component of $G^\prime(V^\prime,E^\prime)$
%the connected components of $G[\,V\setminus (S\setminus \{v\})\,]$
%are $G_{t+1},\ldots,G_k$ plus one whose set of vertices is
%$\{v\}\cup \bigcup_{1\le i\le t} V_i$.
$$
%{\cal B}\left(V\setminus\left(S\setminus \{v\}\right)\right)
{\cal B}\left(\left(V\setminus S\right)\cup \{v\}\right)
\subseteq {\cal B}\left(V\setminus S\right)\cup
%\left\{G\left[\,\{v\}\cup V_1\cup\cdots\cup V_t
%\,\right]\right\}.
\left\{\bar{G}\right\}.
$$
\end{lemma}
\begin{proof}
Let
$\hat{G_1}(\hat{V_1},\hat{E_1}),\ldots,\hat{G_\ell}(\hat{V_\ell},\hat{E_\ell})$ be
the connected components of $G[\,V\setminus S\,]$
and
$t=|\,\{1\le i\le \ell\mid \hat{V}_i\cap N_G(v)\neq\emptyset\}\,|$.
Without loss of generality, suppose that $\hat{V}_i\cap N_G(v)\neq\emptyset$
for $1\le i\le t$ and $\hat{V}_i\cap N_G(v)=\emptyset$ for $t+1\le i\le \ell$.
%Take $G^\prime=G[\,\{v\}\cup V_1\cup\cdots\cup V_t\,]$.
%Let the connected components of $G[\,V\setminus S\,]$ be
%$G_1(V_1,E_1),\ldots,G_k(V_k,E_k),$ where $1\le t\le k$ of them have
%a vertex
%vertices
%in $N_G(v)\setminus S$.
%Without loss of generality, assume that $G_1,\ldots,G_t$ have
%vertices
%a vertex
%in $N_G(v)\setminus S,$ whereas $G_{t+1},\ldots,G_k$ do not.
Clearly,
%$G[\,\{v\}\cup \hat{V}_1\cup\cdots\cup \hat{V}_t\,]$
%is the connected component of $G[\,V\setminus (S\setminus \{v\})\,]$
%containing
%$v$ as a vertex, i.e.,
$\bar{G}=G[\,\{v\}\cup \hat{V}_1\cup\cdots\cup
\hat{V}_t\,]$.
Besides $\bar{G},$
%Furthermore,
the
other
connected components of
%$G[\,V\setminus (S\setminus \{v\})\,]$
$G[\,(V\setminus S)\cup \{v\}\,]$
are
%easily verified to be
%precisely
%$G[\,\{v\}\cup V_1\cup\cdots\cup V_t\,]$ and
$\hat{G}_{t+1},\ldots,\hat{G}_\ell$.
% and $G^\prime$.
%plus one whose set of vertices is $\{v\}\cup
%\bigcup_{1\le i\le t} V_i$.
%Fix arbitrarily $t+1\le i\le k$.
Hence to complete the proof, we only need to show that
for $t+1\le i\le \ell,$
%$G_i$
%is bad w.r.t.\ $(S\setminus \{v\}, V\setminus (S\setminus \{v\}))$
%$\hat{G}_i\in {\cal B}(V\setminus (S\setminus \{v\}))$
$\hat{G}_i\in {\cal B}((V\setminus S)\cup \{v\})$
%if and
only if
%it is bad w.r.t.\ $(S,V\setminus S)$.
$\hat{G}_i\in {\cal B}(V\setminus S)$.
%which is immediate from Lemma~\ref{goodbadremainsifnotadjacenttov}
%with $\hat{V}_i$ playing the role of $H$.
%\comment{ % needless to explain how to invoke the lemma?
By
%item~\ref{badonVSsideremains} of
Lemma~\ref{goodbadremainsifnotadjacenttov}(\ref{badonVSsideremains}) with $\hat{G}_i$ playing the role of
$H,$
we need only check that
$N_G^*(v)\cap \hat{V}_i=\emptyset$ for $t+1\le i\le \ell,$
which is true because
%$G_{t+1},\ldots,G_k$ are connected components of
%$G[V\setminus ]$.
$v\in S=V\setminus (\hat{V}_1\cup\cdots\cup \hat{V}_k)$ and
$\hat{V}_i\cap N_G(v)=\emptyset$ for $t+1\le i\le \ell$.
%} % needless to explain how to invoke the lemma?
%By the premise that
%As
%$V_i\cap N_G(v)=\emptyset$ for $t+1\le i\le k,$
%each $u\in V_i$ satisfies
%$v\notin N_G(u),$ $t+1\le i\le k$.
\comment{ % we have a lemma for the following purpose
For $t+1\le i\le k,$ the assumption that $V_i\cap N_G(v)=\emptyset$
leads to $v\notin N_G(u)$ for every $u\in V_i$.
Hence for all $t+1\le i\le k$ and $u\in V_i,$
$|\,N_G(u)\cap S\,|=|\,N_G(u)\cap (S\setminus \{v\})\,|$
and $|\,N_G(u)\setminus S\,|=|\,N_G(u)\setminus (S\setminus \{v\})\,|$;
so each $u\in V_i$ is bad w.r.t.\
%$(S,V\setminus S)$
$(S\setminus \{v\},V\setminus (S\setminus \{v\}))$
if and only if it is bad w.r.t.\
%$(S\setminus \{v\},V\setminus (S\setminus \{v\})),$
$(S,V\setminus S),$
$t+1\le i\le k$.
Therefore,
$G_i\in {\cal B}(V\setminus (S\setminus \{v\}))$
if and only if $G_i\in {\cal B}(V\setminus S),$
$t+1\le i\le k$.
%As $v\in V\setminus (S\setminus \{v\})$ and each of $G_1,\ldots,G_t$ has a
%vertex in $N_G(v),$
%the vertices $\{v\}\cup \bigcup_{1\le i\le t} V_i$ will be in one connected
%component of $G[\,V\setminus (S\setminus \{v\})\,]$.
%Each of $G_{t+1},\ldots,G_k$ is a connected component of both
%$G[\,V\setminus (S\setminus \{v\})\,]$
%because $$
} % we have a lemma for the following purpose
\end{proof}

\comment{ % no more need for this corollary
\begin{corollary}\label{nomorebadexceptfortheonewithv}
Let $G(V,E)$ be an undirected connected graph,
$(S,V\setminus S)$ be a proper cut,
$G_1(V_1,E_1),\ldots,G_k(V_k,E_k)$ be
the connected components of $G[\,V\setminus S\,],$
$v\in S$ and $t\ge 0$.
Suppose that each of $G_1,\ldots,G_t$ has a vertex
in $N_G(v),$ whereas none of $G_{t+1},\ldots,G_k$ does.
Then
any bad (w.r.t.\ $(S\setminus \{v\},V\setminus (S\setminus \{v\}))$)
connected component of $G[\,V\setminus (S\setminus \{v\})\,]$
that does not contain $v$
is a bad (w.r.t.\ $(S,V\setminus S)$) connected component of $G[\,V\setminus
S\,]$.
%are $G_{t+1},\ldots,G_k$ plus one whose set of vertices is
%$\{v\}\cup \bigcup_{1\le i\le t} V_i$.
\end{corollary}
\begin{proof}
By Lemma~\ref{componentsonVSside},
the bad (w.r.t.\ $(S\setminus \{v\},V\setminus (S\setminus \{v\}))$)
connected components of $G[\,V\setminus (S\setminus \{v\})\,]$
not containing $v$ are precisely $G_{t+1},\ldots,G_k$.
As none of $G_{t+1},\ldots,G_k$ has a vertex in $N(v),$
for $t+1\le i\le k$ and any vertex $u\in V_i,$
$|\,N(u)\cap S\,|=|\,N(u)\cap (S\setminus \{v\})\,|$
and $|\,N(u)\setminus S\,|=|\,N(u)\setminus (S\setminus \{v\})\,|$;
so $u$ is bad w.r.t.\ $(S,V\setminus S)$ if and only if it is bad w.r.t.\
$(S\setminus \{v\},V\setminus (S\setminus \{v\}))$.
So for $t+1\le i\le k,$
$G_i$ is bad w.r.t.\ $(S,V\setminus S)$ if and only if it is bad
w.r.t.\ $(S\setminus \{v\},V\setminus (S\setminus \{v\}))$.
\end{proof}
} % no more need for this corollary

% no more consider the number of bad components

\begin{corollary}
Let $G(V,E)$ be an undirected connected graph,
%$S\subseteq V$ be proper,
$(S,V\setminus S)$ be a proper cut,
$v\in S,$ $v^*\in V$ and
$\bar{G}(\bar{V}, \bar{E})$ be the connected component of
%$G[\,V\setminus (S\setminus \{v\})\,]$
$G[\,(V\setminus S)\cup \{v\}\,]$
that contains $v$.
%with $v\in \bar{V}$.
Then
\begin{eqnarray}
\sum_{\hat{G}(\hat{V},\hat{E})\in
%{\cal B}(V\setminus (S\setminus \{v\}))}
{\cal B}((V\setminus S)\cup \{v\})}
\,
%\min_{v\in \hat{V}}\, d\left(v,v^*\right)
d\left(v^*,\hat{V}\right)
\le \left(\sum_{\hat{G}(\hat{V},\hat{E})\in {\cal B}(V\setminus S)}\,
%\min_{u\in \hat{V}}\, d\left(u,v^*\right)
d\left(v^*,\hat{V}\right)
\right)
+
%\min_{u\in V^\prime}\, d\left(u,v^*\right).
d\left(v^*,\bar{V}\right).
\label{sumofdistancesVSside}
\end{eqnarray}
\end{corollary}
\begin{proof}
Immediate from
Lemma~\ref{componentsonVSside}.
%Corollary~\ref{listthebadcomponentsofVSside}.
\end{proof}

\comment{ % said in some other form
\begin{lemma}
Let $G(V,E)$ be an undirected connected graph,
%$S\subseteq V$ be proper,
$(S,V\setminus S)$ be a proper cut,
$G^\prime(V^\prime, E^\prime)$ be a connected connected component of
$G[\,S\,],$
$v\in V^\prime$ satisfy $d(v^*,u)=\min_{u\in V^\prime}\, d(v^*,u),$
$\bar{G}(\bar{V}, \bar{E})$ be the connected component of
$G[\,V\setminus (S\setminus \{v\})\,]$ that contains $v$ and $v^*\in V$.
Then
$$\sum_{\hat{G}(\hat{V},\hat{E})\in {\cal B}(V\setminus (S\setminus
\{v\}))}\, \min_{v\in \hat{V}}\, d\left(v^*,v\right)
\le \left(\sum_{\hat{G}(\hat{V},\hat{E})\in {\cal B}(V\setminus S)}\,
\min_{v\in \hat{V}}\, d\left(v^*,v\right)\right)
+\min_{v\in \bar{V}}\, d\left(v^*,v\right).$$
\end{lemma}
} % said in some other form

%Combining Corollaries~\ref{corforthenumberofbadonSside}
%and~\ref{listthebadcomponentsofVSside} yields the following.

We now arrive at the following key lemma, which allows us to repeatedly
%modify cuts to
increase the $\psi(\cdot,v^*)$-values of cuts by moving one vertex at a time.
%Namely,
%the $\psi(\cdot,v^*)$-value of
%for
%a bad
%(w.r.t.\ $(S,V\setminus S)$) connected component of $G[\,S\,],$
%can be increased by moving a suitably chosen vertex to the other side
%of the cut.
%moving
%its nearest vertex to $v^*$
% stating that
%moving a vertex of a bad
%(w.r.t.\ $(S,V\setminus S)$) connected component to the other side of the
%cut

\begin{lemma}\label{increasethepotential}
Let $G(V,E)$ be an undirected connected
graph, $(S,V\setminus S)$ be a proper cut,
$G^\prime(V^\prime,E^\prime)\in {\cal B}(S),$ $v\in V^\prime$ and $v^*\in
V\setminus V^\prime$.
If
%satisfy
$d(v^*,v)=d(v^*,V^\prime),$
then
$\psi(S\setminus \{v\},v^*)>\psi(S,v^*)$.
%$e(S,V\setminus S)=e(S\setminus \{v\},V\setminus (S\setminus \{v\}))$
%and
%\begin{eqnarray}
%\left|\,{\cal B}\left(S\setminus \{v\}\right)\,\right|
%+\left|\,{\cal B}\left(V\setminus \left(S\setminus \{v\}\right)\right)\,\right|
%\le \left|\,{\cal B}(S)\,\right| + \left|\,{\cal B}\left(V\setminus S\right)\,\right|.
%\label{justasumup}
%\end{eqnarray}
\end{lemma}
\begin{proof}

\comment{ % already deprived the potential function from this term
Summing up inequalities~(\ref{inequalitycorforthenumberofbadonSside})
and~(\ref{inequalitylistthebadcomponentsofVSside}),
%Inequality~(\ref{justasumup}) follows directly from
%Corollaries~\ref{corforthenumberofbadonSside}
%and~\ref{listthebadcomponentsofVSside}.
\begin{eqnarray}
\left|\,{\cal B}\left(S\setminus \{v\}\right)\,\right|
+\left|\,{\cal B}\left(V\setminus \left(S\setminus \{v\}\right)\right)\,\right|
\le \left|\,{\cal B}(S)\,\right| + \left|\,{\cal B}\left(V\setminus S\right)\,\right|.
\label{badnumberinequality}
\end{eqnarray}
} % already deprived the potential function from this term

As $v\in V^\prime$ and $v^*\notin V^\prime,$ $d(v^*,v)>0$.
%By the connectedness of $G,$
As $G$ is connected,
there exists a vertex $w\in N_G(v)$ with $d(v^*,w)=d(v^*,v)-1$.
We must have
%which implies
$w\notin V^\prime$ because $d(v^*,v)=d(v^*,V^\prime)$
says that
$v$ is among the vertices in $V^\prime$ that are closest to $v^*$.
%(recall
%that $d(v^*,V^\prime)=\min_{u\in V^\prime}\, d(v^*,u)$).
Suppose for contradiction that $w\in S$.
Then
%$w\in V^\prime$
%because
the facts that
$v\in V^\prime,$ $w\in N_G(v)$
and $G^\prime(V^\prime,E^\prime)$ is a connected component of $G[\,S\,]$
%imply
force
$w\in V^\prime,$ a contradiction.
So $w\notin S$
%which
%together with the fact that $w\neq v$
%implies that
and, therefore,
\begin{eqnarray}
w\in \left(V\setminus S\right)\cup \{v\}.\label{thecloserontheotherside1}
\end{eqnarray}
Trivially,
\begin{eqnarray}
v\in \left(V\setminus S\right)\cup \{v\}.\label{thecloserontheotherside2}
\end{eqnarray}
%This
Eqs.~(\ref{thecloserontheotherside1})--(\ref{thecloserontheotherside2})
and the fact that $w\in N_G(v)$ put $w$ and $v$ in the same connected
component of
%$G[\,V\setminus (S\setminus \{v\})\,],$
$G[\,(V\setminus S)\cup \{v\}\,],$
denoted
$\bar{G}(\bar{V},\bar{E})$.
Note that
\begin{eqnarray}
d\left(v^*,\bar{V}\right)
\le d\left(v^*,w\right)
= d\left(v^*,v\right)-1
= d\left(v^*,V^\prime\right)-1.
\label{getonestepclosertwocomponents}
\end{eqnarray}
Summing
inequalities~(\ref{sumofdistancesSside})--(\ref{sumofdistancesVSside}),
we have
\begin{eqnarray}
%\left(
&&\sum_{\hat{G}(\hat{V},\hat{E})\in {\cal B}(S\setminus \{v\})}\,
d\left(v^*,\hat{V}\right)
+\sum_{\hat{G}(\hat{V},\hat{E})\in {\cal B}((V\setminus S)\cup \{v\})}\,
d\left(v^*,\hat{V}\right)\nonumber\\
%\right)
&\le&
\left(\sum_{\hat{G}(\hat{V},\hat{E})\in {\cal B}(S)}\,
d\left(v^*,\hat{V}\right)
+\sum_{\hat{G}(\hat{V},\hat{E})\in {\cal B}(V\setminus S)}\,
d\left(v^*,\hat{V}\right)\right)
- d\left(v^*,V^\prime\right)
+ d\left(v^*,\bar{V}\right)\nonumber\\
&\le&
\left(\sum_{\hat{G}(\hat{V},\hat{E})\in {\cal B}(S)}\,
d\left(v^*,\hat{V}\right)
+\sum_{\hat{G}(\hat{V},\hat{E})\in {\cal B}(V\setminus S)}\,
d\left(v^*,\hat{V}\right)\right)
-1,\label{onestepclosertotal}
\end{eqnarray}
where the last inequality follows from inequality~(\ref{getonestepclosertwocomponents}).
As $G^\prime$ is bad w.r.t.\ $(S,V\setminus S)$ and $v\in V^\prime,$
Lemma~\ref{cutsizeremainsundermovingbadverticeslemma} gives
%, the badness of $G^\prime$ w.r.t.\ $(S,V\setminus S)$ and $v\in V^\prime,$
%As $G^\prime$ is bad w.r.t.\ $(S,V\setminus S)$ and $v\in V^\prime,$
%$|\,N(v)\cap S\,|=|\,N(v)\setminus S\,|,$
%which implies
%implying
\begin{eqnarray*}
e\left(S,V\setminus S\right)
%=e\left(S\setminus \{v\}, V\setminus S\right)+\left|\,N(v)\setminus S\,\right|
%=e\left(S\setminus \{v\}, V\setminus S\right)+\left|\,N(v)\cap S\,\right|
%=e\left(S\setminus \{v\},\left(V\setminus S\right)\cup \{v\}\right)
=e\left(S\setminus \{v\},
%\left(V\setminus S\right)\setminus \{v\}\right).
\left(V\setminus S\right)\cup \{v\}\right).
%\label{cutsizeinequality}
\end{eqnarray*}
%because (1) the edges incident on $v$ contribute $|\,N(v)\setminus S\,|$
%to the lefthand side and $|\,N(v)\cap S\,|$ to the righthand side and (2)
%all other edges contribute the same amount to either side.
%By Eq.~(\ref{cutsizeinequality}) and
This and
inequality~(\ref{onestepclosertotal}) show that
$\psi(S\setminus \{v\},v^*)>\psi(S,v^*)$.
\end{proof}

%Repeatedly modifying cuts to increase their $\psi(\cdot,v^*)$-values, as
%made possible by the above lemma, yields a cut $(S,V\setminus S)$ where $G[\,S\,]$ and
%$G[\,V\setminus S\,]$ have at most one bad (w.r.t.\ $(S,V\setminus S)$) connected component in
%total.
The above lemma allows us to increase the $\psi(\cdot,v^*)$-values of cuts
whenever there is a bad connected component of $G[\,S\,]$
that does not contain $v^*$.
%Such a cut will enable our bound on the minimum size of irreversible
%dynamos.
As the $\psi(\cdot,v^*)$-values are bounded from above,
they cannot be increased forever.
So repeatedly applying the above lemma will finally yield a cut where
%there is
%only one possible bad component of either $G[\,S\,]$ or $G[\,V\setminus S\,],$
%i.e., the one containing $v^*$.
all bad connected components of $G[\,S\,]$ must contain $v^*,$
%essentially running
meaning
$|{\cal B}(S)|=1$
%down to one (
as $v^*$ cannot appear in
two
%distinct
connected components.
Such a result is stated below,
%in a slightly more general form that
which
considers $G[\,V\setminus S\,]$ as well.

\begin{lemma}\label{getthecut}
Given an undirected connected graph $G(V,E),$
%and $v^*\in V,$
a proper cut $(S,V\setminus S)$ with
\begin{eqnarray}
\left|\,{\cal B}(S)
\cup
%\,\right|+\left|\,
{\cal B}\left(V\setminus S\right)\,\right|\le 1
\label{onlyonebad}
\end{eqnarray}
can be found in
polynomial
time.
%polynomial in $|\,V\,|$.
\end{lemma}
\begin{proof}
Fix $v^*\in V$ arbitrarily.
By Fact~\ref{localadjustment}, a proper cut $(S_0,V\setminus S_0)$
can be found in time polynomial $|\,V\,|$.
%Unless
If
$$\left|\,{\cal B}(S_0)
%\,|+|\,
\cup
{\cal B}(V\setminus S_0)\,\right|
\le
1,$$
%then
taking $S=S_0$ satisfies inequality~(\ref{onlyonebad}).

Inductively, let $(S_i,V\setminus S_i)$ be a proper cut with
\begin{eqnarray}
\left|\,{\cal B}(S_i)
\cup
%\,\right|+\left|\,
{\cal B}\left(V\setminus S_i\right)\,\right|
>1,
\label{ithhastoomanybadcomponents}
\end{eqnarray}
where $i\ge 0$.
We show how to compute a proper cut $(S_{i+1},V\setminus S_{i+1})$
with
\begin{eqnarray}
\psi\left(S_{i+1},v^*\right)>\psi\left(S_{i},v^*\right)
\label{improvedcut}
\end{eqnarray}
in time polynomial in $|\,V\,|$.
%Pick
%in polynomial (in $|\,V\,|$) time
%any $G^\prime(V^\prime,E^\prime)\in {\cal B}(S_i)\cup {\cal
%B}(V\setminus S_i)$ with $v^*\notin V^*$.
%Using the depth-first search,
The connected components of $G[\,S_i\,]$ and $G[\,V\setminus S_i\,]$ can
be found in time polynomial in $|\,V\,|$ using the breadth-first search~\cite{CLRS01}.
By inequality~(\ref{ithhastoomanybadcomponents}),
%one can pick
we can pick
an arbitrary $G^\prime(V^\prime,E^\prime)\in {\cal B}(S_i)\cup {\cal B}(V\setminus S_i)$
with $v^*\notin V^\prime$.
Assume without loss of generality that $G^\prime\in {\cal B}(S_i)$;
otherwise
%for, otherwise,
we
%may
switch $S_i$ and $V\setminus S_i$ from the beginning.
By computing $d(v^*,u)$ for every $u\in V^\prime$ using the breadth-first
search, we
%may pick
find
a $v\in V^\prime$ with $d(v^*,v)=d(v^*,V^\prime)$
in time polynomial in $|\,V\,|$.
As $v\in V^\prime$ and $G^\prime\in {\cal B}(S_i)$ is bad w.r.t.\
$(S_i,V\setminus S_i),$
%$|\,N_G(v)\cap S_i\,|=|\,N_G(v)\setminus S_i\,|,$
%implying
Lemma~\ref{cutsizeremainsundermovingbadverticeslemma} implies that
\begin{eqnarray}
e\left(S_i,V\setminus
S_i\right)
=e\left(S_i\setminus \{v\},\left(V\setminus S_i\right)\cup
\{v\}\right).
\label{cutsizeremains}
\end{eqnarray}
%because (1) the edges incident on $v$ contribute $|\,N_G(v)\setminus S_i\,|$
%to the lefthand side and $|\,N_G(v)\cap S_i\,|$ to the righthand side and
%(2) all other edges contribute the same to either side.
%If $(S_i\setminus \{v\}, V\setminus (S_i\setminus \{v\}))$ is proper,
%then
By Lemma~\ref{increasethepotential},
$\psi(S_i\setminus \{v\},v^*)>\psi(S_{i},v^*)$.
% for $S_{i+1}=S_i\setminus \{v\}$.
%So
Therefore, if
%$(S_i\setminus \{v\}, V\setminus (S_i\setminus \{v\}))$
$(S_i\setminus \{v\}, (V\setminus S_i)\cup \{v\})$
is proper,
then
%\begin{eqnarray}
%\psi\left(S_{i+1},v^*\right)>\psi\left(S_{i},v^*\right)
%\label{improvedcut}
%\end{eqnarray}
inequality~(\ref{improvedcut}) holds
for a proper cut $(S_{i+1},V\setminus S_{i+1})$ by taking
$S_{i+1}=S_i\setminus \{v\}$.
Otherwise,
Fact~\ref{localadjustment} implies that a proper cut
%$(T,V\setminus T)$ with $e(T,V\setminus T)> e(S_i\setminus \{v\},V\setminus (S_i\setminus \{v\}))$
$(T,V\setminus T)$ with $e(T,V\setminus T)> e(S_i\setminus \{v\},(V\setminus
S_i)\cup \{v\})$
can be found in time polynomial in $|\,V\,|$.
Hence by Eq.~(\ref{cutsizeremains}),
$e(T,V\setminus T)>e(S_i,V\setminus S_i),$
implying
%So
$\psi(T,v^*)>\psi(S_{i},v^*)$
by Lemma~\ref{cutsizedominate}.
Again, inequality~(\ref{improvedcut}) holds for a proper cut
$(S_{i+1},V\setminus S_{i+1})$ by taking $S_{i+1}=T$.
%for $S_{i+1}=T$
%because the
%$e(S,V\setminus S)$ term is multiplied by $|\,V\,|^2>|\,V\,|\,(|\,V\,|-1)$ in Eq.~(\ref{potential})
%whereas
%$|\,{\cal B}(S)\,|+|\,{\cal B}(V\setminus S)\,|\le |\,V\,|$ and
%$d(U,v^*)\le |\,V\,|-1$ for any $\emptyset\subsetneq U\subseteq V$.
%In either case,
%we obtain
%a proper cut $(S_{i+1},V\setminus S_{i+1})$
%with $\psi(S_{i+1},v^*)>\psi(S_{i},v^*)$.

Now continue computing a proper cut $(S_{i+1},V\setminus S_{i+1})$
with $\psi(S_{i+1},v^*)>\psi(S_{i},v^*)$
from $(S_i,V\setminus S_i)$
%for an increasing $i\ge 0$
until inequality~(\ref{ithhastoomanybadcomponents})
fails for some $i\ge 0$.
%If $\psi(S_{i+1},v^*)>\psi(S_{i},v^*)$ for some $t\in\mathbb{N}$
%and all
%$0\le i< t,$
%then $\psi(S_t,v^*)\ge \psi(S_0,v^*)+t$.
%with
%$\psi(S_{i+1},v^*)>\psi(S_{i},v^*)$
%satisfy $\psi(S_{i+1},v^*)>\psi(S_{i},v^*)$
%for an increasing $i\ge 0$
%until
%inequality~(\ref{ithhastoomanybadcomponents})
%We can now compute proper cuts $(S_i,V\setminus S_i)$ for
%$i\ge 0$
%such that inequality~(\ref{ithhastoomanybadcomponents}) implies that
%$(S_{i+1},V\setminus S_{i+1})$ is a proper cut and
%$\psi(S_{i+1},v^*)>\psi(S_{i},v^*)$.
As $|\,\psi(\cdot,\cdot)\,|$ is at most
polynomial in $|\,V\,|,$
there is a $k\in \mathbb{N},$ which is at most polynomial in $|\,V\,|,$
with
\begin{eqnarray*}
\left|\,{\cal B}(S_k)
\cup
%\,\right|+\left|\,
{\cal B}\left(V\setminus
S_k\right)\,\right|\le
1,
\end{eqnarray*}
completing the proof.
\end{proof}

%Now we have
The above lemma provides us with
a cut $(S,V\setminus S)$ where $G[\,S\,]$ and
$G[\,V\setminus S\,]$
together
have at most one bad (w.r.t.\ $(S,V\setminus S)$)
connected component.
Next, we show that $S$ or $V\setminus S$ plus one vertex from the only bad
component (if it exists) is an irreversible dynamo under the strict-majority
scenario.

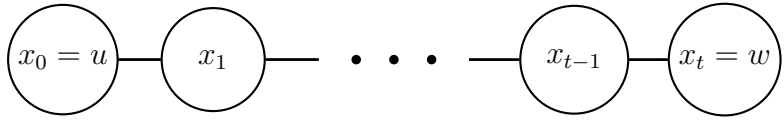
\begin{figure}
\centering
\begin{pspicture}(0,-0.5)(8.8,0)
\cnodeput[](0,0){u}{$x_0=u$}
\cnodeput[](2,0){x1}{$\,\,\,\,\,x_1\,\,\,\,\,$}
\pnode(3.4,0){dummy1}
\psdots(3.9,0)(4.4,0)(4.9,0)
\pnode(5.4,0){dummy2}
\cnodeput[](6.8,0){xtminus1}{$\,\,\,x_{t-1}\,\,\,$}
\cnodeput[](8.8,0){w}{$x_t=w$}

%\definecolor{nearblack}{rgb}{0,1,0}
\ncline[]{u}{x1}
\ncline[]{u}{x1}
\ncline[]{u}{x1}
\ncline[]{x1}{dummy1}
\ncline[]{x1}{dummy1}
\ncline[]{x1}{dummy1}
\ncline[]{dummy2}{xtminus1}
\ncline[]{dummy2}{xtminus1}
\ncline[]{dummy2}{xtminus1}
\ncline[]{xtminus1}{w}
\ncline[]{xtminus1}{w}
\ncline[]{xtminus1}{w}
% a single line doesn't appear completely black, some pixels appear a little
% bit gray. Putting multiple lines that are in fact the same solves it.
\end{pspicture}
\caption{Consider
%a proper cut $(S,V\setminus S)$ and
a path $x_0=u,x_1,\ldots,x_{t-1},x_t=w$ lying in a connected
component of $G[\,V\setminus S\,]$.
If $(S,V\setminus S)$ is a proper cut,
then each of $x_0,\ldots,x_t$
has more or equally many neighbors in $S$ than in $V\setminus S$.
%for $0\le i\le t$.
Thus, when $x_i$ and all the vertices in $S$ are colored white,
$x_{i+1}$ will have strictly more white neighbors than black ones, $0\le i< t$.
%needs at most one
%white neighbor to be colored white.
Consequently,
coloring $x_0$ and the vertices in $S$
white
%will
%also
can
color $x_1,\ldots,x_t$ white, in that order,
under the strict-majority
scenario.
%Theorem~\ref{maintheoremundirectedstrict}
}
\label{activateallfigure}
\end{figure}

\begin{theorem}\label{maintheoremundirectedstrict}
Given an undirected connected graph $G(V,E),$
an irreversible dynamo of
${\cal N}(G,\phi^\text{strict})$ with size
at most $\lceil |\,V\,|/2\rceil$
can be found in polynomial time.
% polynomial in $|\,V\,|$.
\end{theorem}
\begin{proof}
%By
Lemma~\ref{getthecut} says
a proper cut $(S,V\setminus S)$ with
%\begin{eqnarray}
$|\,{\cal B}(S)
\cup
%\,|+|\,
{\cal B}(V\setminus S)\,|\le 1$
%\end{eqnarray}
can be found in time polynomial in $|\,V\,|$.
%Let $G^\prime(V^\prime,E^\prime)$ be the unique member of ${\cal B}(S)\cup
%{\cal B}(V\setminus S)$ if $|\,{\cal B}(S)\,|+|\,{\cal B}(V\setminus
%S)\,|=1$ and an arbitrary connected component of $G[\,S\,]$ or
%$G[\,V\setminus S\,]$ otherwise.
%Take $v\in V^\prime$ arbitrarily.
(1) If $|\,{\cal B}(S)
\cup
%\,|+|\,
{\cal B}(V\setminus S)\,|=1,$
let $x\in V$ be an arbitrary vertex of the unique member of ${\cal B}(S)\cup
{\cal B}(V\setminus S)$.
(2) Otherwise, take any $x\in V$.

Pick
%arbitrarily
any connected component $H(V_H,E_H)$
of $G[\,V\setminus S\,]$.
We
%proceed to
%now
next
show that
%any connected component $G^\prime(V^\prime,E^\prime)$
%of $G[\,V\setminus S\,]$
%satisfies
%has a vertex in
\begin{eqnarray}
V_H\cap c\left(S\cup \{x\},G,\phi^\text{strict}\right)\neq \emptyset.
\label{eachcomponenthasactivevertex}
\end{eqnarray}
If $H\in {\cal B}(S)\cup {\cal B}(V\setminus S),$
then
$x\in V_H$
%$v\in c(S\cup \{v\},G,\phi^\text{strict})$ is a vertex of $G^\prime$
by
our choice of $x$ in case (1) above,
%construction,
proving inequality~(\ref{eachcomponenthasactivevertex}).
Otherwise, $H$ must be a good (w.r.t.\ $(S,V\setminus S)$) connected component of $G[\,V\setminus
S\,]$.
So there exists a vertex $u\in V_H$
with $|\,N_G(u)\setminus S\,|\neq |\,N_G(u)\cap S\,|,$
which together with the
%fact that
properness of
$(S,V\setminus S)$
%is proper
yields
$$\left|\,N_G(u)\setminus S\,\right|< \left|\,N_G(u)\cap S\,\right|.$$
%So by definition,
This gives
%Therefore,
%implying
%$u\in c(S\cup \{v\},G,\phi^\text{strict}),$
$u\in c(S,G,\phi^\text{strict})$ by definition,
which implies
$u\in c(S\cup \{x\},G,\phi^\text{strict})$
by Fact~\ref{monotone}.
%and thus
Again,
%validating
inequality~(\ref{eachcomponenthasactivevertex})
%by Fact~\ref{monotone}.
%is validated.
holds.

Next, we prove that $V\setminus S\subseteq c(S\cup
\{x\},G,\phi^\text{strict})$.
For this purpose,
%To prove that $V\setminus S\subseteq c(S\cup \{v\},G,\phi^\text{strict}),$
we need only show that every $w\in V_H$
%$V^\prime\subseteq
belongs to
$c(S\cup
\{x\},G,\phi^\text{strict})$ because $H$ is an arbitrary
%for each
connected component
%$G^\prime(V^\prime,E^\prime)$
of $G[\,V\setminus S\,]$.
Let $u\in V_H\cap c(S\cup \{x\},G,\phi^\text{strict}),$
%which exists
whose existence is guaranteed
by inequality~(\ref{eachcomponenthasactivevertex}).
As $w,u\in V_H$ and $H$ is a connected component
of $G[\,V\setminus S\,],$
there is a path $x_0=u,\ldots,x_t=w$ whose vertices are in $V_H$.
%To complete the proof, we
We proceed to
%We now
%use mathematical induction to
show that
$w\in c(S\cup \{x\},G,\phi^\text{strict})$ by induction.
%as
%illustrated in
See Fig.~\ref{activateallfigure} for illustration.
The induction base is $x_0\in c(S\cup \{x\},G,\phi^\text{strict}),$
which is true by construction.
Inductively, assume $x_i\in c(S\cup \{x\},G,\phi^\text{strict}),$
$0\le i< t$.
Clearly, $\{x_i\}\cup S\subseteq c(S\cup \{x\},G,\phi^\text{strict})$; hence
\begin{eqnarray}
&&\left|\,N_G\left(x_{i+1}\right)\cap c\left(S\cup
\{x\},G,\phi^\text{strict}\right)\,\right|\nonumber\\
&\ge& \left|\,N_G\left(x_{i+1}\right)\cap \left(\{x_i\}\cup
S\right)\,\right|\nonumber\\
&=& \left|\,N_G\left(x_{i+1}\right)\cap \left\{x_i\right\}\,\right|
+\left|\,N_G\left(x_{i+1}\right)\cap S\,\right|.\nonumber\\
&=& 1 + \left|\,N_G\left(x_{i+1}\right)\cap S\,\right|.\label{theiplusone}
\end{eqnarray}
%Above, the first equality uses the fact that $x_i\notin S,$ which is true
%because $x_i\in V_H$ and $H$ is a connected component of $G[\,V\setminus
%S\,]$.
%The second equality holds by construction of the $x_i$'s.
%where the equalities hold because $x_0,\ldots,x_t$ form a path
%with vertices in $V_H$ and $H$ is a connected component of $G[\,V\setminus S\,]$.
%where the first equality follows from $x_i\in V^\prime$ and
%$V^\prime\subseteq V\setminus S$.
As $S$ is proper,
$|\,N_G(x_{i+1})\setminus S\,|\le |\,N_G(x_{i+1})\cap S\,|,$
which together with inequality~(\ref{theiplusone}) gives
$$\left|\,N_G\left(x_{i+1}\right)\cap c\left(S\cup \{x\},G,\phi^\text{strict}\right)\,\right|
>\frac{N_G\left(x_{i+1}\right)}{2};$$
thus $x_{i+1}\in c(S\cup \{x\},G,\phi^\text{strict})$.

We have shown that
$V\setminus S\subseteq c(S\cup \{x\},G,\phi^\text{strict}),$
which yields $V=c(S\cup \{x\},G,\phi^\text{strict})$.
By symmetry,
$V=
%S\subseteq
c((V\setminus S)\cup \{x\},G,\phi^\text{strict})$.
So both $S\cup \{x\}$ and $(V\setminus S)\cup \{x\}$ are irreversible
dynamos of ${\cal N}(G,\phi^\text{strict})$.
%The smaller of $S\cup \{x\}$ and $(V\setminus S)\cup \{x\}$
%can be easily verified to
%must
%have size at most $\lceil |\,V\,|/2\rceil$.
To complete the proof, it remains to show that the smaller of $S\cup \{x\}$
and $(V\setminus S)\cup \{x\}$ has size at most $\lceil |\,V\,|/2\rceil$.
%If $\min\{|\,S\,|,|\,V\setminus S\,|\}<|\,V\,|/2,$
%then
As $x$ lies in exactly one of $S$ and $V\setminus S,$
$$|\,V\,|=\left|\,S\cup \{x\}\,\right|
+ \left|\,\left(V\setminus S\right)\cup \{x\}\,\right|-1,$$
forcing the smaller of $|\,S\cup \{x\}\,|$ and $|\,(V\setminus S)\cup
\{x\}\,|$ to be at most $\lfloor(|\,V\,|+1)/2\rfloor=\lceil |\,V\,|/2\rceil$.
\end{proof}

The
%$|\,V\,|/2$ upper
bound
of
%$1/2$ constant
%in
%Theorem~\ref{maintheoremundirectedsimple}
Theorem~\ref{maintheoremundirectedstrict}
cannot be lowered because
$\text{min-seed}(G,\phi^\text{strict})=\lceil|\,V\,|/2\rceil$
%when every connected component of $G$ consists of two vertices and an
%undirected edge between them.
when $G$ is the complete graph on $V$.
That is, among all undirected connected graphs on $V,$
%without isolated vertices,
the complete graph attains the maximum value for
$\text{min-seed}(G,\phi^\text{strict})$.
% with an even number of vertices.
%That is,
%$\text{min-seed}(G,\phi^\text{simple})\le |\,V\,|/2$
%is maximized by the complete graph on $V,$
%for all undirected graph without isolated vertices
%the $|\,V\,|/2$ bound of Theorem~\ref{maintheoremundirectedsimple}
%is tight
%An irreversible dynamo
%under either the strict or the simple-majority scenario
%is often interpreted
%as a set of processors whose faulty
%behavior leads all processors to erroneous
%results~\cite{FGS01, FKRRS03, FLLPS04, LPS99, Pel02}.
Under
%this interpretation,
the interpretation of an irreversible dynamo as a set of processors whose
faulty
behavior leads all processors to erroneous
results,
therefore,
%Therefore,
%Theorem~\ref{maintheoremundirectedsimple} and the fact that
%$\text{min-seed}(G,\phi^\text{simple})=|\,V\,|/2$ for a complete graph
fully interconnecting the processors
maximizes the
%minimum
number of
adversarially placed
faulty processors needed to
%produce erroneous results on all processors.
render all processors' results erroneous.

%\section{Upper bounds for the simple-majority
%scenario}\label{simplemajoritysection}
%The results in the previous section can be adapted to
%deal with
%the
%simple-majority scenario.

\comment{ % again, we don't really need this
An immediate Corollary follows.

\begin{corollary}\label{copythecombine}
Let $G(V,E)$ be a digraph,
$u\in V,$
%and
$S\subseteq N^\text{in}(u)$
have
size $\lceil\text{deg}^\text{in}(u)/2\rceil$
and $H=G[\,V\setminus c(S,G,\phi_G^\text{simple})\,]$.
%, $S\subseteq V$ satisfy
%$V\setminus c(S,G,\phi_G^\text{strict})\neq\emptyset$
%and $G^\prime$ be the subgraph of $G$ induced by $V\setminus
%c(S,G,\phi_G^\text{strict})$.
Then for
%and
any irreversible dynamo $T$ of
%$H=G[\,V\setminus
%c(S,G,\phi_G^\text{simple})\,]$ under the simple-majority scenario,
${\cal N}(H,\phi_H^\text{strict}),$
$S\cup T$ is an irreversible dynamo of
%$G$ under the simple-majority scenario.
${\cal N}(G,\phi_G^\text{strict})$.
\end{corollary}
} % again, we don't really need this

\comment{ % Again, not too sure if we want such a corollary if we sell it as tolerating the most faults
\begin{corollary}
For any digraph
$G(V,E),$
${\cal N}(G,\phi^\text{simple})$
has an irreversible dynamo of size at most
$$0.727\cdot \left|\,\left\{v\in V\mid
\text{deg}^\text{in}(v)>0\right\}\,\right|
+\left|\,\left\{v\in V\mid
\text{deg}^\text{in}(v)=0\right\}\,\right|.$$
\end{corollary}
\begin{proof}
%In the proof of Corollary~\ref{simplecorollary},
Replace each occurrence of
``strict,'' ``0.7732'' and ``Theorem~\ref{maintheorem}''
in the proof of Corollary~\ref{simplecorollary}
by ``simple,'' ``0.727'' and ``Theorem~\ref{maintheoremsimplemajority},''
respectively.
\end{proof}
} % Again, not too sure if we want such a corollary if we sell it as tolerating the most faults

\comment{ % the local-adjustment fact is moved earlier
For the undirected case,
%undirected graphs without isolated vertices,
%we prove that every
%network
%has an irreversible dynamo of size at most $|\,V\,|/2$
%under the simple-majority scenario.
%For this purpose,
%we introduce the following fact.
we need the following fact.

\begin{fact}(\cite[pp.\ 303--304]{Pap94})\label{maxcutpartition}
Given
an
%a simple
undirected graph $G(V,E),$
a set $S\subseteq V$
with the following properties
can be found in
%time polynomial in $|\,V\,|$:
polynomial time:
\begin{enumerate}
%\addtolength{\itemsep}{-0.2\baselineskip}
%\addtolength{\itemsep}{-0.5\baselineskip}
\item\label{localoptimalone} $\forall v\in S,$ $e(S,V\setminus S)\ge e(S\setminus\{v\}, (V\setminus
S)\cup \{v\})$.
\item\label{localoptimaltwo} $\forall v\in V\setminus S,$ $e(S,V\setminus S)\ge e(S\cup\{v\}, (V\setminus
S)\setminus \{v\})$.
\end{enumerate}
%such that $e(S,V\setminus S)$ cannot be increased by moving any $v\in S$ to
%$V\setminus S$ or any $u\in V\setminus S$ to $S$.
\end{fact}
} % the local-adjustment fact is moved earlier

\comment{ % why say it? Not really used
In fact, the partition $(S,V\setminus S)$ of $V$ in
Fact~\ref{maxcutpartition}
constitutes a
$0.5$-approximation for the {\sc max cut} problem~\cite{Pap94}.
%Below we show that both $S$ and $V\setminus S$ are irreversible dynamos of
%${\cal N}(G,\phi^\text{simple}),$
%provided that $G$ does not have
%isolated vertices.
} % why say it? Not really used

\comment{ % change to a brief intro of the next lemma
Below
we prove that every
%network
undirected
%connected
graph
without isolated vertices
%with at least two vertices
has an irreversible dynamo of size at most $\lfloor|\,V\,|/2\rfloor$
under the simple-majority scenario.
} % change to a brief intro of the next lemma

% this is weaker than previous results

%an even number of processors
%results in a system that tolerates
%faulty behavior from any set of $|\,V\,|/2$ processors
%the maximum number of arbitrarily placed faulty processors
%before all processors produce erroneous results.
%without inducing erroneous results on all processors.
%a system represented by
%a complete graph can tolerate behavior from half of the processors.
\comment{ % maybe not need such a corollary because we don't do general graphs
An immediate corollary of Theorem~\ref{maintheoremundirectedsimple} follows.

\begin{corollary}
Given
an
%a simple
undirected graph $G(V,E),$
an irreversible dynamo of
${\cal N}(G,\phi^\text{simple})$ with
%$G$ with
size at most
%$|\,V\,|/2$
$$\frac{\left|\,\left\{v\in V\mid \text{deg}(v)\neq 0\right\}\,\right|}{2}
+\left|\,\left\{v\in V\mid \text{deg}(v)=0\right\}\,\right|$$
can be found
%in time polynomial in $|\,V\,|$.
in polynomial time.
\end{corollary}
} % maybe not need such a corollary because we don't do general graphs

% the bounds for regular is no better

\section{Inapproximability}\label{inapproximabilitysection}
\begin{figure}
\centering
\begin{pspicture}(0.5,0.5)(9.7,11.3)
%(0.5,1)(9.7,11.3) % real size of the picture
\cnodeput[fillstyle=solid,fillcolor=lightgray](0.5,1){X1}{${\cal X}$}
\psdots[dotsize=3pt](1,1)(1.2,1)(1.4,1)
\cnodeput[fillstyle=solid,fillcolor=lightgray](1.9,1){X2}{${\cal X}$}

\cnodeput[fillstyle=solid,fillcolor=lightgray](2.9,1){X3}{${\cal X}$}
\psdots[dotsize=3pt](3.4,1)(3.6,1)(3.8,1)
\cnodeput[fillstyle=solid,fillcolor=lightgray](4.3,1){X4}{${\cal X}$}

\cnodeput[fillstyle=solid,fillcolor=lightgray](5.3,1){X5}{${\cal X}$}
\psdots[dotsize=3pt](5.8,1)(6,1)(6.2,1)
\cnodeput[fillstyle=solid,fillcolor=lightgray](6.7,1){X6}{${\cal X}$}

\cnodeput[fillstyle=solid,fillcolor=lightgray](7.7,1){X7}{${\cal X}$}
\cnodeput[fillstyle=solid,fillcolor=lightgray](8.7,1){X8}{${\cal X}$}
\cnodeput[fillstyle=solid,fillcolor=lightgray](9.7,1){X9}{${\cal X}$}

\cnodeput[fillstyle=solid,fillcolor=lightgray,doubleline=true](2.5,3){N1}{$v_1$}
\cnodeput[fillstyle=solid,fillcolor=lightgray,doubleline=true](4.5,3){N2}{$v_2$}
\cnodeput[fillstyle=solid,fillcolor=lightgray,doubleline=true](6.5,3){N3}{$v_3$}
\cnodeput[fillstyle=solid,fillcolor=lightgray,doubleline=true](8.5,3){v}{$\,v\,$}
\psarc[linewidth=0.05](8.5,3){0.7}{-150}{-30}
\rput(10,2.6){\rnode{markedgesofvtoX}{$\text{deg}_G(v)$}}
\psarc[linewidth=0.05](8.5,3){0.7}{70}{170}
\rput(10,3.6){\rnode{markedgesofvtoW}{$\text{deg}_G(v)+1$}}

\cnodeput[](2.5,5){w1}{$w_{v_1}$}
\cnodeput[](4.5,5){w2}{$w_{v_2}$}
\cnodeput[](6.5,5){w3}{$w_{v_3}$}
\cnodeput[fillstyle=solid,fillcolor=lightgray](8.5,5){wv}{$\,w_v\,$}
\psarc[linewidth=0.05](8.5,5){0.7}{30}{140}
\rput(10,5.4){\rnode{markedgesofwvtoY}{$\text{deg}_G(v)$}}
\psarc[linewidth=0.05](8.5,5){0.7}{-170}{-70}
\rput(10,4.4){\rnode{markedgesofwvtoV}{$\text{deg}_G(v)+1$}}

\comment{ % polygon enclosing B_v
\pspolygon[linestyle=dotted,linewidth=0.1](-0.2,0.3)(10.4,0.3)(10.4,6)(7.5,6)(7.5,4.4)(-0.2,4.4)
%(-1,8.5)(-1,-1.7)(5,-1.7)(5,1)(3.4,1)(3.4,8.5)(3.4,8.5)
\rput(0.5,3){\rnode{nondominating}{\large $B_v$}}
} % polygon enclosing B_v

\cnodeput[doubleline=true](0.5,7){Y1}{${\cal Y}$}
\psdots[dotsize=3pt](1,7)(1.2,7)(1.4,7)
\cnodeput[doubleline=true](1.9,7){Y2}{${\cal Y}$}

\cnodeput[doubleline=true](2.9,7){Y3}{${\cal Y}$}
\psdots[dotsize=3pt](3.4,7)(3.6,7)(3.8,7)
\cnodeput[doubleline=true](4.3,7){Y4}{${\cal Y}$}

\cnodeput[doubleline=true](5.3,7){Y5}{${\cal Y}$}
\psdots[dotsize=3pt](5.8,7)(6,7)(6.2,7)
\cnodeput[doubleline=true](6.7,7){Y6}{${\cal Y}$}

\cnodeput[doubleline=true](7.7,7){Y7}{${\cal Y}$}
\cnodeput[doubleline=true](8.7,7){Y8}{${\cal Y}$}
\cnodeput[doubleline=true](9.7,7){Y9}{${\cal Y}$}

%\cnodeput[linestyle=dashed](6,5){Y1}{${\cal Y}$}
%\cnodeput[linestyle=dashed](6,3){Y2}{${\cal Y}$}
%\cnodeput[linestyle=dashed](6,1){Y3}{${\cal Y}$}

%\cnodeput[doubleline=true](8,4){z1}{$z_1$}
%\cnodeput[doubleline=true](8,2){z2}{$z_2$}
\cnodeput[](4.3,9.3){z1}{$z_1$}
\cnodeput[](5.9,9.3){z2}{$z_2$}

\cnodeput[doubleline=true](4.3,11.3){g1}{$g_1$}
\cnodeput[doubleline=true](5.9,11.3){g2}{$g_2$}

%\cnodeput[linestyle=dashed](10,5.7){U11}{${\cal U}_1$}
%\psdots(10,5.2)(10,5)(10,4.8)
%\cnodeput[linestyle=dashed](10,4.3){U12}{${\cal U}_1$}

%\cnodeput[linestyle=dashed](10,2.5){U21}{${\cal U}_2$}
%\psdots(10,2)(10,1.8)(10,1.6)
%\cnodeput[linestyle=dashed](10,1.1){U22}{${\cal U}_2$}

\ncline[]{X1}{N1}
\ncline[]{X2}{N1}

\ncline[]{X3}{N2}
\ncline[]{X4}{N2}

\ncline[]{X5}{N3}
\ncline[]{X6}{N3}

\ncline[]{X7}{v}
\ncline[]{X8}{v}
\ncline[]{X9}{v}

\ncline[]{N1}{w1}
%\ncline{N2}{w1}
%\ncline{N3}{w1}
\ncline[]{v}{w1}

%\ncline{N1}{w2}
\ncline[]{N2}{w2}
%\ncline{N3}{w2}
\ncline[]{v}{w2}

%\ncline{N1}{w3}
%\ncline{N2}{w3}
\ncline[]{N3}{w3}
\ncline[]{v}{w3}

\ncline[]{N1}{wv}
\ncline[]{N2}{wv}
\ncline[]{N3}{wv}
\ncline[]{v}{wv}

\ncline[]{w1}{Y1}
\ncline[]{w1}{Y2}

\ncline[]{w2}{Y3}
\ncline[]{w2}{Y4}

\ncline[]{w3}{Y5}
\ncline[]{w3}{Y6}

\ncline[]{wv}{Y7}
\ncline[]{wv}{Y8}
\ncline[]{wv}{Y9}

\ncline[]{Y1}{z1}
\ncline[]{Y2}{z1}
\ncline[]{Y3}{z1}
\ncline[]{Y4}{z1}
\ncline[]{Y5}{z1}
\ncline[]{Y6}{z1}
\ncline[]{Y7}{z1}
\ncline[]{Y8}{z1}
\ncline[]{Y9}{z1}

\ncline[]{Y1}{z2}
\ncline[]{Y2}{z2}
\ncline[]{Y3}{z2}
\ncline[]{Y4}{z2}
\ncline[]{Y5}{z2}
\ncline[]{Y6}{z2}
\ncline[]{Y7}{z2}
\ncline[]{Y8}{z2}
\ncline[]{Y9}{z2}

%\ncline[linewidth=0.025]{z1}{z2}
\ncline[]{z1}{g1}
\ncline[]{z2}{g2}
%\ncline{z1}{U11}
%\ncline{z1}{U12}

%\ncline{z2}{U21}
%\ncline{z2}{U22}
\end{pspicture}
\caption{Suppose $N_G(v)=\{v_1,v_2,v_3\}$.
%Let $v\in V$ satisfy $\text{deg}_G(v)=3$ (the choice of $3$ is arbitrary).
%be a vertex with $\text{deg}_G(v)=3$.
%We assume $\text{deg}_G(v)=3$ for illustration purposes.
From
%left to right,
bottom to top,
the vertices in
%$\cup_{u\in \{v\}\cup N_G(v)} {\cal X}_u,$
$\cup_{u\in N_G^*(v)} {\cal X}_u,$
%$\{v\}\cup N_G(v),$
$N_G^*(v),$
%$\{w_u\mid u\in \{v\}\cup N_G(v)\},$
$\{w_u\mid u\in N_G^*(v)\},$
%$\cup_{u\in \{v\}\cup N_G(v)} {\cal Y}_u$
$\cup_{u\in N_G^*(v)} {\cal Y}_u,$
%and
$\{z_1,z_2\}$
and $\{g_1,g_2\}$
%and ${\cal U}$
%${\cal U}_1$ and ${\cal U}_2$
are shown.
%Other vertices in ${\cal G}$ are not depicted.
%For visual clarity, other vertices of ${\cal G}$ are not drawn.
%The vertices $v$ and $w_v$ are labeled by $v$ and $w_v,$ respectively.
%Each other vertices are labeled by a set containing it.
%Except for $v,$ $w_v,$ $z_1$ and $z_2,$
%Each vertex
%except for $v,$ $w_v,$
%is either labeled
%by a set containing it.
Lines represent the edges of
%$G^\prime$.
$\boldsymbol{\cal G}$.
A vertex is labeled
%by
${\cal X}$ or ${\cal Y}$
%${\cal U}_1$ or ${\cal U}_2$
if it belongs to the respective sets.
%For illustration,
The vertices in $B_v$ are
%enclosed in a dotted polygon.
filled
with light gray.
As every edge has an endpoint in double circle and the other in single
circle,
%$G^\prime$
$\boldsymbol{\cal G}$
is bipartite.
\comment{ % maybe not explain them here
Visually, Lemma~\ref{normalizeseeds} shows how to modify
an irreversible dynamo
to include only vertices shown in double-lined circles.
%In particular, having $z_1$ and $z_2$
%as seeds leads to all vertices in ${\cal Y}\cup {\cal U}$ being colored
%white.
%So there is no need to have any vertex in
%Then
%Lemma~\ref{dominatingsetextendedtodynamo} shows that
%For
%an irreversible dynamo
%including only vertices shown in double-lined circles,
Lemma~\ref{dominatingsetextendedtodynamo} shows that
any irreversible dynamo must contain at least one vertex in
rectangle $B,$ for otherwise no vertices in $B$ can be colored
white.
} % maybe not explain them here
%The dashed or doubled lining of some circles are for illustration
%purposes
%when we introduce the key lemmas.
%in the lemmas of th
%Some circles are dashed and/or double-lined for illustration
%purposes in later lemmas.
%some circles are dashed and some are enclosed in a dotted polygon.
}
\label{complexreduction}
\end{figure}
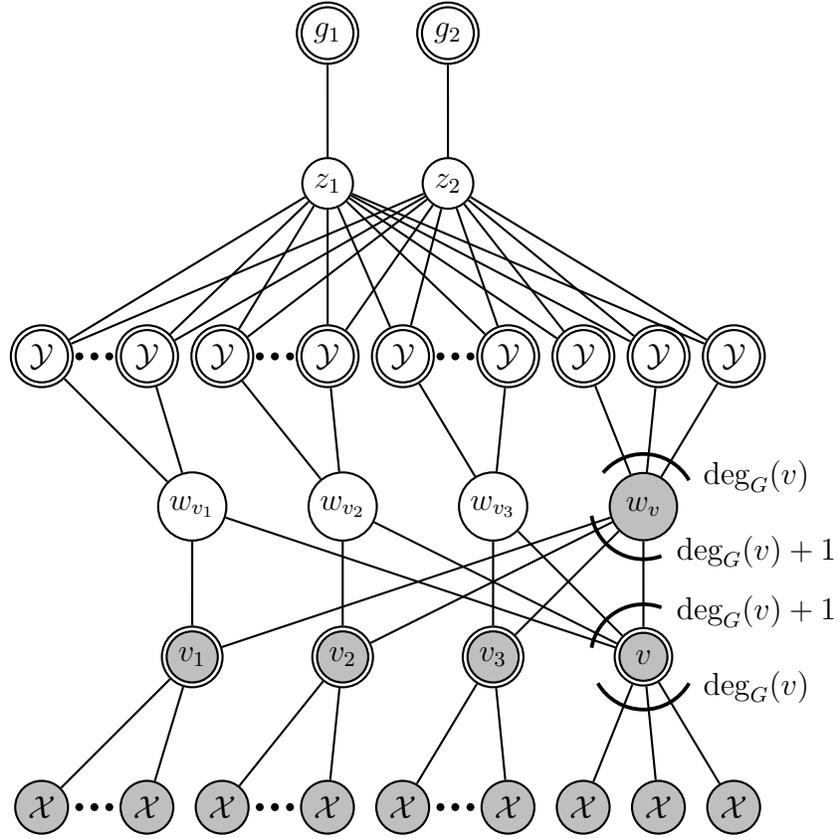

%$G(V,E),$ $\mathbf{\cal \cal G}(\mathbf{\cal V},\mathbf{\cal E}),$
%$\boldsymbol{\cal G}(\boldsymbol{\cal V},\boldsymbol{\cal E}),$
%$\bm{\cal G}(\bm{\cal V},\bm{\cal E})$
In this section, we establish inapproximability results on finding minimum
irreversible dynamos.
Given any undirected graph $G(V,E),$
we define an undirected graph
$\boldsymbol{\cal G}(\boldsymbol{\cal V}, \boldsymbol{\cal E})$ as follows.
First, define
\begin{eqnarray*}
{\cal X}_v &\equiv& \left\{x_{v,i}\mid 1\le i\le \text{deg}_G(v)\right\},
v\in V,\\
{\cal Y}_v &\equiv& \left\{y_{v,i}\mid 1\le i\le \text{deg}_G(v)\right\},
v\in V,\\
{\cal X} &\equiv& \cup_{v\in V} {\cal X}_v,\\
{\cal Y} &\equiv& \cup_{v\in V} {\cal Y}_v,\\
{\cal W} &\equiv& \cup_{v\in V} \{w_v\}.
\end{eqnarray*}
%as illustrated in
%Fig.~\ref{complexreduction},
Then define $\boldsymbol{\cal G}(\boldsymbol{\cal V},\boldsymbol{\cal E})$ by
\begin{eqnarray*}
\boldsymbol{\cal V}
&\equiv&
V
%\\
%&\cup&
\cup {\cal W}
%\left\{w_{v}\mid v\in V\right\}\\
%&\cup&
\cup {\cal X}
%\left\{x_{v,i}\mid v\in V, 1\le i\le \text{deg}_G(v)\right\}\\ % to $v$
%&\cup&
\cup {\cal Y}
%\left\{y_{v,i}\mid v\in V, 1\le i\le \text{deg}_G(v)\right\}\\ % to $w_v$
%&\cup&
\cup
\left\{z_1,z_2\right\}
\cup \left\{g_1,g_2\right\},\\
%{\cal E}
\boldsymbol{\cal E}
%^\prime
&\equiv&
\left\{(v, x)\mid v\in V, x\in {\cal X}_v\right\}\\
%\left\{(v, x_{v,i})\mid v\in V, 1\le i\le \text{deg}_G(v)\right\}\\
&\cup&
\left\{(w_v,u)\mid v\in V, u\in N_G^*(v)\right\}\\
%\left\{(u, w_v)\mid v\in V, u\in N_G^*(v)\right\}\\
&\cup&
\left\{(w_v, y)\mid v\in V, y\in {\cal Y}_v\right\}\\
%\left\{(w_v, y_{v,i})\mid v\in V, 1\le i\le \text{deg}_G(v)\right\}\\
&\cup&
\left\{(y, z_1)\mid y\in {\cal Y}\right\}\\
%\left\{(y_{v,i}, z_1)\mid v\in V, 1\le i\le \text{deg}_G(v)\right\}\\
&\cup&
\left\{(y, z_2)\mid y\in {\cal Y}\right\}\\
%\left\{(y_{v,i}, z_2)\mid v\in V, 1\le i\le \text{deg}_G(v)\right\}\\
&\cup&
%\left\{(z_1,z_2)\right\}.
\left\{(z_1,g_1)\right\}\\
&\cup&
\left\{(z_2,g_2)\right\}.
\end{eqnarray*}
%by letting
%as illustrated in
%Fig.~\ref{complexreduction}.
%Clearly,
%$$\left\{(u, w_v)\mid v\in V, u\in N_G^*(v) \right\}
%=\left\{(v, w_u)\mid v\in V, u\in N_G^*(v) \right\}.$$
%For $v\in V,$
For convenience, define
\begin{eqnarray*}
%B_v &\equiv& \{w_v\}\cup \{v\}\cup N_G(v)\cup \bigcup_{u\in \{v\}\cup N_G(v)} {\cal X}_u,\,\, v\in V,\\
B_v &\equiv& \{w_v\}\cup N_G^*(v)\cup \left(\bigcup_{u\in N_G^*(v)} {\cal X}_u\right),\,\, v\in
V.
%{\cal U}_1 &\equiv& \left\{u_{1,k}\mid 1\le k\le 2\cdot |\,E\,|+1\right\},\\
%{\cal U}_2 &\equiv& \left\{u_{2,k}\mid 1\le k\le 2\cdot |\,E\,|+1\right\},\\
%{\cal U} &\equiv& {\cal U}_1\cup {\cal U}_2.
\end{eqnarray*}
%$v\in V$.
As every edge in
%${\cal E}$
%$E^\prime$
$\boldsymbol{\cal E}$
has an endpoint in $V\cup {\cal Y}\cup
\{g_1,g_2\}$ and the other in ${\cal X}\cup {\cal W}\cup \{z_1,z_2\},$
%${\cal G}$
%$G^\prime$
$\boldsymbol{\cal G}$
is bipartite~\cite{Wes01}.
%, i.e., ${\cal V}$ can be partitioned into two
%nonempty subsets such that every edge has an endpoint in either
%subset~\cite{Wes01}.
%We conveniently write
See
Fig.~\ref{complexreduction}
for illustration.
\comment{ % new description of the graph
\begin{eqnarray*}
%{\cal V}
%V^\prime
\boldsymbol{\cal V}
&\equiv& V\\
&\cup& \left\{w_{v}\mid v\in V\right\}\\
&\cup& \left\{x_{v,i}\mid v\in V, 1\le i\le \text{deg}_G(v)\right\}\\ % to $v$
&\cup& \left\{y_{v,i}\mid v\in V, 1\le i\le \text{deg}_G(v)\right\}\\ % to $w_v$
&\cup& \left\{z_1,z_2\right\},\\
%{\cal E}
%E^\prime
\boldsymbol{\cal E}
&\equiv&
\left\{(v, x_{v,i})\mid v\in V, 1\le i\le \text{deg}_G(v)\right\}\\
&\cup& \left\{(u, w_v)\mid v\in V, u\in
N_G^*(v)
\right\}\\
&\cup& \left\{(w_v, y_{v,i})\mid v\in V, 1\le i\le \text{deg}_G(v)\right\}\\
&\cup& \left\{(y_{v,i}, z_1)\mid v\in V, 1\le i\le \text{deg}_G(v)\right\}\\
&\cup& \left\{(y_{v,i}, z_2)\mid v\in V, 1\le i\le \text{deg}_G(v)\right\}\\
&\cup& \left\{(z_1,z_2)\right\}.
\end{eqnarray*}
} % new description of the graph

%The graph ${\cal G}$ is illustrated in
Clearly,
%Given $G,$
%${\cal G}$
%$G^\prime$
$\boldsymbol{\cal G}$
can
%clearly
be constructed in polynomial time from $G$.
%Furthermore,
As
%it is easy to see that
%${\cal G}$
%$G^\prime$
$\boldsymbol{\cal G}$
clearly
has no isolated vertices,
%Hence
the networks
%${\cal N}(G^\prime,\phi^\text{strict}_{G^\prime})$
${\cal N}(\boldsymbol{\cal G},\phi^\text{strict}_{\boldsymbol{\cal G}})$
and
%${\cal N}(G^\prime,\phi^\text{simple}_{G^\prime})$
${\cal N}(\boldsymbol{\cal G},\phi^\text{simple}_{\boldsymbol{\cal G}})$
as well as their coloring
processes are all well-defined.
%illustrates ${\cal G}$ and is helpful in
%visualizing the facts and lemmas in this section.
%In particular, for $v\in {\cal V},$
%denote
%$\phi^\text{strict}_{\cal G}(v)=\lceil (\text{deg}_{\cal G}(v)+1)/2\rceil$
%and $\phi^\text{simple}_{\cal G}(v)=\lceil \text{deg}_{\cal
%G}(v)/2\rceil$.
Below are some easy facts about
%$G^\prime$.
$\boldsymbol{\cal G}$.

%\comment{ % maybe we don't need it
%Below are some immediate facts about $G$ and ${\cal G}$.

\begin{lemma}\label{easytwo}
For
%an undirected graph $G(V,E)$ and
any $v\in V,$
\begin{enumerate}
%\addtolength{\itemsep}{-0.2\baselineskip}
%\item\label{numberofverticesofnewgraph} $|\,{V}\,|=O(\,|\,V\,|^2)$.
\item\label{domsetguaranteerneighbors}
%$N_{G^\prime}(w_v)={\cal Y}_v\cup N_G^*(v)$.
$N_{\boldsymbol{\cal G}}(w_v)={\cal Y}_v\cup N_G^*(v)$.
%$N_{\cal G}(w_v)={\cal Y}_v\cup \{v\}\cup N_G(v)$.
%\item\label{justbydef} $|\,N_G(v)\,|=\text{deg}_G(v)$.
\item\label{sizeofYv} $|\,{\cal Y}_v\,|=\text{deg}_G(v)$.
\item\label{domsetguaranteerthreshold}
%$\text{deg}_G(v)+1=\phi_{G^\prime}^\text{strict}(w_v)$.
$\text{deg}_G(v)+1=\phi_{\boldsymbol{\cal G}}^\text{strict}(w_v)$.
\item\label{colorY}
%${\cal Y}\cup\{g_1,g_2\}\subseteq c(\{z_1,z_2\}, G^\prime, \phi_{G^\prime}^\text{strict})$.
${\cal Y}\cup\{g_1,g_2\}\subseteq c(\{z_1,z_2\}, \boldsymbol{\cal G},
\phi_{\boldsymbol{\cal G}}^\text{strict})$.
\item\label{noneedforWandX} ${\cal X}_v\cup\{w_v\}
\subseteq c(\{v,z_1,z_2\}, \boldsymbol{\cal G}, \phi_{\boldsymbol{\cal G}}^\text{strict})$.
\item\label{originalvertexfactone}
%$N_{G^\prime}(v)={\cal X}_v\cup \,(\,\bigcup_{u\in N_G^*(v)} \{w_u\}\,)$.
$N_{\boldsymbol{\cal G}}(v)={\cal X}_v\cup \,(\,\bigcup_{u\in N_G^*(v)} \{w_u\}\,)$.
%$N_{\cal G}(v)={\cal X}_v\cup \bigcup_{u\in \{v\}\cup N_G(v)} \{w_u\}$.
\item\label{originalvertexfacttwo} $|\,{\cal X}_v\,|=\text{deg}_G(v)$.
\item\label{originalvertexfactthree} $|\,\cup_{u\in N_G^*(v)} \{w_u\}\,|=\text{deg}_G(v)+1$.
%$|\,\cup_{u\in \{v\}\cup N_G(v)} \{w_u\}\,|=\text{deg}_G(v)+1$.
\item\label{originalvertexfactfour}
%$\text{deg}_G(v)+1=\phi_{G^\prime}^\text{strict}(v)$.
$\text{deg}_G(v)+1=\phi_{\boldsymbol{\cal G}}^\text{strict}(v)$.
%$\phi_{\cal G}^\text{strict}(v)=\text{deg}_G(v)+1$.
\end{enumerate}
%$$\left|\,N_{\cal G}(w_v)\cap \left(\{v\}\cup N_G(v)\right)\,\right|
%>\frac{\text{deg}_{\cal G}(w_v)}{2}.
%=\text{deg}_G(v)+1>\frac{\text{deg}_{\cal G}(w_v)}{2}.
%\ge \phi_{\cal G}^\text{simple}(w_v)=\text{deg}_G(v)+1.
%$$
\end{lemma}
\begin{proof}
Items~\ref{domsetguaranteerneighbors}--\ref{colorY} are
%straightforward.
immediate from the definitions.
%We only prove item~\ref{noneedforWandX} because the others are
%straightforward.
So we prove item~\ref{noneedforWandX} next.
%Trivial.
By
%Then
item~\ref{colorY},
%of
%item~\ref{domsetguaranteerthreshold} of
%Fact~\ref{easytwo},
%implies that
${\cal Y}_v\subseteq c(\{v,z_1,z_2\}, \boldsymbol{\cal G}, \phi_{\boldsymbol{\cal G}}^\text{strict})$;
%and
%completes the proof.
thus, trivially,

\begin{eqnarray}
{\cal Y}_v\cup \{v\}\subseteq c(\{v,z_1,z_2\}, \boldsymbol{\cal G},
\phi_{\boldsymbol{\cal G}}^\text{strict}).\label{goonelevel}
\end{eqnarray}
By
%items~\ref{domsetguaranteerneighbors} and \ref{domsetguaranteerthreshold} of
%Fact~\ref{easytwo},
items~\ref{domsetguaranteerneighbors}--\ref{domsetguaranteerthreshold},
${\cal Y}_v\cup \{v\}$ is a
%$\phi_{\cal G}^\text{simple}(w_v)$-size
subset of $N_{\boldsymbol{\cal G}}(w_v)$ and has size
$\phi_{\boldsymbol{\cal G}}^\text{strict}(w_v),$
which together with relation~(\ref{goonelevel}) implies
%So
$$w_v\in c(\{v,z_1,z_2\}, \boldsymbol{\cal G},
\phi_{\boldsymbol{\cal G}}^\text{strict})$$
by the coloring process.
%$w_v\in c(\{v,z\}, {\cal
%G}, \phi_{\cal G}^\text{simple})$.
%Now item~\ref{noneedforWandX} is proved
%Finally,
This and the
trivial
fact
%Clearly,
%Trivially,
${\cal X}_v\subseteq c(\{v\}, \boldsymbol{\cal G},
\phi_{\boldsymbol{\cal G}}^\text{strict}),$
which holds because
%$N_{\cal G}(x_{v,i})=\{v\}$ for $1\le i\le \text{deg}_G(v)$
$N_{\boldsymbol{\cal G}}(x)=\{v\}$ for $x\in {\cal X}_v,$
prove item~\ref{noneedforWandX}.
%is trivial.
%So item~\ref{noneedforWandX} is proved.

Now
take
%an arbitrary $u\in \{v\}\cup N_G(v)$.
any $u\in N_G^*(v)$.
Equivalently,
%$v\in \{u\}\cup N_G(u)$.
$v\in N_G^*(u),$
which implies $v\in N_{\boldsymbol{\cal G}}(w_u)$ by
item~\ref{domsetguaranteerneighbors}.
%with $u$ and $v$ switched.
%By item~\ref{domsetguaranteerneighbors} with the roles of $u$ and $v$
%switched,
%$$\{u\}\cup N_G(u)\subseteq N_{\cal G}(w_u).$$
%This and
%the
%fact that $v\in \{u\}\cup N_G(u)$
%give
%$v\in N_{\cal G}(w_u),$ or
Equivalently, $w_u\in N_{\boldsymbol{\cal G}}(v)$.
%We have shown
Now that
%$\cup_{u\in \{v\}\cup N_G(v)} \{w_u\}\subseteq N_{\cal G}(v),$
$\cup_{u\in N_G^*(v)} \{w_u\}\subseteq N_{\boldsymbol{\cal G}}(v),$
%from which
item~\ref{originalvertexfactone}
%is
%easily verified.
obtains.
%All
%the rest
The remaining items
%are straightforward.
follow immediately.
%can be easily verified.
\comment{ % more plain English proof
In the graph
%$G^\prime,$
$\boldsymbol{\cal G},$
%By definition,
$w_u$ is adjacent
%(in ${\cal G}$)
to every vertex in $\{u\}\cup N_G(u)$
%$\{u\}\cup N_G(u)\subseteq \{u\}\cup N_G(u)\cup {\cal Y}_u
%=N_{\cal G}(w_u)$.
%by definition.
by item~\ref{domsetguaranteerneighbors}.
In particular,
$w_u$ is adjacent
%(in ${\cal G}$)
%$v\in N_{\cal G}(w_u)$
to $v$
because $v\in \{u\}\cup N_G(u)$.
%So item~\ref{originalvertexfactone} follows.
%We only prove item~\ref{originalvertexfactone}, which implies
%items~\ref{originalvertexfacttwo}--\ref{originalvertexfactfour}.
%For every $u\in \{v\}\cup N_G(v),$
%$w_u$ is adjacent to every vertex in $\{u\}\cup N_G(u)$ by definition.
%In particular, $w_u$ is adjacent to $v$ because $v\in \{u\}\cup N(u)$.
%Furthermore, it is easy to verify that no vertices
%So ${\cal X}_v\cup \bigcup_{u\in \{v\}\cup N_G(v)}
%\{w_u\}\subseteq N_{\cal G}(v)$.
} % more plain English proof
\end{proof}

% combined to the preceding fact

% % remove this fact

% separated version

% proof chunk
% original, maybe better proof.

A set $D\subseteq V$ is called a dominating set of $G$ if it shares at least one
vertex with $N_G^*(v)$ for each $v\in V$~\cite{CLRS01}.
The next lemma shows that adding $z_1$ and $z_2$ to a dominating set of $G$
produces an
irreversible dynamo of
%$G^\prime$
$\boldsymbol{\cal G}$
%${\cal N}({\cal G},\phi_{\cal G}^\text{simple})$.
under the strict-majority scenario.
\comment{ % remove the sketch of proof
In the proof, we observe that with $z_1$ and $z_2$ as seeds, all the vertices in
${\cal Y}$ will be white.
Then, with at least one seed in $N_G^*(v)$ for each $v\in V,$
every vertex in ${\cal W}$ will also be white.
%Finally, the vertices in $V$ and then those in ${\cal X}$ will be white.
Coloring all the vertices in ${\cal W}$ white will color
the vertices in $V$ and then those in ${\cal X}$ white.
%Finally, those in ${\cal X}$ are colored white.
} % remove the sketch of proof

\begin{lemma}\label{dominatingsetextendedtodynamo}
If $D\subseteq V$ is a dominating set of
%an undirected graph
$G(V,E),$
then
%$c(D\cup \{z_1,z_2\}, G^\prime, \phi_{G^\prime}^\text{strict})=V^\prime$.
$c(D\cup \{z_1,z_2\}, \boldsymbol{\cal G}, \phi_{\boldsymbol{\cal G}}^\text{strict})
=\boldsymbol{\cal V}$.
\end{lemma}
\begin{proof}
Consider the coloring process in
%${\cal N}(G^\prime, \phi_{G^\prime}^\text{strict})$
${\cal N}(\boldsymbol{\cal G}, \phi_{\boldsymbol{\cal G}}^\text{strict})$
with $D\cup \{z_1,z_2\}$ as the set of seeds.
Pick $v\in V$ arbitrarily.
%Let $v\in V$ be arbitrary.
%Item~\ref{domsetguaranteerneighbors} of Fact~\ref{easytwo} says
%%$N_{\cal G}(w_v)={\cal Y}_v\cup\{v\}\cup N_G(v)$.
%$N_{\cal G}(w_v)={\cal Y}_v\cup N_G^*(v)$.
%$v\in V$.
%By
%item~\ref{colorY} of Fact~\ref{easytwo},
%${\cal Y}\subseteq c(D\cup \{z_1,z_2\},
%{\cal G}, \phi_{\cal G}^\text{strict}),$
%so
All
the
%the $\text{deg}_G(v)$
%neighbors of $w_v$ in ${\cal Y}$
vertices
in
%${\cal Y}_v$
${\cal Y}\cup \{g_1,g_2\}$
%$N_{\cal G}(w_v)\cap {\cal Y}={\cal Y}_v$
will be white
%, i.e., ${\cal Y}\subseteq c(D\cup \{z_1,z_2\}, {\cal G}, \phi_{\cal G}^\text{strict}),$
by
% item~\ref{colorY} of
Lemma~\ref{easytwo}(\ref{colorY}).
In particular, all the vertices in ${\cal Y}_v$ will be white.
%during the coloring process,
%before the coloring process ends.
%$v\in V$.
%For any $v\in V,$ $D\cap (\{v\}\cup N_G(v))\neq \emptyset$ because $D$ is a
%dominating set of $G$.
%Therefore $w_v$ has at least one white neighbor in $\{v\}\cup N_G(v),$ $v\in
%V$.
Since
%As
%$D\cap (\{v\}\cup N_G(v))\neq \emptyset$
$D\cap N_G^*(v)\neq \emptyset$
by the definition of dominating
sets,
%and $\{v\}\cup N_G(v)\subseteq N_{\cal G}(w_v),$
%there exists a $u\in D\cap (\{v\}\cup N_G(v))$.
%$w_v$ has at least one white neighbor in $\{v\}\cup N_G(v)$ at the
%beginning, $v\in V$.
at least one vertex in
%$N_{\cal G}(w_v)\cap(\{v\}\cup N_G(v))$
%$\{v\}\cup N_G(v)$
$N_G^*(v)$
is
%white at the beginning.
a seed, i.e., a white vertex initially.
%from the beginning of the coloring process.
%$v\in V$.
%In total, for
%Hence for each $v\in V,$
%$w_v$
In total,
%$w_v$
%Hence for $v\in V,$ $w_v$
%will have
at least
$|\,{\cal Y}_v\,|+1$
%$|\,{\cal Y}_v\,|+1=\text{deg}_{G}(v)+1$
%$\text{deg}_{G}(v)+1$
%\ge \phi_{\cal G}^\text{simple}(w_v)$
%white
%neighbors in
vertices in
%${\cal Y}_v\cup \{v\}\cup N_G(v)=N_{\cal G}(w_v)$
%${\cal Y}_v\cup N_G^*(v)=N_{\cal G}(w_v)$
${\cal Y}_v\cup N_G^*(v)$
%{\cal Y}\cup \{v\}\cup N_G(v)$
will be white.
In other words,
%Equivalently,
%i.e.,
%items~\ref{domsetguaranteerneighbors}--\ref{domsetguaranteerthreshold}
%of Fact~\ref{easytwo},
at least
%$\phi_{G^\prime}^\text{strict}(w_v)$
$\phi_{\boldsymbol{\cal G}}^\text{strict}(w_v)$
vertices in
%$N_{G^\prime}(w_v)$
$N_{\boldsymbol{\cal G}}(w_v)$
will be white
%during the coloring process,
%before the coloring process ends.
%$v\in V$.
because
%$|\,{\cal Y}_v\,|+1=\phi_{G^\prime}^\text{strict}(w_v)$
$|\,{\cal Y}_v\,|+1=\phi_{\boldsymbol{\cal G}}^\text{strict}(w_v)$
%and ${\cal Y}_v\cup N_G^*(v)=N_{\cal G}(w_v)$
%$\text{deg}_G(v)+1=\phi_{\cal G}^\text{strict}(w_v)$
%for each $v\in V$
by
%items~\ref{sizeofYv}--\ref{domsetguaranteerthreshold}
%items~\ref{domsetguaranteerneighbors}--\ref{domsetguaranteerthreshold},
%items~\ref{sizeofYv}--\ref{domsetguaranteerthreshold} of
Lemma~\ref{easytwo}(\ref{sizeofYv}) and (\ref{domsetguaranteerthreshold}),
and
%As
%${\cal Y}_v\cup N_G^*(v)=N_{G^\prime}(w_v)$
${\cal Y}_v\cup N_G^*(v)=N_{\boldsymbol{\cal G}}(w_v)$
by Lemma~\ref{easytwo}(\ref{domsetguaranteerneighbors}).
%and
%of
%Fact~\ref{easytwo},
%Therefore,
%implying
%\begin{eqnarray*}
%\cup_{v\in V}
%\{w_v\}\subseteq
%\begin{eqnarray}
%By definition,
So
%$w_v\in c\left(D\cup \{z_1,z_2\}, G^\prime, \phi_{G^\prime}^\text{strict}\right),$
$w_v\in c\left(D\cup \{z_1,z_2\}, \boldsymbol{\cal G},
\phi_{\boldsymbol{\cal G}}^\text{strict}\right),$
%\label{guaranteerswhite}
%\end{eqnarray}
%by definition,
%\label{domset}
%\end{eqnarray*}
%for
%all
%an arbitrary
%$v\in V$.
%\comment{ % maybe not in display form
%As the picking of $v\in V$
%has been
%chosen
%picked
%arbitrarily,
%Consequently,
%is arbitrary,
implying
\begin{eqnarray}
%{\cal W}\subseteq c\left(D\cup \{z_1,z_2\}, G^\prime, \phi_{G^\prime}^\text{strict}\right).
{\cal W}\subseteq c\left(D\cup \{z_1,z_2\}, \boldsymbol{\cal G},
\phi_{\boldsymbol{\cal G}}^\text{strict}\right).
\label{guaranteerswhite}
\end{eqnarray}
%because
%the picking of $v\in V$ is arbitrary.
%we have picked
%$v\in V$
%arbitrarily.
%is arbitrary.
%which together with
%} % maybe not in display form

%We have shown that $w_v\in
%c\left(D\cup \{z_1,z_2\}, {\cal G}, \phi_{\cal
%G}^\text{strict}\right)$ for all $v\in V,$
For each $v\in V,$
relation~(\ref{guaranteerswhite})
and
%item~\ref{originalvertexfactone} of
Lemma~\ref{easytwo}(\ref{originalvertexfactone})
imply
at least $|\,\cup_{u\in N_G^*(v)}\{w_u\}\,|$ vertices in
%$N_{G^\prime}(v)$
$N_{\boldsymbol{\cal G}}(v)$
will be white.
Furthermore,
%by
%Thus, by
%This
%and
%which together with
%items~\ref{originalvertexfactone}--\ref{originalvertexfactfour}
%items~\ref{originalvertexfactthree}--\ref{originalvertexfactfour} of Fact~\ref{easytwo},
%$|\,\cup_{u\in N_G^*(v)}\{w_u\}\,|=\phi_{G^\prime}^\text{strict}(v)$
$|\,\cup_{u\in N_G^*(v)}\{w_u\}\,|=\phi_{\boldsymbol{\cal G}}^\text{strict}(v)$
by
% items~\ref{originalvertexfactthree}--\ref{originalvertexfactfour} of
Lemma~\ref{easytwo}(\ref{originalvertexfactthree}) and
(\ref{originalvertexfactfour}).
In summary,
%Therefore,
%imply
%implies
%that
at least
%$\phi_{G^\prime}^\text{strict}(v)$
$\phi_{\boldsymbol{\cal G}}^\text{strict}(v)$
vertices in
%$N_{G^\prime}(v)$
$N_{\boldsymbol{\cal G}}(v)$
will be white, and as a result,
% before the coloring process ends.
%, $v\in V$.
%So
%all
%the
%vertices in $V$
every
$v\in V$
%$v\in V$
will be
%colored
white.
%
%For $v\in V,$
%the definition of ${\cal G}$ implies
%it is easy to see that
%$N_{\cal G}(v)={\cal X}_v\cup \bigcup_{u\in \{v\}\cup N_G(v)} \{w_u\}$
%and
%$$\left|\,{\cal X}_v\,\right|
%=\left|\,\cup_{u\in \{v\}\cup N_G(v)} \{w_u\}\,\right|
%=\text{deg}_{\cal G}(v)+1.
%=\phi_{\cal G}^\text{simple}(v).
%$$
%This and Eq.~(\ref{domset}) imply that at least
%$\text{deg}_{\cal G}(v)/2=\phi_{\cal
%G}^\text{simple}(v)$
%half of the
%neighbors of
%$v\in V$ are in $c(D\cup\{z\}, {\cal G}, \phi_{\cal G}^\text{simple}),$
%any
%$v\in V$ will become white during
%the coloring process,
%$v\in V$.
%So $V\subseteq c(D\cup \{z\}, {\cal G},
%\phi_{\cal G}^\text{simple})$.
Finally, all
the
vertices in ${\cal X}$ will be
%colored
white once all those
in $V$ are white.
%After $w_v$ is colored white for all $v\in V,$
%Hence by Lemma~\ref{normalseed}, ${\cal X}$
\end{proof}

%Now we show that Lemma~\ref{dominatingsetextendedtodynamo}
Below we show that
%non-dominating sets of $G$ cannot become an
every
irreversible dynamo
of
%${\cal N}(G^\prime, \phi_{G^\prime}^\text{strict})$
${\cal N}(\boldsymbol{\cal G}, \phi_{\boldsymbol{\cal G}}^\text{strict})$
%${\cal G}$ under the simple-majority scenario
%by
%simply
%including $z_1$ and $z_2$.
has a non-empty intersection with $B_v$ for every $v\in V$.
\comment{ % remove the sketch of proof again
In the proof,
we show that every vertex in $B_v$ has strictly more neighbors
in $B_v$ than out of it;
so a set of seeds with empty intersection with $B_v$
cannot color any vertex in $B_v$ white.
} % remove the sketch of proof again

\begin{lemma}\label{nondominatingset}
For each $v\in V,$
every
irreversible dynamo $S$ of
%${\cal N}(G^\prime, \phi_{G^\prime}^\text{strict})$
${\cal N}(\boldsymbol{\cal G}, \phi_{\boldsymbol{\cal G}}^\text{strict})$
%${\cal G}$ under the simple-majority scenario
%by
%simply
%including $z_1$ and $z_2$.
%has non-empty intersection with $B_v,$ $v\in V$.
satisfies $S\cap B_v\neq\emptyset$.
% for all $v\in V$.
\end{lemma}
\begin{proof}
%\bigcup_{a\in \{u\}\cup N(u)} w_a$.
%As
%$$\left|\,\cup_{a\in \{u\}\cup N(u)} w_a\,\right|=\text_{\cal G}(u)+1
%=\left|\,{\cal X}_u\,\right|,$$
%$u$ cannot become white until at least
%\comment{ % original proof
%Let $v\in V$ be arbitrary.
Recall that
%$B_v\equiv \{v\}\cup N_G(v)\cup \{w_v\} \cup \bigcup_{u\in \{v\}\cup N_G(v)} {\cal X}_u,$
$B_v= \{w_v\}\cup N_G^*(v)\cup\, (\,\bigcup_{u\in N_G^*(v)} {\cal
X}_u\,)$.
%as
%enclosed
%shown
%in
%the dotted polygon of
%light gray in
%drawn in doubly lined circles in
%Fig.~\ref{complexreduction}.
%To complete the proof,
%we only need to show that
%we only need to show that
%$$c\left({\cal V}
%\setminus \left(\{v\}\cup N_G(v)\cup \{w_v\}
%\cup \bigcup_{u\in \{v\}\cup N_G(v)} {\cal X}_u\right),
%{\cal G}, \phi_{\cal G}^\text{simple}\right)\neq {\cal V},$$
%$c({\cal V}\setminus B_v, {\cal G}, \phi_{\cal G}^\text{strict})\neq {\cal V},$
%which is true
%for the following reasons:
%because
We
%first
proceed to
show that
%The following three observations show that
every
%$x\in B_v$
$\alpha\in B_v$
satisfies
\begin{eqnarray}
%\left|\,N_{\cal G}(x)\cap B_v\,\right|>\frac{\text{deg}_{\cal G}(x)}{2}\label{solidset}
%\left|\,N_{G^\prime}(\alpha)\cap B_v\,\right|>\frac{\text{deg}_{G^\prime}(\alpha)}{2}\label{solidset}
\left|\,N_{\boldsymbol{\cal G}}(\alpha)\cap B_v\,\right|
>\frac{\text{deg}_{\boldsymbol{\cal G}}(\alpha)}{2}\label{solidset}
\end{eqnarray}
%vertex in $B_v$ has strictly more neighbors in $B_v$ than
%in ${\cal V}\setminus B_v,$
%as explained
in three cases below
according to whether
%$x$
$\alpha$
%belongs to $\{w_v\},$
is $w_v,$ a member of
$N_G^*(v)$ or a member of
%one of
$\cup_{u\in N_G^*(v)} {\cal X}_u$:
\begin{itemize}
%\addtolength{\itemsep}{-0.2\baselineskip}
\item $\alpha=w_v$:
By
%item~\ref{domsetguaranteerneighbors} of
Lemma~\ref{easytwo}(\ref{domsetguaranteerneighbors}),
%$N_{\cal G}(w_v)\cap B_v=N_G^*(v)$.
%$N_{G^\prime}(\alpha)\cap B_v=N_G^*(v)$.
$N_{\boldsymbol{\cal G}}(\alpha)\cap B_v=N_G^*(v)$.
So
%$|\,N_{\cal G}(w_v)\cap B_v\,| =|\,\{v\}\cup N_G(v)\,|$.
%$|\,N_{\cal G}(w_v)\cap B_v\,| =|\,N_G^*(v)\,|=\text{deg}_G(v)+1$.
%$|\,N_{G^\prime}(\alpha)\cap B_v\,| =|\,N_G^*(v)\,|=\text{deg}_G(v)+1$.
$|\,N_{\boldsymbol{\cal G}}(\alpha)\cap B_v\,| =|\,N_G^*(v)\,|=\text{deg}_G(v)+1$.
By
%items~\ref{sizeofYv}--
%item~\ref{domsetguaranteerthreshold} of
Lemma~\ref{easytwo}(\ref{domsetguaranteerthreshold}),
%$|\,\{v\}\cup N_G(v)\,|=\phi_{\cal G}^\text{strict}(w_v)
%>\text{deg}_{\cal G}(w_v)/2$.
%$|\,N_G^*(v)\,|=1+\text{deg}_G(v)=\phi_{\cal G}^\text{strict}(w_v)>\text{deg}_{\cal G}(w_v)/2$.
%$\text{deg}_G(v)+1=\phi_{\cal G}^\text{strict}(w_v)>\text{deg}_{\cal G}(w_v)/2$.
%$\text{deg}_G(v)+1=\phi_{G^\prime}^\text{strict}(\alpha)>\text{deg}_{G^\prime}(\alpha)/2$.
$\text{deg}_G(v)+1=\phi_{\boldsymbol{\cal G}}^\text{strict}(\alpha)
>\text{deg}_{\boldsymbol{\cal G}}(\alpha)/2$.
Therefore,
%In summary,
%$|\,N_{\cal G}(w_v)\cap B_v\,|=|\,\{v\}\cup N_G(v)\,|>\text{deg}_{\cal G}(w_v)/2$.
%$|\,N_{\cal G}(w_v)\cap B_v\,|=|\,N_G^*(v)\,|>\text{deg}_{\cal G}(w_v)/2$.
%$|\,N_{\cal G}(w_v)\cap B_v\,|>\text{deg}_{\cal G}(w_v)/2$.
%$|\,N_{G^\prime}(\alpha)\cap B_v\,| =\text{deg}_G(v)+1 >\text{deg}_{G^\prime}(\alpha)/2$.
$|\,N_{\boldsymbol{\cal G}}(\alpha)\cap B_v\,| =\text{deg}_G(v)+1
>\text{deg}_{\boldsymbol{\cal G}}(\alpha)/2$.
%\item In the graph ${\cal G},$ each vertex in $\cup_{u\in \{v\}\cup N_G(v)}
%{\cal X}_u$
%has exactly one neighbor, which is
%in $\{v\}\cup N_G(v)\subseteq B_v$.
\item $\alpha\in N_G^*(v)$:
%For every
%$u\in \{v\}\cup N_G(v),$
%$u\in N_G^*(v),$
%it is clear that
Clearly,
%$v\in \{u\}\cup N_G(u)$.
%$v\in N_G^*(u)$.
$\alpha\in V$ and $v\in N_G^*(\alpha)$.
By
%item~\ref{originalvertexfactone} of
Lemma~\ref{easytwo}(\ref{originalvertexfactone}),
%with
%the roles of
%$u$ and $v$ switched,
%${\cal X}_u\cup \{w_v\}\subseteq N_{\cal G}(u)$.
%${\cal X}_\alpha\cup \{w_v\}\subseteq N_{G^\prime}(\alpha)$.
${\cal X}_\alpha\cup \{w_v\}\subseteq N_{\boldsymbol{\cal G}}(\alpha)$.
So
\begin{eqnarray}
%{\cal X}_u\cup \{w_v\}\subseteq N_{\cal G}(u)\cap B_v \label{theneighborsinBv}
%{\cal X}_\alpha\cup \{w_v\}\subseteq N_{G^\prime}(\alpha)\cap B_v \label{theneighborsinBv}
{\cal X}_\alpha\cup \{w_v\}\subseteq N_{\boldsymbol{\cal G}}(\alpha)\cap B_v \label{theneighborsinBv}
\end{eqnarray}
by the
definition of $B_v$.
By
% items~\ref{originalvertexfacttwo} and \ref{originalvertexfactfour} of
Lemma~\ref{easytwo}(\ref{originalvertexfacttwo})~and~(\ref{originalvertexfactfour}),
% with
%the roles of
%$u$ and $v$ switched,
%$$\left|\,{\cal X}_u\cup \{w_v\}\,\right| =\text{deg}_G(u)+1 =\phi_{\cal G}^\text{strict}(u)>\frac{\text{deg}_{\cal G}(u)}{2}.$$
%$$\left|\,{\cal X}_\alpha\cup \{w_v\}\,\right|=\text{deg}_G(\alpha)+1=\phi_{G^\prime}^\text{strict}(\alpha)>\frac{\text{deg}_{G^\prime}(\alpha)}{2}.$$
$$\left|\,{\cal X}_\alpha\cup \{w_v\}\,\right|
=\text{deg}_G(\alpha)+1
=\phi_{\boldsymbol{\cal G}}^\text{strict}(\alpha)
>\frac{\text{deg}_{\boldsymbol{\cal G}}(\alpha)}{2}.$$
This and relation~(\ref{theneighborsinBv}) give
%$|\,N_{\cal G}(u)\cap B_v\,|\ge |\,{\cal X}_u\cup \{w_v\}\,| >\text{deg}_{\cal G}(u)/2$.
%$|\,N_{G^\prime}(\alpha)\cap B_v\,| \ge |\,{\cal X}_\alpha\cup \{w_v\}\,| >\text{deg}_{G^\prime}(\alpha)/2$.
$|\,N_{\boldsymbol{\cal G}}(\alpha)\cap B_v\,| \ge |\,{\cal X}_\alpha\cup \{w_v\}\,|
>\text{deg}_{\boldsymbol{\cal G}}(\alpha)/2$.
%\item For every $u\in \{v\}\cup N_G(v),$
%it is clear that $v\in \{u\}\cup N_G(u)$.
%By item~\ref{originalvertexfactone}
%of Fact~\ref{easytwo}
%with the roles of $u$ and $v$ switched,
%${\cal X}_u\cup \{w_v\}\subseteq N_{\cal G}(u)$.
%So
%${\cal X}_u\cup \{w_v\}\subseteq N_{\cal G}(u)\cap B_v$
%by the
%definition of $B_v$.
%Moreover, $|\,{\cal X}_u\cup \{w_v\}\,|=\phi_{\cal
%G}^\text{strict}(u)>\text{deg}_{\cal G}(u)/2$ by
%items~\ref{originalvertexfacttwo}--\ref{originalvertexfactfour} of
%Fact~\ref{easytwo}.
%In summary, $|\,N_{\cal G}(u)\cap B_v\,|\ge |\,{\cal X}_u\cup \{w_v\}\,|>\text{deg}_{\cal G}(u)/2$.
%Furthermore, Fact~\ref{easytwo} implies that
%${\cal X}_u\cup \{w_v\}$ constitutes more than half of the vertices in
%$N_{\cal G}(u)$.
%$u$ is adjacent to $w_v$ and all vertices in ${\cal X}_u$
%and (2)
%$\{w_v\}\cup {\cal X}_u$ constitutes more than half of the vertices in
%$N_{\cal G}(u)$.
%$$\frac{deg_{\cal G}(v)}{2}
%<\left|\,\{w_v\}\cap {\cal X}_u\,\right|$$
%Each $u\in \{v\}\cup N_G(v)$ is adjacent to $w_v$ and all vertices in
%${\cal X}_u$.
%For every $s\in \{v\}\cup N_G(v),$ $N_{\cal G}(u)={\cal X}_s\cup
%\bigcup_{t\in \{s\}\cup N(s)} \{w_t\}$ and $|\,{\cal X}_s\,|=|\,\cup_{t\in
%\{s\}\cup N(s)} \{w_t\}\,|=\text{deg}_{\cal G}s)+1$.
%So
%$\text{deg}_{\cal G}(s)=2\cdot (\text{deg}_G(s)+1)$.
%Furthermore,
%${\cal X}_s\cup \{w_v\}\subseteq N_{\cal G}(s)\cap
%(\{v\}\cup N_G(v)\cup {\cal X}\cup \{w_v\}),$ i.e., more than half of the
%neighbors of $s$ are in $\{v\}\cup N_G(v)\cup \bigcup_{u\in \{v\}\cup N_G(v)} {\cal
%X}_u\cup \{w_v\}$.
\item $\alpha\in {\cal X}_u$
%for a
where
$u\in N_G^*(v)$:
%For any
%$u\in \{v\}\cup N_G(v)$
%$u\in N_G^*(v)$
%and
%$x_{u,i}$ where
%$1\le i\le \text{deg}_G(u),$ $N_{\cal G}(x_{u,i})=\{u\}\subseteq B_v$.
%$x\in {\cal X}_u,$
Clearly,
%$N_{\cal G}(x)=\{u\}\subseteq B_v$.
%$N_{G^\prime}(\alpha)=\{u\}\subseteq B_v$.
$N_{\boldsymbol{\cal G}}(\alpha)=\{u\}\subseteq B_v$.
So
%$|\,N_{\cal G}(x_{u,i})\cap B_v\,|=1>1/2=\text{deg}_{\cal G}(x_{u,i})/2$.
%$|\,N_{\cal G}(x)\cap B_v\,|=1>1/2=\text{deg}_{\cal G}(x)/2$.
%$|\,N_{G^\prime}(\alpha)\cap B_v\,|=1>1/2=\text{deg}_{G^\prime}(\alpha)/2$.
$|\,N_{\boldsymbol{\cal G}}(\alpha)\cap B_v\,|=1>1/2
=\text{deg}_{\boldsymbol{\cal G}}(\alpha)/2$.
%$\{v\}\cup N_G(v)\cup \bigcup_{u\in \{v\}\cup N_G(v)} {\cal X}_u\cup \{w_v\}$.
%$N_{\cal G}(w_v)=\{v\}\cup N_G(v)\cup {\cal Y}_v$ and $|\,\{v\}\cup
%N_G(v)\,|=|\,{\cal Y}_v\,|+1$. So more than half of the neighbors of $w_v$
%are in $\{v\}\cup N_G(v)\cup \bigcup_{u\in \{v\}\cup N_G(v)} {\cal
%X}_u\cup \{w_v\}$.
%Among the vertices $N_{\cal G}(s),$ those in
%${\cal X}_s\cup \{w_v\}$
\end{itemize}
%} % original proof
%The above
%three items
%completes our verification of
Having verified
inequality~(\ref{solidset}) for all
%$x\in B_v$.
$\alpha\in B_v,$
%Now for $\alpha\in B_v,$
%By inequality~(\ref{solidset}),
\begin{eqnarray}
%&& \left|\,N_{\cal G}(x)\cap ({\cal V}\setminus B_v)\,\right|\nonumber\\
%&=&\left|\,N_{\cal G}(x)\,\right|-\left|\,N_{\cal G}(x)\cap
%B_v\,\right|\nonumber\\
%&=&\text{deg}_{\cal G}(x)-\left|\,N_{\cal G}(x)\cap B_v\,\right|\nonumber\\
%&<&\frac{\text{deg}_{\cal G}(x)}{2}\nonumber\\
%&<&\phi_{\cal G}^\text{strict}(x),\,\,\,\, x\in B_v.\label{deadlocked}
%&& \left|\,N_{G^\prime}(\alpha)\cap ({V^\prime}\setminus B_v)\,\right|\nonumber\\
%&=&\left|\,N_{G^\prime}(\alpha)\,\right|-\left|\,N_{G^\prime}(\alpha)\cap
%B_v\,\right|\nonumber\\
%&=&\text{deg}_{G^\prime}(\alpha)-\left|\,N_{G^\prime}(\alpha)\cap B_v\,\right|\nonumber\\
%&\stackrel{\text{inequality~(\ref{solidset})}}{<}&\frac{\text{deg}_{G^\prime}(\alpha)}{2}\nonumber\\
%&<&\phi_{G^\prime}^\text{strict}(\alpha).
&& \left|\,N_{\boldsymbol{\cal G}}(\alpha)\cap (\boldsymbol{\cal V}\setminus B_v)\,\right|\nonumber\\
&=&\left|\,N_{\boldsymbol{\cal G}}(\alpha)\,\right|-\left|\,N_{\boldsymbol{\cal
G}}(\alpha)\cap B_v\,\right|\nonumber\\
&=&\text{deg}_{\boldsymbol{\cal G}}(\alpha)-\left|\,N_{\boldsymbol{\cal G}}(\alpha)\cap B_v\,\right|\nonumber\\
&\stackrel{\text{inequality~(\ref{solidset})}}{<}&\frac{\text{deg}_{\boldsymbol{\cal
G}}(\alpha)}{2}\nonumber\\
&<&\phi_{\boldsymbol{\cal G}}^\text{strict}(\alpha).
%\,\,\,\, \alpha\in B_v.
\label{deadlocked}
\end{eqnarray}
%for all
%$\alpha\in B_v$.
%where the first inequality follows from inequality~(\ref{solidset}).
%for all $x\in B_v$.
%Now consider

Next, suppose for contradiction that
at least one vertex in $B_v$ ends up white
in
the coloring process in
%${\cal N}(G^\prime,\phi_{G^\prime}^\text{strict})$
${\cal N}(\boldsymbol{\cal G},\phi_{\boldsymbol{\cal G}}^\text{strict})$
with
%$V^\prime\setminus B_v$
$\boldsymbol{\cal V}\setminus B_v$
as the set of seeds.
%Suppose for contradiction that a nonzero number of vertices $B_v$ are
%Then
%Fixing an order according to which at least one vertex in $B_v$ can be colored
%there is a vertex
Let
$\alpha^*\in B_v$
%that can
be colored white first
%, i.e., $\alpha^*$
%can be colored white before all the other
among all
vertices in $B_v$.
%the first vertex in $B_v$ that is colored white, denoted $\alpha^*,$
%, if one exists,
Then $\alpha^*$
%has
must have
%strictly more neighbors in ${\cal V}\setminus B_v$ than in $B_v$
at least
%$\phi^\text{strict}_{G^\prime}(\alpha^*)$
$\phi^\text{strict}_{\boldsymbol{\cal G}}(\alpha^*)$
vertices in
%$V^\prime\setminus B_v$
$\boldsymbol{\cal V}\setminus B_v$
by the coloring process,
%which is impossible by
contradicting
inequality~(\ref{deadlocked}).
Consequently,
%Suppose for contradiction that
\begin{eqnarray}
%c\left(V^\prime\setminus B_v, G^\prime, \phi_{G^\prime}^\text{strict}\right)
%= V^\prime\setminus B_v
%\neq V^\prime.\label{Bvisneeded}
c\left(\boldsymbol{\cal V}\setminus B_v, \boldsymbol{\cal G},
\phi_{\boldsymbol{\cal G}}^\text{strict}\right)
= \boldsymbol{\cal V}\setminus B_v
\neq \boldsymbol{\cal V}.\label{Bvisneeded}
\end{eqnarray}
%is false.
%Consider the coloring process in ${\cal N}({\cal G},\phi_{\cal
%G}^\text{strict})$ with ${\cal V}\setminus B_v$ as the set of seeds,.
%The first vertex in $B_v$ that is colored white
%because no vertex in $B_v$ can be colored white given that ${\cal
%V}\setminus B_v$ is the set of seeds.
%But

Now
if $S\cap B_v=\emptyset,$
then
%$S\subseteq V^\prime\setminus B_v$
$S\subseteq \boldsymbol{\cal V}\setminus B_v$
and
$$
%{\cal V}=
%c\left(S, G^\prime, \phi_{G^\prime}^\text{strict}\right)
%\subseteq c\left(V^\prime\setminus B_v, G^\prime, \phi_{G^\prime}^\text{strict}\right)
c\left(S, \boldsymbol{\cal G}, \phi_{\boldsymbol{\cal G}}^\text{strict}\right)
\subseteq c\left(\boldsymbol{\cal V}\setminus B_v, \boldsymbol{\cal G},
\phi_{\boldsymbol{\cal G}}^\text{strict}\right)
%= {\cal V}\setminus B_v
%\neq {\cal V},
$$
by Fact~\ref{monotone}.
%a contradiction.
%This contradicts
%contradicting
This and
inequality~(\ref{Bvisneeded})
%because $S$ is an irreversible dynamo of
%${\cal N}({\cal G},\phi_{\cal G}^\text{strict}),$
%i.e., $c(S, {\cal G}, \phi_{\cal G}^\text{strict})={\cal V}$.
contradict the premise that $S$ is an irreversible dynamo of
%${\cal N} (G^\prime,\phi_{G^\prime}^\text{strict})$.
${\cal N}(\boldsymbol{\cal G},\phi_{\boldsymbol{\cal G}}^\text{strict})$.
\end{proof}

The following Lemma shows that
%if $G$ has no isolated vertices,
%then
%$G^\prime$
$\boldsymbol{\cal G}$
%has
is a bipartite graph with
%an $O(1)$
diameter
%$6$
%$O(1)$
at most $8$
if $G$ has no isolated vertices.
\comment{ % Omit explaining what is diameter
That is,
%$G^\prime$
$\boldsymbol{\cal G}$
is bipartite and
every two vertices of
%$G^\prime$
$\boldsymbol{\cal G}$
are connected by a path of
length at most
$8$.
} % Omit explaining what is diameter

\begin{lemma}\label{lowdiameter}
Assume that
$G$ has no isolated vertices.
Then
%$G^\prime$
$\boldsymbol{\cal G}$
%has
is a bipartite graph with
%an $O(1)$
diameter
%$O(1)$.
at most $8$.
\end{lemma}
\begin{proof}
Partition
%$V^\prime$
$\boldsymbol{\cal V}$
into
%$V^\prime_1=V\cup {\cal Y}\cup \{g_1,g_2\}$
$\boldsymbol{\cal V}_1=V\cup {\cal Y}\cup \{g_1,g_2\}$
and
%$V^\prime_2={\cal X}\cup {\cal W}\cup \{z_1,z_2\}$.
$\boldsymbol{\cal V}_2={\cal X}\cup {\cal W}\cup \{z_1,z_2\}$.
It is immediate from the definition of
%$G^\prime$
$\boldsymbol{\cal G}$
that
each edge in
%$E^\prime$
$\boldsymbol{\cal E}$
has an endpoint in
%$V^\prime_1$
$\boldsymbol{\cal V}_1$
and the other in
%$V^\prime_2$.
$\boldsymbol{\cal V}_2$.
So
%$G^\prime$
$\boldsymbol{\cal G}$
is bipartite.

%For each $v,$ $x\in {\cal X}_v,$ $y\in {\cal Y}_v$ and $i\in\{1,2\},$
%$x,v,w_v,y,z_i,g_i$
%is a path by the definition of ${\cal G}$.
As $G$ has no isolated vertices, $\text{deg}_G(v)>0$ for all $v\in V$.
Hence ${\cal Y}_v\neq\emptyset$
and ${\cal X}_v\neq\emptyset$
for all $v\in V$
by
%items~\ref{sizeofYv} and \ref{originalvertexfacttwo} of
Lemma~\ref{easytwo}(\ref{sizeofYv})~and~(\ref{originalvertexfacttwo}),
respectively.
To show that
%$G^\prime$
$\boldsymbol{\cal G}$
has diameter at most $8,$
it suffices to establish
%every vertex in ${\cal V}$ has distance at most $4$ to both $z_1$ and $z_2$.
%$d_{G^\prime}(u,z_1)\le 4$
$d_{\boldsymbol{\cal G}}(u,z_1)\le 4$
for all
%$u\in V^\prime,$
$u\in \boldsymbol{\cal V},$
% and $i\in\{1,2\},$
which is true because
for each $v\in V,$ $x\in {\cal X}_v$ and $y\in {\cal Y}_v,$
% as explained below:
%For this purpose,
%simply
%observe that by
%By
%the definition of ${\cal G},$
%both
\begin{eqnarray*}
P_1(v,x,y)&\equiv& \left(x,v,w_v,y,z_1\right),\\
P_2(v,x,y)&\equiv& \left(g_1,z_1\right),\\
P_3(v,x,y)&\equiv& \left(g_2,z_2,y,z_1\right)
\end{eqnarray*}
%$$x,v,w_v,y,z_1$$
%and
%$$g_1,z_1,y,z_2,g_2$$
are all
paths of
%$G^\prime$
$\boldsymbol{\cal G}$
by definition.
%is a path of ${\cal G}$
%for each
%where
%$v\in V,$ $x\in {\cal X}_v$ and $y\in {\cal
%Y}_v$.
% and $i\in \{1,2\}$.
%On these paths,
%Furthermore,
%$$g_1,z_1,y,z_2,g_2$$
%is also a path.
\comment{ % maybe not list these one by one
observe the following:
\begin{enumerate}
%\addtolength{\itemsep}{-0.2\baselineskip}
\item\label{countdiameterYtoz} Every vertex in ${\cal Y}$ is adjacent to both $z_1$ and $z_2$.
\item Every vertex
in
${\cal W}$ is adjacent to at least one vertex in ${\cal Y}$
because ${\cal Y}_v\neq\emptyset$ and
%$(w_v,y)\in E^\prime$
$(w_v,y)\in \boldsymbol{\cal E}$
for all $v\in V$ and $y\in {\cal Y}_v$.
% and ${\cal Y}_v\neq\emptyset$.
\item Every vertex in $V$ is adjacent to at least one vertex in ${\cal W}$.
\item\label{countdiameterXtoV} Every vertex in ${\cal X}$ is adjacent to exactly one vertex in
$V$.
\item\label{countdiametergtoz} For $i\in\{1,2\},$ $g_i$ is adjacent to $z_i$.
\item\label{countdiameterztoz} By item~\ref{countdiameterYtoz},
%$z_1$ and $z_2$ have distance at most $2$ to each other.
%$d_{G^\prime}(z_1,z_2)\le 2$.
$d_{\boldsymbol{\cal G}}(z_1,z_2)\le 2$.
%For $i,j\in\{1,2\}$ with $i\neq j,$ $g_i$ is adjacent
\end{enumerate}
Above, items~\ref{countdiameterYtoz}--\ref{countdiameterXtoV} successively
show that every vertex in ${\cal Y},$ ${\cal Y}\cup{\cal W},$ ${\cal
Y}\cup{\cal W}\cup V$ and ${\cal Y}\cup{\cal W}\cup V\cup {\cal X}$
has distance at most $1,2,3$ and $4$ to both $z_1$ and $z_2$.
Items~\ref{countdiametergtoz}--\ref{countdiameterztoz} show that
$g_i$ has distance at most $3$ to both $z_1$ and $z_2,$ $i\in \{1,2\}$.
} % maybe not list these one by one
% come on, need to be so ugly?
% maybe no need to be cumbersome for diameter
\end{proof}

The following fact is due to Feige~\cite{Fei98}.

\begin{fact}(\cite{Fei98})\label{inapproximabilitydominatingset}
Let $\epsilon>0$ be any constant.
If {\sc dominating set} has a polynomial-time, $((1-\epsilon)\ln N)$-approximation algorithm for
$N$-vertex graphs without isolated vertices, then $\text{NP}\subseteq
\text{TIME}(n^{O(\ln \ln n)})$.
\comment{ % original statement
Given an undirected graph
$G(V,E)$ without isolated vertices,
if there exists
%a constant $\epsilon>0$ and
a polynomial-time,
%$o(\ln |\,V\,|)$-approximation
$((1-\epsilon)\ln |\,V\,|)$-approximation
algorithm for {\sc dominating set},
then $\text{NP}\subseteq \text{TIME}(n^{O(\ln \ln n)})$.
} % original statement
\end{fact}

We now
%prove
%equalize
relate
%link
the inapproximability of
%{\sc dominating set} with that of
{\sc irreversible dynamo (strict majority)}
%for bipartite graphs with diameter at most $8$.
with that of {\sc dominating set}.

\begin{theorem}\label{inapproximabilitystrict}
Let $\epsilon>0$ be any constant.
If {\sc irreversible dynamo (strict majority)} has a polynomial-time,
$((1/2-\epsilon)\ln N)$-approximation algorithm for
$N$-vertex graphs,
%undirected
%bipartite
%graphs with $N$ vertices and diameter at most $8,$
then $\text{NP}\subseteq \text{TIME}(n^{O(\ln \ln n)})$.
\comment{ % original statement
Given an
undirected
bipartite
graph
%$G(V,E),$
with $N$ vertices and diameter at most $8,$
if there exists a polynomial-time,
%$o(\ln N)$-approximation
$((1/2-\epsilon)\ln N)$-approximation
algorithm
for {\sc irreversible dynamo (strict majority)},
then $\text{NP}\subseteq \text{TIME}(n^{O(\ln \ln
n)})$.
} % original statement
\end{theorem}
\begin{proof}
We will prove the stronger statement that, if
%$\text{NP}\subseteq \text{TIME}(n^{O(\ln \ln n)})$
%given that
{\sc irreversible dynamo (strict majority)} has a
polynomial-time, $((1/2-\epsilon)\ln N)$-approximation algorithm ALG
for bipartite graphs with $N$ vertices and diameter at most $8,$
then $\text{NP}\subseteq \text{TIME}(n^{O(\ln \ln n)})$.
%Assume
%Suppose
%for contradiction
%that
%Let {\sc irreversible dynamo (strict majority)} have a polynomial-time,
%$((1/2-\epsilon)\ln N)$-approximation algorithm ALG for bipartite graphs with
%$N$ vertices and diameter at most $8$.
Given an undirected graph $G(V,E)$ without isolated vertices,
$\boldsymbol{\cal G}$ is a bipartite graph with diameter at most $8$
by Lemma~\ref{lowdiameter}.
%$S=\text{ALG}(\boldsymbol{\cal G})$
%can be obtained in time polynomial in $|\,V\,|$
%By constructing
The construction of
$\boldsymbol{\cal G}$
followed by
the calculation of
%and then feeding it to ALG,
$S=\text{ALG}(\boldsymbol{\cal G})$
can be
%obtained
done
in time polynomial in $|\,V\,|$.
%By Lemma~\ref{lowdiameter}, $\boldsymbol{\cal G}$ is a bipartite graph with diameter at most $8$.
%So our
Our
assumption on ALG implies that
$S$ is an irreversible dynamo of
%find an irreversible dynamo $S$ of
%${\cal G}$ under the strict-majority scenario
%${\cal N}(G^\prime,\phi_{G^\prime}^\text{strict})$
${\cal N}(\boldsymbol{\cal G},\phi_{\boldsymbol{\cal G}}^\text{strict})$
with
\begin{eqnarray}
&& |\,S\,|\nonumber\\
&\le&
%o\left(\ln |\,{\cal V}\,|\right)\cdot \text{min-seed}\left({\cal
%G},\phi_{\cal G}^\text{strict}\right)
%\left[
\left(\frac{1}{2}-\epsilon\right)\cdot
%\left(\,
%\ln |\,V^\prime\,|
\ln \left|\,\boldsymbol{\cal V}\,\right|
%\,\right)
%\right]
\cdot
%\text{min-seed}\left(G^\prime,\phi_{G^\prime}^\text{strict}\right)\nonumber\\
\text{min-seed}\left(\boldsymbol{\cal G},\phi_{\boldsymbol{\cal G}}^\text{strict}\right)\nonumber\\
%=o\left(\ln |\,V\,|\right)\cdot \text{min-seed}\left({\cal
%G},\phi_{\cal G}^\text{strict}\right),
&\le& \left[\,\left(1-2\epsilon\right)\cdot\ln{|\,V\,|}+O(1)\,\right]
\cdot
%\text{min-seed}\left(G^\prime,\phi_{G^\prime}^\text{strict}\right).
\text{min-seed}\left(\boldsymbol{\cal G},\phi_{\boldsymbol{\cal G}}^\text{strict}\right).
\label{computinganirreversibledynamo}
\end{eqnarray}
%in
%time polynomial in $|\,V\,|,$
%polynomial time.
%where
Above,
the
%equality
second inequality
follows
%because $|\,{\cal V}\,|$ is at most polynomial in $|\,V\,|$.
from
%is immediate from
%the easily verifiable fact
%$|\,V^\prime\,|=O(\,|\,V\,|^2),$
$|\,\boldsymbol{\cal V}\,|=O(\,|\,V\,|^2),$
which is easily verified given items~\ref{sizeofYv}
and \ref{originalvertexfacttwo}
of Lemma~\ref{easytwo}.
%item~\ref{numberofverticesofnewgraph} of Fact~\ref{easytwo}.
%Fact~\ref{newgraphsize}
%${\cal G}$ can be found in time polynomial in $|\,V\,|$.
%Suppose for contradiction that the theorem is false.
%Then an irreversible dynamo of ${\cal N}({\cal})$
Denote by $\gamma(G)$ the
%minimum
size of
any minimum
dominating set of $G$.
%Clearly, $\text{OPT}\ge 1$.
By Lemma~\ref{dominatingsetextendedtodynamo},
\begin{eqnarray}
%\text{min-seed}\left(G^\prime,\phi_{G^\prime}^\text{strict}\right)
\text{min-seed}\left(\boldsymbol{\cal G},\phi_{\boldsymbol{\cal G}}^\text{strict}\right)
\le
%\gamma(G)+\left|\,\{z_1,z_2\}\,\right|
%=
\gamma(G)+2.
%\le 3\cdot \text{OPT}.
\label{relatingthesizes}
\end{eqnarray}

%After one has
%found
%computed
%Having obtained
With
%$G^\prime$
$\boldsymbol{\cal G}$
and $S$
in hand,
%Clearly,
\begin{eqnarray}
%D
\tilde{D}
%\left(G^\prime,S\right)
\equiv\{u\in V\mid S\cap \left(
\{w_u\}\cup \{u\}\cup
{\cal X}_u
%\cup \{u\}\cup \{w_u\}
\right)\neq
\emptyset\}\label{creatingdomset}
\end{eqnarray}
can
clearly
%easily
be
%found
constructed
in time polynomial in $|\,V\,|$.
%after one has computed
%${\cal G}$ and $S$.
%satisfies all the required properties.
%Recall that
As
%$B_v\equiv \{w_v\}\cup \{v\}\cup N_G(v)\cup \bigcup_{u\in \{v\}\cup N_G(v)}\, {\cal X}_u$
$B_v= \{w_v\}\cup N_G^*(v)\cup\, (\,\bigcup_{u\in N_G^*(v)}\,
{\cal X}_u),$
%By the definition of $B_v,$
\begin{eqnarray}
B_v\subseteq \bigcup_{u\in
%\{v\}\cup N_G(v)
N_G^*(v)
} \left(\{w_u\}\cup \{u\}\cup {\cal X}_u
%\cup \{u\}\cup \{w_u\}
\right).
\label{enlarge}
\end{eqnarray}
%$v\in V$.
%If $D\cap (\{v\}\cup N_G(v))=\emptyset$ for some $v\in V,$
%then the righthand side of Eq.~(\ref{enlarge}) must have empty intersection
%with $S$.
For each $v\in V,$
%By
Lemma~\ref{nondominatingset}
%implies that
says
$S\cap B_v\neq \emptyset$.
%for every $v\in
%V$;
Hence
%for each $v\in V,$
relation~(\ref{enlarge})
%Eq.~(\ref{enlarge})
%establishes
implies
%that
the existence of a
%$u^*\in \{v\}\cup N_G(v)$
$u^*\in N_G^*(v)$
with
%$S\cap ({\cal X}_{u^*}\cup\{u^*\}\cup \{w_{u^*}\})\neq\emptyset,$
$S\cap (\{w_{u^*}\}\cup \{u^*\}\cup {\cal X}_{u^*})\neq\emptyset,$
%for some $u^*\in \{v\}\cup N_G(v),$
%or
equivalently,
%i.e.,
%or
%yielding
$u^*\in \tilde{D}$.
%$u^*\in D(G^\prime,S)$
%by Eq.~(\ref{creatingdomset}).
Consequently,
%$(\{v\}\cup N_G(v))\cap D\neq\emptyset$
%$D\cap N_G^*(v)\neq\emptyset$
%$D(G^\prime,S)\cap N_G^*(v)\neq\emptyset$
$\tilde{D}\cap N_G^*(v)\neq\emptyset$
for all $v\in V,$
i.e.,
%$D\cap (\{v\}\cup N_G(v))\neq \emptyset,$
%for otherwise the righthand side of
%relation~(\ref{enlarge})
%Eq.~(\ref{enlarge})
%would have an empty
%intersection with $S,$ $v\in V$.
%the existence of a $u\in \{v\}\cup N_G(v)$ with
%$$
%\cup_{u\in \{v\}\cup N_G(v)}
%\left(S\cap \left({\cal X}_u\cup \{u\}\cup \{w_u\}\right)\right)\neq
%\emptyset.$$
%$$
%\left(S\cap \left({\cal X}_{u}\cup \{u\}\cup \{w_{u}\}\right)\right)\neq
%\emptyset.$$
%for each
%$v\in V$.
%This and the definition of $D$ show that $u^*\in D$.
%which in turn implies the existence of a $u\in D$
%by the definition of $D$.
%Consequently,
%So
%$D$
%$D(G^\prime,S)$
$\tilde{D}$
is a dominating set of $G$.

%We have
Now,
%We have shown that $D$ is a dominating set of $G$.
%Therefore,
\begin{eqnarray}
%\text{OPT}\le
%|\,D\,|
\left|\,
%D\left(G^\prime,S\right)
\tilde{D}
\,\right|
=
%\sum_{v\in D}\, 1
\sum_{v\in
%D\left(G^\prime,S\right)
\tilde{D}
}\, 1
\le
%\sum_{u\in D}\,
\sum_{u\in
%D\left(G^\prime,S\right)
\tilde{D}
}\,
%\left|\,S\cap \left({\cal X}_u\cup \{u\}\cup \{w_u\}\right)\,\right|
\left|\,S\cap \left(\{w_u\}\cup \{u\}\cup {\cal X}_u\right)\,\right|
%= \sum_{v\in D} \left|\,S\cap \left({\cal X}_v\cup \{v\}\cup \{w_v\}\right)\,\right|
%\le \left|\,S\cap \left({\cal X}\cup V\cup {\cal W}\right)\,\right|
\le |\,S\,|,\label{justmovingseeds}
\end{eqnarray}
where the first inequality follows from Eq.~(\ref{creatingdomset}).
%and the
%second from the disjointness of
%$(({\cal X}_u\cup \{u\}\cup \{w_u\})$
%and
%$({\cal X}_v\cup \{v\}\cup \{w_v\})$
%for distinct $u,v\in V$.
%This and
%Inequalities~(\ref{justmovingseeds}),
Inequalities~(\ref{computinganirreversibledynamo})--(\ref{relatingthesizes})
and (\ref{justmovingseeds}) yield
\begin{eqnarray}
%|\,D\,|
\left|\,
%D\left(G^\prime,S\right)
\tilde{D}
\,\right|
\le
\left[\,\left(1-2\epsilon\right)\cdot\ln{|\,V\,|}+O(1)\,\right]
\cdot \left(\gamma(G)+2\right),
%\label{sizeboundonD}
\end{eqnarray}
%As
%$$
%\left[\,\left(1-2\epsilon\right)\cdot\ln{|\,V\,|}+O(1)\,\right] \cdot
%\left(\gamma(G)+2\right)
%= \left(1-2\epsilon\right)\cdot \ln{|\,V\,|}\cdot \gamma(G)
%+O\left(\gamma(G)+\ln{|\,V\,|}\,\right),
%$$
%$$\lim_{|\,V\,|\to\infty,\gamma(G)\to\infty}\,
%\frac{\left[\,\left(1-2\epsilon\right)\cdot\ln{|\,V\,|}+O(1)\,\right] \cdot
%\left(\gamma(G)+2\right)}
%{\left(1-\epsilon\right)\cdot \ln{|\,V\,|}\cdot \gamma(G)}
%=\frac{1-2\epsilon}{1-\epsilon}
%<1,$$
%inequality~(\ref{sizeboundonD})
%which implies
implying
%that
%there exists a constant
%which shows
the existence of a constant
%$f(\epsilon)$ depending only on $\epsilon$
$C$
%such that inequality
with
%$\min\{|\,V\,|,\gamma(G)\}>f(\epsilon)$ implies
%For a sufficiently large constant $f(\epsilon)$ depending only on $\epsilon,$
%Eqs.~(\ref{computinganirreversibledynamo})--(\ref{relatingthesizes})
%show that
%$|\,D\,|=o(\ln |\,V\,|)\cdot \text{OPT}$.
%$|\,D\,|$
$|\,
%D(G^\prime,S)
\tilde{D}
\,|
\le (1-\epsilon)\cdot
%(\,
\ln |\,V\,|
%\,)
\cdot \gamma(G)$
%when both $|\,V\,|$ and $\gamma(G)$ are larger than $f(\epsilon)$.
for
%$\min\{\,|\,V\,|,\gamma(G)\,\}>f(\epsilon)$.
$\min\{\,|\,V\,|,\gamma(G)\,\}>C$.

%Having
We have
shown that
(1)
%$D(G^\prime,S)$
$\tilde{D}$
%$D$
can be found in time polynomial in $|\,V\,|,$
(2)
%$D(G^\prime,S)$
$\tilde{D}$
%$D$
is a dominating set of $G$ and
(3)
%$|\,D\,|=o(\ln |\,V\,|)\cdot \text{OPT},$
%$|\,D\,|\le (1-\epsilon)\,(\ln |\,V\,|)\cdot \gamma(G)$
$|\,
%D(G^\prime,S)
\tilde{D}
\,|\le (1-\epsilon)\cdot\ln |\,V\,|\cdot \gamma(G)$
for
%$\min\{\,|\,V\,|,\gamma(G)\,\}>f(\epsilon)$.
$\min\{\,|\,V\,|,\gamma(G)\,\}>C$.
%Furthermore,
%On the other hand,
When
%$\min\{\,|\,V\,|,\gamma(G)\,\}\le f(\epsilon),$
$\min\{\,|\,V\,|,\gamma(G)\,\}\le C,$
a minimum dominating set of $G$ can be found
by brute force
in time polynomial in
$|\,V\,|$.
%implying
%$\text{NP}\subseteq \text{TIME}(n^{O(\ln \ln n)})$
%by
%contradicting
%Consequently,
%{\sc dominating set} has a $$
Hence
%Fact~\ref{inapproximabilitydominatingset}
%, with $\epsilon$ replaced by $3\epsilon,$
%concludes that
$\text{NP}\subseteq
\text{TIME}(n^{O(\ln \ln n)})$ by Fact~\ref{inapproximabilitydominatingset}.
\end{proof}

\comment{ % just says that the general case cannot be easier
An immediate corollary follows.

\begin{corollary}
Let $\epsilon>0$ be any constant.
If {\sc irreversible dynamo (strict majority)} has a polynomial-time,
$((1/2-\epsilon)\ln N)$-approximation algorithm for $N$-vertex graphs,
% for undirected bipartite
%graphs with $N$ vertices and diameter at most $8,$
then $\text{NP}\subseteq \text{TIME}(n^{O(\ln \ln n)})$.
\comment{ % original statement
Given an undirected
graph
%$G(V,E),$
with $N$ vertices,
% and diameter at most $8,$
if there exists a polynomial-time,
%$o(\ln N)$-approximation
$((1/2-\epsilon)\ln N)$-approximation
algorithm
for {\sc irreversible dynamo (strict majority)}, then $\text{NP}\subseteq \text{TIME}(n^{O(\ln \ln
n)})$.
} % original statement
\end{corollary}
\begin{proof}
%Any approximation algorithm for
%{\sc irreversible dynamo (strict majority)}
Immediate from Theorem~\ref{inapproximabilitystrict}.
\end{proof}
} % just says that the general case cannot be easier

Analogous to the strict-majority case,
the following result can be proved for
{\sc irreversible dynamo (simple
majority)}.

\begin{theorem}\label{inapproximabilitysimple}
Let $\epsilon>0$ be any constant.
If {\sc irreversible dynamo (simple majority)} has a polynomial-time,
$((1/2-\epsilon)\ln N)$-approximation algorithm
for
$N$-vertex graphs,
%undirected
%bipartite graphs with $N$ vertices and diameter at most $8,$
then $\text{NP}\subseteq \text{TIME}(n^{O(\ln \ln n)})$.
\comment{ % original statement
Given
an undirected
bipartite
graph
%$G(V,E)$
with $N$ vertices and diameter at most $8,$
if there exists a polynomial-time,
%$o(\ln |\,V\,|)$-approximation
%$o(\ln N)$-approximation
$((1/2-\epsilon)\ln N)$-approximation
algorithm
for {\sc irreversible dynamo (simple majority)},
then $\text{NP}\subseteq \text{TIME}(n^{O(\ln \ln n)})$.
} % original statement
\end{theorem}
\begin{proof}
We will show $\text{NP}\subseteq \text{TIME}(n^{O(\ln \ln n)})$
%given that
if
{\sc irreversible dynamo (simple majority)} has a polynomial-time,
$((1/2-\epsilon)\ln N)$-approximation algorithm
for
bipartite graphs with $N$ vertices and diameter at most $8$.
%[Sketch of proof.]
%Analogous to the development in this section, except that we
By Theorem~\ref{inapproximabilitystrict},
we need only show that
every vertex of
%$G^\prime$
$\boldsymbol{\cal G}$
has an odd degree, so that the strict and the
simple-majority scenarios coincide.
%Then Theorem~\ref{inapproximabilitystrict} completes the proof.
By
%items~\ref{originalvertexfactone}--\ref{originalvertexfactthree} of
Lemma~\ref{easytwo}(\ref{originalvertexfactone})--(\ref{originalvertexfactthree}),
%$\text{deg}_{G^\prime}(v)=2\cdot\text{deg}_G(v)+1$
$\text{deg}_{\boldsymbol{\cal G}}(v)=2\cdot\text{deg}_G(v)+1$
is odd for each
$v\in V$.
By
% items~\ref{domsetguaranteerneighbors}--\ref{sizeofYv} of
Lemma~\ref{easytwo}(\ref{domsetguaranteerneighbors})--(\ref{sizeofYv}),
%$\text{deg}_{G^\prime}(w_v) =\text{deg}_G(v)+|\,N_G^*(v)\,| =2\cdot\text{deg}_G(v)+1$
$\text{deg}_{\boldsymbol{\cal G}}(w_v)
=\text{deg}_G(v)+|\,N_G^*(v)\,| =2\cdot\text{deg}_G(v)+1,$
also
odd for each $v\in V$.
The vertices in
$\{g_1,g_2\},$
${\cal X}$ and ${\cal Y}$ have odd degrees of $1,1$ and $3$
in
%$G^\prime,$
$\boldsymbol{\cal G},$
respectively.
By definition,
%By Fact~\ref{easythree},
$$
%\text{deg}_{G^\prime}(z_1)
%=\left|\,\{g_1\}\cup {\cal Y}\,\right|
%=1+\sum_{v\in V}\, \left|\,{\cal Y}_v\,\right|
%=1+\sum_{v\in V}\, \text{deg}_G(v)
%=1+2\cdot |\,E\,|
\text{deg}_{\boldsymbol{\cal G}}(z_1)
=\left|\,\{g_1\}\cup {\cal Y}\,\right|
=1+\sum_{v\in V}\, \left|\,{\cal Y}_v\,\right|
=1+\sum_{v\in V}\, \text{deg}_G(v)
=1+2\cdot |\,E\,|
$$
is odd, where the last equality
holds because each edge in $E$ is counted twice in $\sum_{v\in V}\,
\text{deg}_G(v)$.
%can be found in most
%books on graph theory.
Finally,
%By symmetry,
%$\text{deg}_{G^\prime}(z_2)$
$\text{deg}_{\boldsymbol{\cal G}}(z_2)$
is odd by symmetry.
%Furthermore,
%It is easy to verify that all other vertices also
%have an odd degree
%in ${\cal G}$.
%As all
%other vertices in ${\cal V}$ are easily verified
%to have an odd degree
%in ${\cal G},$
%So
%the simple-majority scenario coincides with the strict-majority one.
%the strict and the simple-majority scenarios coincide.
\comment{ % obselete, old
Add to $G^\prime$ another
vertex $z^\prime$ and let
%to ${\cal G}$.
$z^\prime$ be adjacent to all vertices in $N_{G^\prime}(z)$.
Then for each $v\in V,$
delete a vertex in ${\cal X}_v$ from $G^\prime$.
The proofs are similar to those developed in this section.
} % obselete, old
\end{proof}

\comment{ % this corollary just says the general case can't be easier
\begin{corollary}
Let $\epsilon>0$ be any constant.
If {\sc irreversible dynamo (simple majority)} has a polynomial-time,
$((1/2-\epsilon)\ln N)$-approximation algorithm for $N$-vertex graphs,
%for undirected bipartite
%graphs with $N$ vertices and diameter at most $8,$
then $\text{NP}\subseteq \text{TIME}(n^{O(\ln \ln n)})$.
\comment{ % original statement
Given
an undirected
graph
%$G(V,E)$
with $N$ vertices,
% and diameter at most $8,$
if there exists a polynomial-time,
%$o(\ln |\,V\,|)$-approximation
%$o(\ln N)$-approximation
$((1/2-\epsilon)\ln N)$-approximation
algorithm
for {\sc irreversible dynamo (simple majority)},
then $\text{NP}\subseteq \text{TIME}(n^{O(\ln \ln n)})$.
} % original statement
\end{corollary}
\begin{proof}
Immediate from Theorem~\ref{inapproximabilitysimple}.
\end{proof}
} % this corollary just says the general case can't be easier

\comment{ % kind of don't want to compare our results with KKT03
Let $\epsilon>0$ be an arbitrary constant.
Given a digraph $G(V,E)$ and $k\in \mathbb{N},$
Kempe, Kleinberg and Tardos~\cite{KKT03}
show a $(1-(1/e)-\epsilon)$-approximation algorithm for finding a set
$S\subseteq V$ of size $k$ maximizing $E[\,c(S,G,\phi)\,],$
where
the expectation is taken over
$\phi(v)$
being $\text{deg}^\text{in}(v)$ multiplied by an independent,
uniformly random member of $[\,0,1\,]$ for each $v\in V$.
%$\rho_v$ being an independent,
%uniformly random member of $[\,0,1\,]$ for each $v\in V$
%and
%$\phi(v)=\rho_v\cdot \text{deg}^\text{in}(v)$.
%and $\rho_v$ is
%a uniformly random member of $[\,0,1\,]$ picked independently for $v\in V$.
In contrast,
Theorems~\ref{inapproximabilitystrict}--\ref{inapproximabilitysimple}
show that
no
$o(\ln |\,V\,|)$-approximations algorithms can exist for
finding the minimum irreversible dynamos, unless
$\text{NP}\subseteq \text{TIME}(n^{O(\ln \ln n)})$.
Given a network ${\cal N}(G,\phi)$ and $k\in \mathbb{N},$
Kempe, Kleinberg and Tardos~\cite{KKT03} also show that no
$n^{1-\epsilon}$-approximation algorithms exist for finding a set $S\subseteq V$
of size $k$ maximizing $|\,c(S,G,\phi)\,|$.
Unlike
Theorems~\ref{inapproximabilitystrict}--\ref{inapproximabilitysimple},
their proof requires $G$ to be directed and $\phi$ to be neither the
strict nor the simple-majority scenario.
} % kind of don't want to compare our results with KKT03

\comment{ % here we argue that ${\cal G}$ has diameter O(1)
We remark that Fact~\ref{inapproximabilitydominatingset}
holds even for an undirected graphs without isolated vertices.
But if $G$ has no isolated vertices,
then the following statements hold for $G^\prime$:
\begin{itemize}
%\addtolength{\itemsep}{-0.2\baselineskip}
\item Every vertex in ${\cal X}$ is adjacent to at least one vertex in
$V$.
\item Every vertex in $V$ is adjacent to at least one vertex in ${\cal W}$.
\item Every vertex in ${\cal W}$ is adjacent to at least one vertex in ${\cal
Y}$.
\item Every vertex in ${\cal Y}$ is adjacent to both $z_1$ and $z_2$.
\end{itemize}
Consequently,
each vertex in $G^\prime$ is connected with $z_i$ by a path of
length at most $4,$ $i\in \{1,2\}$.
So $G^\prime$ has diameter at most $8$.
%then every vertex of ${\cal G}$ has distance at most $4$ from $z_i,$
%$i\in \{1,2\}$:
%every vertex in ${\cal Y}$ has distance one from $z_i$;
%every vertex in ${\cal W}$ has distance one from 
Therefore,
Theorems~\ref{inapproximabilitystrict}--\ref{inapproximabilitysimple}
hold for
} % here we argue that ${\cal G}$ has diameter O(1)
\comment{ % it is actually not submodular
%As a remark,
%many problems similar to {\sc irreversible dynamo (simple majority)} and
%{\sc irreversible dynamo (strict majority)} have
As noted by Peleg~\cite{Pel02}, many variants of the minimum
monopoly problem have
polynomial-time,
$O(\ln|\,V\,|)$-approximation algorithms.
We remark that {\sc irreversible dynamo (strict majority)} and {\sc
irreversible dynamo (simple majority)} are no exceptions.
Given an undirected graph $G(V,E),$
Wolsey's~\cite{Wol82} results imply a
polynomial-time,
%many graph-theoretic problems concerning monopolies have
$O(\ln|\,V\,|)$-approximation algorithm for
{\sc irreversible dynamo (strict majority)}, as easily seen
by taking
$$z(A)=\sum_{v\in c\left(A,G,\phi^\text{strict}\right)}\, \phi^\text{strict}(v)
+\sum_{v\in V\setminus c\left(A,G,\phi^\text{strict}\right)}\, \left|\,N(v)\cap
c\left(A,G,\phi^\text{strict}\right)\,\right|$$
for $A\subseteq V$
in Wolsey's analysis.
The same holds for {\sc irreversible dynamo (simple majority)} by replacing
each occurrence of ``strict'' with ``simple'' in the definition of $z(\cdot)$.
%In fact, the same
%is true for digraphs
%by modifying ``$N(v)$'' to
%``$N^\text{in}(v)$'' in the definition of $z(\cdot)$.
} % it is actually not submodular

\section{Conclusions}
We
improve
%the results of
Chang and Lyuu's~\cite{CL09}
$(23/27)\,|\,V\,|$ upper bound to $(2/3)\,|\,V\,|$
%by showing
%show
%have shown
%upper bounds of
%$0.7732\,|\,V\,|$ and $0.727\,|\,V\,|$
on the
minimum
%sizes
size
of irreversible dynamos under the strict-majority
%and the simple-majority scenarios, respectively.
scenario.
Our technique also gives a $|\,V\,|/2$ upper bound on the minimum size
of irreversible dynamos under the simple-majority scenario.
The upper bound under the strict-majority scenario can be lowered to
$\lceil |\,V\,|/2\rceil$
%$(2/3)\,|\,V\,|$
%and $\lfloor|\,V\,|/2\rfloor,$
%respectively,
for undirected connected graphs.

\comment{ % no more this result
While the
%constants $2/3$ and
$1/2$
constant
cannot be improved,
it is open whether the constants of $0.7732$ and $0.727$
can be lowered for digraphs.
%While the constants $2/3$ and $1/2$
} % no more this result

We have proved inapproximability results on
{\sc irreversible dynamo (strict majority)}
and
{\sc irreversible dynamo (simple majority)}.
An interesting direction of research is to design approximation algorithms
%with better guarantees
for special
types of graphs.
%Despite our inapproximability results do not forbid 

\comment{ % this does not seem to really help
In deriving bounds on the minimum size of irreversible dynamos,
it is
practically meaningful
%ideal
that the obtained bounds be efficiently computable
and nearly tight for all graphs.
%Otherwise one would be
%unaware of the minimum size of irreversible dynamos
%Another desirable property is for the bounds to be tight or nearly tight for
%all graphs.
However,
our inapproximability results imply that polynomial-time computable
bounds can not be tight within any multiplicative factor of $o(\ln |\,V\,|)$
unless
$\text{NP}\subseteq \text{TIME}(n^{O(\ln \ln n)})$.
In other words, one has to sacrifice either efficient computability or
tightness in obtaining bounds on the minimum size of irreversible dynamos.
Our upper bounds of $(2/3)\,|\,V\,|$ and $|\,V\,|/2$ for undirected graphs
are certainly polynomial-time computable, paying the price of being loose in
some graphs.
Nonetheless, these bounds are tight for infinitely many graphs.
It is also
interesting
%worth research
to derive bounds that are nearly tight for all graphs,
sacrificing efficient computability.
} % this does not seem to really help

% Obsolete

% should-be inapproximability result

\appendix
\section{Acknowledgments}
The authors are grateful to
Shou-De Lin
%Kun-Mao Chao,
%Gen-Huey Chen and
%Cheng-Yuan Liou
for his helpful comments and suggestions.

% this results in something weaker

% the old result of $(2/3)\,|\,V\,|$

\bibliographystyle{plain}
\bibliography{arXiv_irrdynamo}

\noindent

\end{document}